\documentclass[10pt,aps,prx,twocolumn,footinbib,superscriptaddress]{revtex4-2}

\pdfoutput=1

\usepackage[utf8]{inputenc} 
\usepackage{amsmath}
\usepackage{amssymb}
\usepackage{xcolor}
\usepackage{graphicx}
\usepackage{dcolumn}
\usepackage{bm}
\usepackage{microtype}
\usepackage{threeparttable}
\usepackage{mathtools}
\usepackage[colorlinks=true]{hyperref}
\usepackage{physics}
\usepackage{comment}
\usepackage{soul} 
\usepackage{bbold} 
\usepackage{ragged2e}
\usepackage{framed}
\usepackage[caption=false]{subfig} 
\usepackage{float} 

\newcommand{\be}{\begin{eqnarray}}
\newcommand{\ee}{\end{eqnarray}}

\newcommand{\cO}{\mathcal O}

\newcommand{\mP}{\mathcal P}

\begin{document}

\title{The quadratic growth of Krylov spread complexity in the BTZ black hole}

\author{Aranya Bhattacharya}
\email{aranya.bhattacharya@bristol.ac.uk}
\affiliation{School of Mathematics, University of Bristol,\\ Fry Building, Woodland Road, Bristol BS8 1UG, UK}
\affiliation{Institute of Theoretical Physics, Jagiellonian University, \\ Łojasiewicza 11, 30-348 Kraków, Poland}

\author{Mario Flory}
\email{mflory@th.if.uj.edu.pl}
\affiliation{Institute of Theoretical Physics, Jagiellonian University, \\ Łojasiewicza 11, 30-348 Kraków, Poland}

\author{Michal P. Heller}
\email{michal.p.heller@ugent.be}
\affiliation{Department of Physics and Astronomy, Ghent University, \\ Krijgslaan 299, 9000 Ghent, Belgium}
\affiliation{Institute of Theoretical Physics, Jagiellonian University, \\ Łojasiewicza 11, 30-348 Kraków, Poland}
\affiliation{Mark Kac Center for Complex Systems Research, Jagiellonian University, 30-348 Cracow, Poland}

\author{Emiliano Rizza}
\email{emiliano.rizza@doctoral.uj.edu.pl}
\affiliation{Institute of Theoretical Physics, Jagiellonian University, \\ Łojasiewicza 11, 30-348 Kraków, Poland}
\affiliation{Jagiellonian University, Doctoral School of Exact and Natural Sciences, \\
Łojasiewicza 11, 30-348 Kraków, Poland}

\author{Tim Schuhmann}
\email{tim.schuhmann@ugent.be}
\affiliation{Department of Physics and Astronomy, Ghent University, \\ Krijgslaan 299, 9000 Ghent, Belgium}

\begin{abstract}
The boundary quantity that captures the growth of black-hole interiors quantified by holographic complexity remains unknown beyond 2d dilaton gravity. We provide a critical analysis of a partition-function construction of Krylov spread complexity for thermofield-double states that provides a dimension-independent boundary
reconstruction from semiclassical holographic
partition functions, while developing the present dynamical and
bulk construction for the BTZ
saddle. In the double-scaled Sachdev–Ye–Kitaev model, where exact and semiclassical results can be compared, we show that the classical limit is reliable only when taken after the complexity has been reconstructed; taking this limit at the level of individual Lanczos coefficients discards essential information. Applying the construction to a large-central-charge two-dimensional conformal field theory  above the Hawking-Page temperature dual to a Ba{\~n}ados–Teitelboim–Zanelli black hole, we find an intermediate departure from early-time
quadratic growth followed by behavior compatible with a return toward
asymptotically quadratic growth, rather than the linear late-time
behavior of the volume and the
standard finite-functional complexity = anything class. We then match this boundary behavior to a generalized complexity = anything bulk object built from an infinite series of extrinsic-curvature invariants. The construction provides a systematic route from black-hole thermodynamics to Krylov dynamics and can naturally be extended to higher-dimensional holographic black holes.
\end{abstract}

\maketitle

\section{Introduction and summary}

Quantum-information observables have provided some of the sharpest probes of quantum gravity within the holographic duality \cite{Maldacena:1997re, Gubser:1998bc, Witten:1998qj}. While the entanglement entropy possesses a precise geometric representation \cite{Ryu:2006bv,Casini:2011kv,Lewkowycz:2013nqa,Dong:2016hjy}, the microscopic meaning of holographic complexity remains unsettled. Geometric candidates, most prominently the maximal volume of a bulk time slice (CV) \cite{Susskind:2014moa, Stanford:2014jda, Susskind:2014rva} and the wider complexity = anything (CAny) \cite{Belin:2021bga, Belin:2022xmt} class, which includes complexity = action~\cite{Brown:2015bva,Brown:2015lvg} and complexity = volume 2.0~\cite{Couch:2016exn}, exhibit an extended regime of late-time linear growth in eternal black holes. Which boundary quantity, if any, these observables compute is not known in general.

A controlled clue has emerged from the relation between the Sachdev–Ye–Kitaev (SYK) model \cite{Sachdev:1992fk,kitaevvideo} and two-dimensional dilaton gravity \cite{Maldacena:2016hyu, Harlow:2018tqv}. In this setting, the growth of the Einstein–Rosen bridge can be related to Krylov spread complexity \cite{Lin:2022rbf,Rabinovici:2023yex, Heller:2024ldz} originally introduced in \cite{Balasubramanian:2022tpr}. The latter employs the mean position of a time-evolved state in the Krylov basis, constructed from a reference state and the Hamiltonian, as a measure of quantum state complexity, and has been an active area of research recently \cite{Parker:2018yvk, Caputa:2021ori, Dymarsky:2021bjq, Kundu:2023hbk, Erdmenger:2023wjg,Bhattacharya:2023zqt, Baggioli:2024wbz, Caputa:2024sux, Jeong:2024jjn, Bhattacharya:2023yec, Das:2024tnw, Basu:2024tgg, Aguilar-Gutierrez:2025kmw,Ambrosini:2024sre,Fu:2025kkh, Angelinos:2025drf,  Ambrosini:2025hvo,  Fatemiabhari:2025cyy, Heller:2025ddj}, see also \cite{Nandy:2024evd,Baiguera:2025dkc, Rabinovici:2025otw} for reviews. This result raises a basic question: Is the relation between Krylov spread complexity and bulk geometry special to two-dimensional gravity, or does it persist in higher-dimensional holography?\\

The existing derivation relies strongly on the transfer-matrix structure of double-scaled SYK (DSSYK) \cite{Berkooz:2018qkz,Berkooz:2018jqr,Berkooz:2024lgq} and has no evident analogue in a generic higher-dimensional holographic field theory. We therefore take a different route. For a thermofield-double (TFD) reference state, all moments of the Hamiltonian are derivatives of the thermal partition function. These moments determine the Lanczos coefficients and hence the Krylov spread complexity \cite{Balasubramanian:2022tpr}. In semiclassical holography, the same partition function is controlled by the on-shell action of the dominant black-hole saddle. This observation provides a direct route from black-hole thermodynamics to boundary Krylov dynamics.\\

We first test this prescription in double-scaled SYK, where both the exact quantum partition function \cite{Erdos:2014zgc,Berkooz:2018qkz} and its semiclassical saddle \cite{Goel:2023svz} are available. Their comparison reveals an essential order-of-limits rule. Although individual Lanczos coefficients possess expansions in the parameter controlling the classical limit, subleading terms in those expansions contribute to the leading classical Krylov complexity through nontrivial cancellations. The classical limit must therefore be taken only after the complexity has been assembled.\\ 

We then apply this prescription to the black-hole saddle of a holographic conformal field theory (CFT) in two dimensions, described by the  Ba{\~n}ados–Teitelboim–Zanelli (BTZ) black hole  \cite{Banados:1992wn}. The resulting early-time series can be continued using the Pad{\'e} method and corroborated with earlier studies \cite{Balasubramanian:2022tpr,Balasubramanian:2026azk} of structurally similar saddle point partition functions. The analysis indicates a nontrivial crossover: The instantaneous
growth exponent departs from its universal early-time value of two
and turns upward toward two over the accessible continuation range. This behavior is qualitatively incompatible with the linear late-time growth of the standard finite-functional CAny class,
including CV.\\

Finally, we construct a new bulk complexity observable by breaking one key assumption of CAny that ensures late-time linear growth, namely finiteness of the evaluated functional on the accumulation surface, while otherwise retaining the CAny structure. Based on this, we find a candidate bulk observable in the BTZ geometry dual to CFT$_2$ Krylov spread complexity in the classical limit. It is evaluated on maximal-volume slices but weighted by an infinite series of extrinsic-curvature invariants whose coefficients are fixed order by order by the boundary complexity early-time series. Every finite truncation ultimately returns to linear growth, whereas the fully resummed infinite series becomes singular on the final slice and thereby can support quadratic asymptotics. Both the partition-function construction and the bulk counterpart should be naturally expected to generalize to higher-dimensional holographic theories.\\

\emph{Note on abbreviations:} Throughout the text, we introduced several abbreviations, and since our audience comes from several disciplines, we explain them for definiteness here. AdS: Anti-de Sitter. BTZ: Ba{\~n}ados–Teitelboim–Zanelli (black hole). CAny: complexity = anything. CV: complexity = volume. CFT: conformal field theory. \emph{CfZ}: complexity from partition function. ERB: Einstein-Rosen-Bridge. JT: Jackiw-Teitelboim gravity. MM: moment method. SYK: Sachdev-Ye-Kitaev model, whereas DSSYK: double-scaled SYK. TFD: thermofield-double state. \\

\emph{Note on time regimes:} We use the following time-regime terminology throughout. ``Short
time'' refers to the Taylor expansion around \(t=0\);
the ``direct-convergence regime'' is the interval inside its
convergence radius; the ``thermal crossover'' occurs at
\(t/\beta=O(1)\); the ``semiclassical late-time window'' denotes
\(1\ll t/\beta\ll c\); ``pre-saturation'' refers more generally to
\(t\ll t_{\mathrm{sat}}\); and ``finite-\(c\) saturation'' refers to the
parametrically later completion of the evolution.

\section{Krylov spread complexity and\\ the moment method}

Krylov spread complexity~\cite{Balasubramanian:2022tpr} is specified by two pieces of data: a reference state $\ket{R}$ and a time-independent Hamiltonian $H$. The repeated action of $H$ on $\ket{R}$ generates an orthonormal Krylov basis $\{\ket{K_n}\}$ through the Lanczos recursion
\begin{align}
    &\ket{A_{n+1}}=(H-a_n)\ket{K_{n}}-b_n\ket{K_{n-1}}\,\nonumber\\&\ket{K_{n}}=b_{n}^{-1}\ket{A_{n}}
\end{align}
with Lanczos coefficients
\begin{align}
    b_n=\braket{A_n}^{1/2}\,,\quad a_n=\bra{K_n}H\ket{K_n}
\end{align}
and initializations $\ket{K_0}=\ket{R}$ and $b_0=0$. If $b_{n=D_\mathcal{K}}$ vanishes at a finite order $D_\mathcal{K}$, the recursion terminates and the Krylov space is finite-dimensional. Otherwise, the Lanczos algorithm generates an infinite Krylov chain.

For the time-evolved state $\ket{\psi(t)}=e^{-i H t}\ket{R}$, Krylov spread complexity is the mean position of the state along this chain,
\begin{align}
    C_K(t)=\sum_{n=0}^{D_\mathcal{K}} n \abs{\bra{K_n}\ket{\psi(t)}}^2\,.
\end{align}
It therefore measures how far Hamiltonian evolution transports the state away from the initial Krylov vector $\ket{K_0}$. Unlike circuit complexity, the first quantity conjectured to be dual to black hole volume, spread complexity
requires no choice of gate set, tolerance, or cost
function: Once $|R\rangle$ and $H$ are specified, the Krylov
basis and $C_K(t)$ are fixed uniquely.\\

The first ingredient central to our construction in this work is the \emph{moment method} (MM). The Hamiltonian moments in the reference state,
$m_k=\bra{R}H^k\ket{R}$, through order $2n+1$ determine the Lanczos data through order $n$~\cite{Balasubramanian:2022tpr},
\begin{align}
   \{m_0,\dots,m_{2n+1}\}\overset{\text{MM}}{\longrightarrow}\{a_0,\dots,a_n\}\,\text{and}\,\{b_1,\dots,b_n\}. 
\end{align}
A finite segment of the Krylov chain is thus encoded in a finite set of moments. This observation will provide the bridge between thermal partition functions and Krylov dynamics.

The second ingredient is the early-time expansion of the spread complexity. For any $H$ and $\ket{R}$, expanding the evolved state in the Krylov basis gives
\begin{align}
    C_K(t)=\sum_{n=1}^{\infty}c_{2n}t^{2n}
\end{align}
around $t=0$. Only even powers of time appear. More importantly, the coefficient $c_{2n}$ depends only on the first $n$ off-diagonal Lanczos coefficients $\{b_1,\dots,b_n\}$ and the first $n$ diagonal coefficients $\{a_0,\dots,a_{n-1}\}$ (for more details on this, see Appendix \ref{app:moment method details}). The lowest three nontrivial orders are
\begin{align}
\label{eq:C_K_early_time_series_explicitly}
   &C_K(t)= b_1^2 t^2-\frac{b_1^2 t^4}{12}  \left[2(2b_1^2- b_2^2)+(a_0-a_1)^2\right]\nonumber\\&+\frac{b_1^2t^6}{360}  \Big[a_0^2 \left(6 a_1^2+8 b_1^2+b_2^2\right)+a_0 \big(-a_1 \left(16 b_1^2+3 b_2^2\right)\nonumber\\
   &\quad\quad\quad\quad+a_2 b_2^2-4 a_1^3\big)+7 a_1 a_2 b_2^2+a_1^2 \left(8 b_1^2-2 b_2^2\right)\nonumber\\
   &\quad\quad\quad\quad+2 \left(b_2^2 \left(-2 a_2^2-7 b_2^2+3 b_3^2\right)+8 b_1^4+b_2^2 b_1^2\right)\nonumber\\
   &\quad\quad\quad\quad+a_0^4-4 a_1 a_0^3+a_1^4\Big]+\mathcal{O}(t^8)\,.
\end{align}
The particular combination of sums and differences in \eqref{eq:C_K_early_time_series_explicitly} will be essential below. In the presence of a parameter that warrants a semiclassical limit, such as a large central charge for example, individual Lanczos coefficients can contain parametrically large contributions that cancel only after the coefficients are combined into $c_{2n}$. Consequently, terms that appear subleading at the level of $a_n$ and $b_n$ can contribute to the leading classical spread complexity. This cancellation structure is the origin of the order-of-limits problem investigated in the following sections.

The Lanczos data also determine the large-order behavior of the early-time coefficients and, through the asymptotic limit of the ratio of these coefficients, the convergence radius $R$ of the series. For a series written as
\(C_K(x)=\sum_{n\geq1}c_{2n}x^{2n}\), if
\[
\rho=\lim_{n\to\infty}
\left|\frac{c_{2n}}{c_{2n+2}}\right|,
\]
then \(\rho\) is the convergence radius in the variable \(x^2\),
whereas the radius in \(x\) is \(R_x=\sqrt{\rho}\). Depending on the Krylov problem, this scale can be zero, finite, or infinite. In practice, we will compute a large but finite number of coefficients $c_{2n}$, use them to determine the dynamics within the convergence domain, and employ them as input for the continuation methods developed later. The combination of the moment method and the finite-data property of the early-time series is the technical basis of our construction.

\section{Krylov spread complexity from the partition function}

For holographic eternal black holes, the natural reference state is the thermofield-double state $\ket{TFD(\beta)}$ associated with the two asymptotic boundaries~\cite{Maldacena:2001kr}. For this choice of reference state, the thermal partition function directly generates the Hamiltonian moments that enter the Lanczos construction~\cite{Balasubramanian:2022tpr}:
\begin{align}
\label{eq:momentsfromZ_generalformula}
m_k(\beta)=\bra{TFD(\beta)}H^k\ket{TFD(\beta)}=\frac{(-\partial_\beta)^k Z(\beta)}{Z(\beta)}\,.
\end{align}
A derivation of this is given in Appendix~\ref{app:TFD-moments}. The inverse temperature $\beta$ of the reference state is thereby inherited by the moments and, through the moment method, by the Lanczos coefficients $a_n(\beta)$ and $b_n(\beta)$. These coefficients in turn determine the temperature-dependent Krylov spread complexity $C_K(t)_\beta$.

Combining these steps gives the reconstruction that underlies this work, which we refer to as \emph{Krylov spread complexity from the partition function}, short \textit{CfZ}, originally established by \cite{Balasubramanian:2022tpr}:
\begin{align}
\label{eq.CfZ}
    \textit{CfZ: }Z(\beta)\rightarrow m_n(\beta)\rightarrow a_n(\beta),b_n(\beta)\rightarrow C_K(t)_\beta\,.
\end{align}
When the exact partition function is known, each arrow in this chain is exact. A nontrivial question arises when $Z(\beta)$ is known only asymptotically. This is precisely the situation in semiclassical holography, where the partition function is typically determined by the on-shell action of a dominant gravitational saddle in the classical limit.
Although the individual ingredients of 
\textit{CfZ} are well-established, its use in semiclassical holography raises an essential structural question: The successive steps in \textit{CfZ} involve arbitrarily high derivatives of the partition function and nonlinear combinations of the resulting moments and Lanczos coefficients. Taking an asymptotic limit at an intermediate stage of the construction therefore need not give the same answer as completing the construction first and taking the limit only at the level of $C_K(t)_\beta$.\\

\textbf{Our first objective} in this work is therefore to determine the correct order of operations in the context of a semiclassical limit. We formulate this problem through the question:
\begin{itemize}
    \item Suppose that $Z(\beta)$ depends on a parameter controlling the approach to a classical limit. At which stage of \textit{CfZ} can or must this limit be taken in order to retain all contributions to the leading classical Krylov spread complexity?
\end{itemize}

The double-scaled SYK model provides a controlled benchmark for this question. In this model, the exact quantum partition function, its semiclassical saddle, and an independently established holographic dictionary for Krylov complexity are simultaneously available~\cite{Lin:2022rbf,Rabinovici:2023yex,Heller:2024ldz}. 
A satisfactory prescription must explain how these results, in the regimes in which they are applicable, emerge from the semiclassical partition function.

The analysis below will show that the semiclassical saddle does contain sufficient information to determine the leading classical Krylov spread complexity only when the classical limit is taken after the complexity has been assembled. Truncating the expansion of the individual Lanczos coefficients can discard terms that participate in the cancellations described in the preceding section.\\

\textbf{Our second objective} is to apply the resulting prescription to partition-function saddles describing holographic black holes. The principal application we choose to investigate in this work is the high-temperature thermofield-double state of a two-dimensional holographic CFT, whose bulk dual is the eternal BTZ geometry. This takes the construction beyond two-dimensional bulk gravity into three bulk dimensions. It yields a quantitative boundary prediction for the Krylov spread complexity, including the early-time series from which its intermediate- and late-time behavior can be reconstructed, and thereby supplies a concrete target for a bulk dual. The partition-function construction itself is not restricted to the BTZ case and extends naturally to higher-dimensional holographic black holes, which we defer to future work.

\section{Method validation: DSSYK}

We begin our investigations by establishing validity of the \textit{CfZ} method~\eqref{eq.CfZ} in a controlled example. We choose for this reason application to the double-scaled SYK model \cite{Berkooz:2018qkz,Berkooz:2018jqr,Berkooz:2024lgq}, where we have access to the full quantum partition function, as well as its semiclassical saddle. The goal of this section is to understand \textit{CfZ} applied to the semiclassical saddle, benchmarked by the known precise holographic dictionary in the double-scaled SYK model. As a byproduct of this, we work out novel analytic results for classical Krylov spread complexity of finite temperature reference states in this framework and emphasize again the earlier result by two of us~\cite{Heller:2024ldz} that the match between the Jackiw-Teitelboim (JT) gravity volume and Krylov spread complexity~\cite{Lin:2022rbf,Rabinovici:2023yex} requires taking $|TFD(\beta = 0)\rangle$ rather than $|TFD(\beta)\rangle$ as $|R\rangle$.

\subsection{Framework}

\subsubsection{The double-scaled SYK model}

The main model of study in the first part of this work is~SYK \cite{Sachdev:1992fk,kitaevvideo} in its double-scaling limit \cite{Berkooz:2018qkz,Berkooz:2018jqr,Berkooz:2024lgq}. The SYK model is a $(0+1)$-dimensional quantum system of $N$ Majorana fermions with $p$-local interactions, governed by the Hamiltonian
\begin{align}
    H_{\mathrm{SYK}}=
i^{p/2} \sum_{1 \leq i_1 < \cdots < i_p \leq N} J_{i_1 \cdots i_p} \psi_{i_1} \cdots \psi_{i_p} \,.
\end{align}
The couplings $J_{i_1 \cdots i_p}$ are assumed to be Gaussian random with zero mean and appropriate variance, parametrized by a number $J$ which we set to one for convenience and that can be restored through dimensional analysis if needed \footnote{Restoring the coupling scale, the leading half-integer
behavior takes the dimensional form
\(b_n\sim J/\sqrt{|\log q|}\). This estimate makes the overall scale
transparent, but does not explain the full Laurent hierarchy or the
cancellations that occur after the complexity is assembled.}. In the double-scaled SYK (DSSYK) limit, where we send both $p$ and $N$ to infinity while keeping the ratio $\abs{\log q}=2p^2/N$ with $0\leq q\leq1$ fixed, this model is analytically solvable via combinatorial techniques \cite{Berkooz:2018qkz,Berkooz:2018jqr,Berkooz:2024lgq}. These make use of an auxiliary quantum system, namely a Hilbert space $\{\ket{n}:n\in\mathbb{N}_0\}$ of orthonormal states with $n$ open chords (representing a Wick contraction in the original theory) and a transfer matrix $T$ that allows one to create and annihilate them $2\sqrt{\abs{\log q}}\,T\ket{n}=\sqrt{[n+1]_q}\ket{n+1}+\sqrt{[n]_q}\ket{n-1}$ weighted by the $q$-deformed numbers $[n]_q=(1-q^n)/(1-q)$.\\
In particular, coupling-averaged even moments of the Hamiltonian in the infinite-temperature TFD state of the original SYK theory are computed as moments of $T$ in the zero-chord state $\ket{0}$ as $m_{2k}=\langle \Tr (H_{\mathrm{SYK}})^{2k}\rangle_J=\bra{0}T^{2k}\ket{0}$
while odd moments vanish. Note that this is the reason why commonly, by slight abuse of notation, the zero-chord state $\ket{0}$ is called the infinite-temperature TFD state of the auxiliary chord theory. This result for the moments allows us to express the partition function of DSSYK as
\begin{align}
    Z_\mathrm{DSSYK}=\langle \Tr (e^{-\beta H_\mathrm{SYK}})\rangle_J=\bra{0}e^{-\beta T}\ket{0}\,.
\end{align}
A closed form analytic expression can be obtained by diagonalizing $T$ \cite{Berkooz:2018qkz} and reads in our conventions (which follow \cite{Blommaert:2024ydx} up to the replacement $q^2\to q$)
\begin{align}
\label{eq:Z_DSSYK}
    Z_\mathrm{DSSYK}=\int_0^{\pi}d\theta\, \abs{(e^{2i\theta},q)_\infty}^2 e^{\frac{\beta \cos \theta}{\abs{\log q}}}
\end{align}
where $(x;q)_n$ denotes the $q$-Pochhammer symbol with $(x;q)_\infty=\lim_{n\to\infty}(x;q)_n$ and $\theta\in[0,\pi]$ parametrizes the continuous spectrum of eigenvalues $E(\theta)$ of $T$.

\subsubsection{DSSYK semiclassics}

Essential to our investigation now are the insights of \cite{Goel:2023svz} into the semiclassical limit of DSSYK. We restate their key results here. The core claim is that at all energies $E(\theta)$ there exists a semiclassical approximation of DSSYK governed by the coupling $\abs{\log q}$. In the limit $q\to1$, integrands become sharply peaked and are dominated by a saddle point. On this saddle point, the DSSYK temperature $\beta$ is related to the energy parametrized by $\theta$
\begin{align}
\label{eq:dssyk_beta}
    \beta=\frac{2\pi-4\theta}{\sin \theta}\,.
\end{align}
Notice that this cannot be solved in closed form for $\theta(\beta)$. The semiclassical limit $q\to1$ at general $\theta$ of the partition function $Z_\mathrm{DSSYK}(\beta)$ is given by \cite{Goel:2023svz}  
\begin{align}
\label{eq:Zsc}
    Z_\mathrm{sc}=\exp \left[\frac{-2}{\abs{\log q}}\left(\left(\theta-\frac{\pi}{2} \right)^2 +2\left(\theta-\frac{\pi}{2} \right) \cot \theta  \right)\right]
\end{align}
in agreement with the earlier results of \cite{Maldacena:2016hyu}.

\subsubsection{Wormhole length as DSSYK Krylov complexity}

A decade ago, it was realized that the low-energy sector of SYK at fixed $p$ and large $N$ is governed by Schwarzian quantum mechanics \cite{kitaevvideo,Maldacena:2016hyu} and is thereby dual to JT gravity~\cite{Harlow:2018tqv}, see also~\cite{Mertens:2022irh} for a review. That this holds true also for the low-energy sector of the double-scaled model was shown in \cite{Berkooz:2018qkz}. In a recent series of works, an all-energy bulk dual of DSSYK was proposed \cite{Lin:2022rbf,Blommaert:2023opb,Blommaert:2024ydx,Blommaert:2024whf,Blommaert:2025avl}. This dual dilaton gravity theory has a sine potential for the dilaton, is therefore called sine dilaton gravity, and contains the duality between JT gravity and low-energy DSSYK as a limit.\\
For the purpose of this work, we are merely interested in one simple geometric aspect of this rich framework, namely the classical length of the Einstein-Rosen-Bridge (ERB), the spacelike wormhole that extends between the two asymptotic boundaries of the dual double-sided eternal black hole geometry. This is CV in two dimensions. Hence, we will not have to introduce JT and sine dilaton gravity at quantum level, but just mention the classical limit as it will be analogous to our target in higher dimensions. For our purposes, the key statement of~\cite{Blommaert:2024ydx}  is that there is an effective AdS$_2$ black hole geometry at Hawking temperature $\beta_\mathrm{BH}=2\pi/\sin\theta$ 
\begin{align}
\label{eq:ds^2}
    \mathrm{d}s^2_\mathrm{eff}=-\left(\rho^2-\sin(\theta)^2 \right)\mathrm{d} t_s^2+\frac{\mathrm{d}\rho^2}{\rho^2-\sin(\theta)^2},
\end{align}
that governs sine dilaton gravity in its classical limit. The horizon radius is set by $\theta$ in the DSSYK temperature \eqref{eq:dssyk_beta}. Following the arguments in \cite{Blommaert:2024ydx}, the expectation value of the classical wormhole length in sine dilaton gravity $L_\mathrm{SD}$ is just the ERB length in this geometry. Hence, following for example the well-known JT derivation \cite{Harlow:2018tqv}, the classical sine dilaton wormhole length computed in the Kruskal extension of the geometry reads
\begin{align}
\label{eq:L_SD}
    L_\mathrm{SD}=2 \log \left(\cosh \left(t\sin (\theta )/2 \right)\right)-2\log (\sin (\theta ))\,,
\end{align}
which at late times grows linearly in $t/\beta_{\mathrm{BH}}$. The low temperature limit $\theta\ll1$ of this result reproduces the known answer for the wormhole length in JT gravity
\begin{align}
\label{eq:L_JT}
    L_\mathrm{JT}=2 \log \left(\cosh \left(t \theta /2 \right)\right)-2\log (\theta)\,.
\end{align}
In this limit, $\beta_{BH}\approx\beta$ and hence we find the length transition from quadratic to linear at $t/\beta$ of $\mathcal{O}(1)$. In the high temperature limit $\theta=\pi/2$ where $\beta\to0$ and $\beta_{BH}=2\pi$ the result is
\begin{align}
\label{eq:L_beta_zero}
    L_\mathrm{\beta=0}=2 \log \left(\cosh \left(t /2 \right)\right)\,.
\end{align}
These expressions will be used to benchmark Krylov spread complexity predictions from the semiclassical DSSYK partition function. This is a benchmark for holographic complexity, because in \cite{Heller:2024ldz} two of us identified the expectation value of the wormhole length in sine dilaton gravity (this is CV in two dimensions) at finite temperature with Krylov spread complexity of transfer matrix evolution from the zero chord state with analytically continued time $t\to t+i\beta/2$. While this result was previously established making use of the analytic structure of the DSSYK two-point function, we will now investigate which part of this identification, if any, can be reproduced via the moment method from $Z_\mathrm{DSSYK}$ \eqref{eq:Z_DSSYK} and its semiclassical approximation $Z_\mathrm{sc}$  \eqref{eq:Zsc} introduced in the previous paragraphs.

\subsection{Complexity from the semiclassical saddle} 

\subsubsection{Setup and comparison to previous works}

Let us study DSSYK Krylov spread complexity for finite-temperature reference states via \textit{CfZ}. Concretely, we study the Krylov problem defined by the two ingredients
\begin{align}
\label{eq:dssyk_krylov_problem}
    H=T\quad\text{and}\quad \ket{R}=Z(\beta)^{-1/2}e^{-\beta T/2}\ket{0}.
\end{align}
The Hamiltonian that generates time evolution of the state $\ket{\psi(t)}=e^{-iHt}\ket{R}$ is the DSSYK transfer matrix $T$ and the reference state is the DSSYK version of the finite-temperature TFD state. Moments of the Hamiltonian in the reference state and from that Lanczos coefficients via the moment method can be obtained following \eqref{eq:momentsfromZ_generalformula} as
\begin{align}
    m_{k}=\frac{(-\partial_\beta)^k Z(\beta)}{Z(\beta)}\quad\overset{\text{MM}}{\Rightarrow}\quad a_n(\beta)\,,\ b_n(\beta)\,.
\end{align}
with the idea of investigating what Lanczos coefficients and Krylov spread complexity characteristics we find depending on whether we use the full partition function \eqref{eq:Z_DSSYK} or its semiclassical saddle \eqref{eq:Zsc}.\\

\begin{figure}[htb]

\includegraphics[width=0.9\columnwidth]{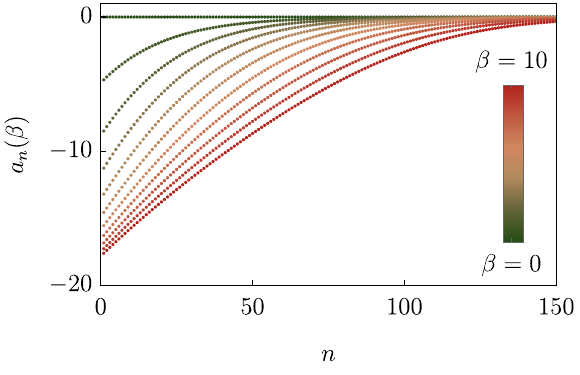}
  \includegraphics[width=0.9\columnwidth]{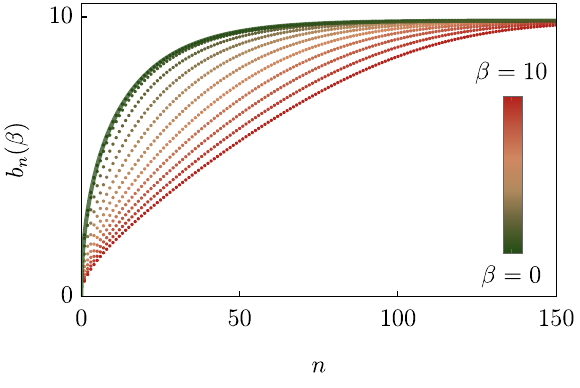}%
    
\caption{Lanczos coefficients $a_n(\beta)$ (top) and $b_n(\beta)$ (bottom) for Krylov spread complexity of finite-temperature TFD state in DSSYK: analytic predictions at $\beta=0$ (solid line) and numerics at $\beta$ (dots) in steps of one from zero to ten. All data are for \(q=0.95\). In both coefficients, an emergent linear regime at intermediate $n$ is visible for nonzero $\beta$.} 
\label{fig:state_of_art_lanczos}
\end{figure}

Let us highlight two important works \cite{Rabinovici:2023yex,Balasubramanian:2024lqk} on the Lanczos coefficients for this setup that have previously appeared, that are methodically independent from deriving complexity from the partition function, and that combine to the current state of the art for spread complexity Lanczos coefficients in DSSYK~\footnote{Lanczos coefficients of related models, for example Random Matrix Theory, are widely studied and well-known. An initial analysis appeared in \cite{Balasubramanian:2022tpr} and for further works we refer to the excellent reviews \cite{Nandy:2024evd,Baiguera:2025dkc,Rabinovici:2025otw} and the references on Krylov spread complexity of TFD states therein.}. From the pioneering work in~\cite{Rabinovici:2023yex} it is known that at infinite temperature $\beta=0$ for DSSYK, without any semiclassical approximation and hence valid for any $0< q<1$, one finds
\begin{align}
\label{eq:an_bn_fullDSSYK_betaZero}
    a_n(\beta=0)=0\,,\quad b_n(\beta=0)=\frac{1}{2}\sqrt{\frac{[n]_q}{\abs{\log q}}}\,.
\end{align}
As discussed in \cite{Heller:2024ldz}, the associated Krylov spread complexity $C_K(t)_{\beta=0}$ to these coefficients does, in the appropriate classical limit that involves rescaling the complexity by $\abs{\log q}$ to keep it finite, match the sine dilaton gravity ERB length at $\beta=0$ given in \eqref{eq:L_beta_zero},
\begin{align}
\label{eq:CK_betazero_equals_Lbetazero}
    \lim_{q\to1}\abs{\log q}C_K(t)_{\beta=0}=2 \log \left(\cosh \left(t /2 \right)\right)=L_{\beta=0}\,.
\end{align}
Furthermore, numerical evaluations of $a_n(\beta)$ and $b_n(\beta)$ for full DSSYK at any $q$ have already appeared in \cite{Balasubramanian:2024lqk} (cf. lower row of Fig.~2 therein). As compared to the infinite temperature result \eqref{eq:an_bn_fullDSSYK_betaZero}, these numerical results show that $a_n(\beta)$ acquires a temperature dependent negative onset and asymptotes to zero with an intermediate linear growth that is more pronounced for larger $\beta$. For the $b_n(\beta)$ coefficients, larger $\beta$ decreases their onset, also makes an intermediate regime of linear growth appear, and leaves the finite constant saturation value unaffected. For completeness, we summarize these known results in Fig.~\ref{fig:state_of_art_lanczos} for $q=0.95$, employing our own numerics independent from the moment method, namely an explicit run of the Lanczos algorithm with $H$ and $\ket{R}$ specified in \eqref{eq:dssyk_krylov_problem} as input expressed in the chord basis with truncation of its dimension that is sufficiently large to not affect the displayed regime. We choose $q$ close to one because $q\to1$ is the regime of DSSYK that we will be interested in for the rest of this work.

Making use of the moment method numerically applied to the full partition function $Z_\mathrm{DSSYK}(\beta)$ given in \eqref{eq:Z_DSSYK}, we can precisely reproduce these results, serving as a consistency check between both numerical implementations.

There are two intertwined knowledge gaps concerning the state of the art that we now illuminate analytically in the following parts:
\begin{itemize}
    \item We have predictions for full DSSYK governed by $Z_\mathrm{DSSYK}$. How are these affected by the semiclassical approximation \eqref{eq:Zsc}? 
    \item Can we find analytic expressions for $a_n(\beta)$, $b_n(\beta)$ in the finite-temperature case that are robust in the classical limit $q\to1$, at least for some $n$? At the level of $C_K(t)_\beta$, this amounts to finding a robust complexity early-time series \eqref{eq:C_K_early_time_series_explicitly} analytically in the $q\to1$ limit.
\end{itemize}
To tackle both of these points, the early-time expansion of Krylov complexity \eqref{eq:C_K_early_time_series_explicitly} plays a key role for us and one goal of this work is to highlight its power and implications.

\subsubsection{From the early-time series for full $Z_{\mathrm{DSSYK}}$ at $\beta=0$ in the classical limit to late times}
\label{sec:warmup_resumming_dssyk_betazero}

As a first illustration of the methods we will employ for the rest of this work, notice that we could have also gained intuition for the conclusion \eqref{eq:CK_betazero_equals_Lbetazero} that $\lim_{q\to1}\abs{\log q}C_K(t)_{\beta=0}$ of DSSYK equals $L_{\beta=0}$ of sine dilaton gravity by studying the early-time series \eqref{eq:C_K_early_time_series_explicitly} derived from the Lanczos coefficients \eqref{eq:an_bn_fullDSSYK_betaZero}. Generically, the early-time series is a considerably simpler object to obtain than the full complexity result, yet contains a lot of information. Explicitly, plugging \eqref{eq:an_bn_fullDSSYK_betaZero} into the early-time expansion of Krylov spread complexity \eqref{eq:C_K_early_time_series_explicitly} and taking the same classical limit as before that required multiplying by a factor of $\abs{\log q}$ to keep the result finite, we find
\begin{align}
\label{eq:series_ckfull_betazero}
    \lim_{q\to1}\abs{\log q}C_K(t)_{\beta=0}=\frac{t^2}{4}-\frac{t^4}{96}+\frac{t^6}{1440}-\frac{17\, t^8}{322560}\nonumber\\+\frac{31\, t^{10}}{7257600}-\frac{691\, t^{12}}{1916006400}+\frac{5461 \,t^{14}}{174356582400}+\dots
\end{align}
The radius of convergence of this early-time series is $R=\pi$. Hence, naively, we cannot make statements for $t>\pi$. But let us now present two approaches to go beyond early times starting purely from a large but finite number of terms of the early-time series.\\

Firstly, we could recognize the coefficients $c_{2n}^{(\beta=0)}$ of this series in $t^{2n}$ as
\begin{align}
\label{eq:log_cosh_series_coeffs_cn}
    c_{2n}^{(\beta=0)}=\frac{\left(2^{2 n}-1\right) B_{2 n}}{n (2 n)!}\,,
\end{align}
where $B_n$ are the Bernoulli numbers. This then allows us to perform the resummation of the infinite sum
\begin{align}
    \lim_{q\to1}\abs{\log q}C_K(t)_{\beta=0}=\sum_{n=1}^{\infty}c_{2n}^{(\beta=0)}t^{2n}=2 \log \left(\cosh \left(t /2 \right)\right)\,,
\end{align}
which recovers $L_{\beta=0}$ of sine dilaton gravity purely from an early-time analysis.\\

\begin{figure}[htb]
    \centering
    \includegraphics[width=0.95\columnwidth]{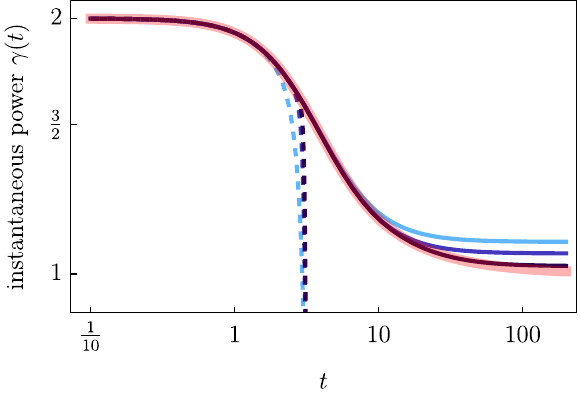}
    \caption{Instantaneous power $\gamma(t)=t \partial_t \log C_K(t)$ of the complexity early-time series in the $q\to1$ limit up to order \{8,24,40\} in $t$ (dashed; light to dark blue) and its diagonal Padé approximations of order \{4,12,20\} (solid; light to dark blue) compared to the exact answer $\gamma(t)=t\,\tanh{(t/2)}/(2\log \cosh (t/2))$ (light red). We see rapid convergence of the Padé approximation towards the exact answer.}
    \label{fig:log_cosh_pade}
\end{figure}

Alternatively, instead of this analytic resummation, we could have also arrived at the conclusion of behavior proportional to $\log \cosh (t/2)$ by analytically continuing the early-time series \eqref{eq:series_ckfull_betazero} via a Padé approximation
\begin{align}
\label{eq:pade_approx_general}
    s(t)=\sum_{k=0}^{M+N}\alpha_k t^k\quad\overset{\text{Padé}}{\Rightarrow}\quad p(t)=\frac{\sum_{k=0}^{M}\rho_k t^k}{\sum_{k=0}^{N}\sigma_k t^k}\,.
\end{align}
To be precise, we carry out a diagonal Padé approximation $M=N$, not of the early-time series \eqref{eq:series_ckfull_betazero} itself, but of the early-time series of its instantaneous power $\gamma$ defined as 
\begin{align}
    \gamma(t)=t\frac{d}{dt}\log C_K(t)=\frac{t\, C_K(t)'}{C_K(t)}\,.
    \label{gammafunction}
\end{align} 
The results are given in Fig.~\ref{fig:log_cosh_pade} and we see rapid convergence towards the exact instantaneous power $\gamma(t)=t\,\tanh{(t/2)}/(2\log \cosh (t/2))$. \\

This method of analytical continuation has a crucial upshot for its predictive ability: in a Padé approximation, the qualitative late-time behavior is fixed by hand, specified by the order of the numerator $M$ and denominator $N$ polynomial we choose in the Padé approximation, as the late-time behavior of the Padé approximant \eqref{eq:pade_approx_general} is by definition $p(t)\sim\rho_M\sigma_N^{-1}t^{M-N}$ for $t\gg1$. In our case, we opt for a diagonal Padé approximation ($\gamma(t)\rightarrow const. $ as $t\rightarrow\infty$), because we want to investigate consistency with a power law growth of $C_{K}(t)$ at late times, as typical for such complexities. Crucially, specifying that the instantaneous power is a constant at late times fixes the complexity to follow power law growth in that limit, but not the numerical value of the power law exponent. Hence, such a Padé approximation of $\gamma$ is a key tool for the remainder of this work that lets us decide between late-time linear and quadratic growth as the approximation picks this coefficient by itself, whenever we find sufficient convergence, purely from the early-time series as input.\\

This concludes the brief pedagogical example reviewing the utility of the early-time expansion, and how to find or approximate the full result at all times from this series. In this specific case we outlined, other strong analytical methods exist that make it unnecessary to resort to these approaches. What we establish in the next section is that one can make constructive progress with these methods based on the early-time series in other examples, where we only have access to partition function saddles and no other available methods. 

\subsubsection{The classical partition function and Lanczos coefficients}

Overall, we are interested in how to make meaningful spread complexity predictions from classical saddles. Hence, in this part, we study the Lanczos coefficients that arise from $Z_{\mathrm{sc}}$ by application of \textit{CfZ}. Because of its full analytic solvability, DSSYK is an invaluable tool to benchmark such an analysis: We can compare to the outcome of the calculation without approximation where we take the saddle point limit only as the very last step.\\

Before we begin, let us mention for completeness one more limit other than the semiclassical limit to be careful about in general in such an analysis: When applying \textit{CfZ} to a partition function, the first step is obtaining moments by taking derivatives with respect to the inverse temperature. This implies immediately that limits in temperature, and in particular keeping only a finite number of terms in a temperature expansion of the full temperature dependence of the partition function, need not commute with \textit{CfZ}. 
Generically, taking limits in temperature at the level of the partition function instead of the assembled complexity will alter the results. For example, for the DSSYK semiclassical partition function $Z_\mathrm{sc}$ in \eqref{eq:Zsc}, we find that expanding the exponent around $\beta\to0$ and keeping only the first $k_\mathrm{cut}$ subleading terms in $\beta$ alters the complexity early-time series prediction starting from order $t^{2(k_\mathrm{cut}+2)}$ when comparing to the prediction without temperature approximations. If instead we are interested in $\beta\to\infty$ and again keep only the first $k_\mathrm{cut}$ subleading terms in $\beta$, we instead find that as long as $k_\mathrm{cut}>1$, we reproduce the expected result at all orders. The details of this investigation on the example of semiclassical DSSYK is outlined in Appendix \ref{sec:app_temp_limits}. It clearly highlights the peculiarities associated with taking finite truncations in a temperature expansion in the partition function. Ultimately, it has to be checked case-by-case whether a certain temperature limit or truncation at the level of the partition function commutes with \textit{CfZ}. For the sake of this work, we will work with the partition function \eqref{eq:Zsc} that entails the full temperature dependence so we do not have to worry about this issue for the rest of this section.
\\ 

Let us now begin our study by applying \textit{CfZ} to the semiclassical partition function $Z_\mathrm{sc}$ given in \eqref{eq:Zsc}. We get analytic results for $a_n^{(\mathrm{sc})}(\beta)$ and $b_n^{(\mathrm{sc})}(\beta)$ for several lowest lying coefficients on a contemporary laptop. The first ones are
\begin{equation}
\label{eq:lanczos_semicl_prediction}
    \begin{aligned}
    &a_0^{(\mathrm{sc})}(\beta)=-\frac{\cos (\theta )}{\abs{\log q}}\,,\\
    &b_1^{(\mathrm{sc})}(\beta)=\sqrt{\frac{\sin ^2(\theta )}{2((\pi -2 \theta ) \cot (\theta )+2) \abs{\log q}}}\,,
    \end{aligned}
\end{equation}
where we express temperature dependence in $\theta$ for convenience. We remind the reader that for full DSSYK, analytic results for $a_n(\beta)$ and $b_n(\beta)$ are not known. \textit{CfZ} applied to $Z_\mathrm{sc}$ hence makes novel predictions for the finite temperature Lanczos coefficients of DSSYK in the semiclassical limit. Computationally, the restriction to the semiclassical saddle makes it much more tractable to find the Lanczos coefficients. The caveat is to understand whether these predictions are meaningful. Hence, let us investigate this question in detail, divided into three key aspects \textbf{i)} Comparison to DSSYK numerics, \textbf{ii)} Analytic temperature behavior of the classical early-time series, and \textbf{iii)} Analytic structure of Lanczos coefficients and early-time complexity as series in $\abs{\log q}$.\\

\begin{figure}[htb]
    \centering
    \includegraphics[width=0.95\columnwidth]{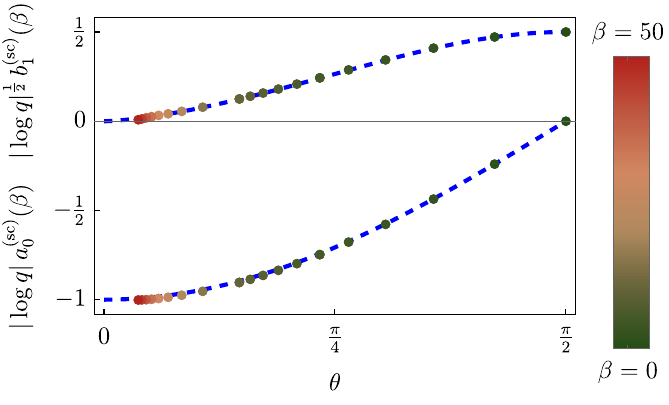}
    \caption{Comparison of Lanczos coefficients $a_0(\beta)$ and $b_1(\beta)$ (normalized by appropriate powers of $\abs{\log q}$) from numerics at various temperatures $\beta(\theta)$ (dots) with analytical predictions $a_0^{(\mathrm{sc})}(\beta)$ and $b_1^{(\mathrm{sc})}(\beta)$ (blue dashed) from the semiclassical partition function of DSSYK via \textit{CfZ}. All data are for \(q=0.95\). The inverse temperature $\beta$ goes from zero to ten in steps of one and from ten to fifty in steps of five.}
    \label{fig:semclass_lanzos_formula_vs_numerics}
\end{figure}

\textbf{i) Comparison to DSSYK numerics.} We begin by comparing these Lanczos coefficient predictions \eqref{eq:lanczos_semicl_prediction} from the semiclassical partition function to the well-established Lanczos numerics for DSSYK with $q$ close to one that we have already introduced previously. This can give us a first understanding whether the classical limit commutes with $\textit{CfZ}$. The result is displayed in Fig.~\ref{fig:semclass_lanzos_formula_vs_numerics}.
We see that the numerical predictions at all temperatures follow the semiclassical predictions well. This is a good first hint that there is meaning to the prediction of Lanczos coefficients and spread complexity from $Z_{\mathrm{sc}}$ in the semiclassical limit. The slight deviation of the numerical data from the semiclassical prediction at larger $\beta$ is due to the fact that we work at $q=0.95$ and finite artificial truncation $\mathcal{D}_\mathcal{K}=n_\mathrm{cut}$ of the Lanczos basis when executing the numerics. The accuracy of the match can be increased by increasing $n_\mathrm{cut}$ and setting $q$ even closer to one. In any case, we do not want to rely only on numerics to claim that \textit{CfZ} commutes with the semiclassical limit. Let us therefore continue with analytics.\\

\textbf{ii) Analytic temperature behavior of the classical early-time series.} Motivated by the numerical agreement, let us investigate analytically the early-time complexity prediction these Lanczos coefficients make in the classical limit. The quantity that we obtain from plugging in $a_n^{(\mathrm{sc})}(\beta)$ and $b_n^{(\mathrm{sc})}(\beta)$ into \eqref{eq:C_K_early_time_series_explicitly} we denote as $C^{(\mathrm{sc})}_K(t)_\beta$. Now for this quantity we again take the limit $q\to1$ after multiplication with $\abs{\log q}$ as before. A first reassuring aspect is that in this classical limit, these Krylov spread complexity series coefficients are finite for $0<\theta<\pi$, by which we mean they precisely scale as $\abs{\log q}^{-1}$ at leading order in $q\to1$. Concretely,
\begin{align}
    C_K^{(\mathrm{sc})}(t)_\beta=\sum_{n=1}^\infty c_{2n}^{(\mathrm{sc})}(\beta)\,t^{2n}\,,
    \label{semiclassicalkrylovcomplexity}
\end{align}
with coefficients $c_{2n}^{(\mathrm{sc})}(\beta)$ that are a series in integer powers of $\abs{\log q}$ starting at order $\abs{\log q}^{-1}$
\begin{align}
\label{eq:c_2n_sc_logq_expansion_starting_minusone}
    c_{2n}^{(\mathrm{sc})}(\beta)=\abs{\log q}^{-1}\sum_{k=0}^\infty\abs{\log q}^kc_{2n}^{(\mathrm{sc})[k]}(\beta)\,,
\end{align}
such that the classical limit of $C^{(\mathrm{sc})}_K(t)_\beta$ picks precisely the leading order of all these coefficients
\begin{align}
    \lim_{q\to1}\abs{\log q}C_K^{(\mathrm{sc})}(t)_\beta=\sum_{n=1}^\infty c_{2n}^{(\mathrm{sc})[0]}(\beta)\,t^{2n}\,.
\end{align}

Let us suppress the superscript $[0]$ for notational convenience from now on when talking about temperature behavior of the dominant contribution $c_{2n}^{(\mathrm{sc})[0]}(\beta)$ in the classical limit. Nontrivially, we find that in the resulting Krylov spread complexity early-time series these coefficients $c_{2n}^{(\mathrm{sc})}(\beta)$ exhibit factors of $\beta_\mathrm{BH}^{-2n}=(2\pi/\sin\theta)^{-2n}$ that we expect at order $t^{2n}$ naturally. This is an incarnation of the fact that DSSYK correlators decay in $t/\beta_{\mathrm{BH}}$ and not $t/\beta$ as expected from DSSYK correlators \cite{Blommaert:2024ydx}. The early-time series of $ \lim_{q\to1}\abs{\log q}C^{(\mathrm{sc})}_K(t)_\beta$ has coefficients
\begin{align}
    c_2^{(\mathrm{sc})}(\beta)&=\frac{2 \pi ^2 \sin (\theta )}{2 \sin (\theta )+(\pi -2 \theta ) \cos (\theta )}\left(\frac{\sin\theta}{2\pi}\right)^2\,,\label{eq:complexitycoeff_sc}\\
    c_4^{(\mathrm{sc})}(\beta)&=\frac{-\pi ^4 \sin (\theta ) }{12 (2 \sin (\theta )-2 \theta  \cos (\theta )+\pi  \cos (\theta ))^5}\left(\frac{\sin\theta}{2\pi}\right)^4\nonumber\\
    &\quad\times \big(\left(6 (\pi -2 \theta )^2-32\right) \cos (2 \theta )+5 \cos (4 \theta )\nonumber\\
    &\quad\quad-2 (\pi -2 \theta ) \sin (2 \theta ) (\cos (2 \theta )-7)+27\big)\,,
    \nonumber
\end{align}
where we display only the first two explicitly as an example.
Their behavior is evidence for the fact that the complexity prediction from $Z_{\mathrm{sc}}$ via \textit{CfZ} is meaningful, as we will explain now. One key observation is that in the infinite temperature limit $\beta\to0$ or equivalently $\theta\to\pi/2$, they precisely agree with the coefficients of the series expansion of the full result in the classical limit $\lim_{q\to1}\abs{\log q}C_K(t)_{\beta=0}$ previously stated in \eqref{eq:series_ckfull_betazero}. Explicitly, again on the example of the first two coefficients
 \begin{align}
    &\lim_{\beta\to0} c_2^{(\mathrm{sc})}(\beta)=\frac{1}{4}\,,\quad\lim_{\beta\to0} c_4^{(\mathrm{sc})}(\beta)=-\frac{1}{96}\,,
 \end{align}
 and so on for higher orders. This was one of our benchmark demands and Krylov spread complexity from the semiclassical saddle successfully passes this check. \\
 
In the classical limit $q\to1$ but away from $\beta=0$, neither directly implemented numerics for the Krylov problem \eqref{eq:dssyk_krylov_problem} nor (equivalently) a \textit{CfZ} prediction from \eqref{eq:Z_DSSYK} or \eqref{eq:Zsc} do reproduce the sine dilaton gravity wormhole volume prediction $L_{\mathrm{SD}}$ given in \eqref{eq:L_SD}. Let us state this clearly once more: DSSYK Krylov spread complexity for the finite temperature reference state $\ket{R}=Z(\beta)^{-1/2}e^{-\beta T/2}\ket{0}$ \textit{does not equal} the wormhole volume in sine dilaton gravity away from $\beta=0$. In particular, this contains a disagreement between the complexity prediction in the low-temperature JT limit $\beta\to\infty$ and the JT wormhole length \eqref{eq:L_JT}. Let us immediately be clear that agreement was never expected in the first place, based on the results of two of us in \cite{Heller:2024ldz}, and therefore does not compromise the validity of \textit{CfZ}: In that work, we found that the wormhole volume is not reproduced by the classical limit of the Krylov spread complexity assigned to the spread of the finite temperature reference state in its (temperature-dependent) Krylov basis, but instead by the classical limit of the analytic continuation $t\to t+i\beta/2$ of the complexity for the reference state at infinite temperature spreading in the infinite temperature Krylov basis. Because \textit{CfZ} derives complexity for the finite temperature reference state by definition, only the wormhole volume at $\beta=0$, where both approaches coincide, can serve as a sanity check based on previously established results. This one, as demonstrated before, is passed by \textit{CfZ} applied to the semiclassical saddle \eqref{eq:Zsc}. \\

\begin{figure}[h]
    \centering
    \includegraphics[width=0.95\columnwidth]{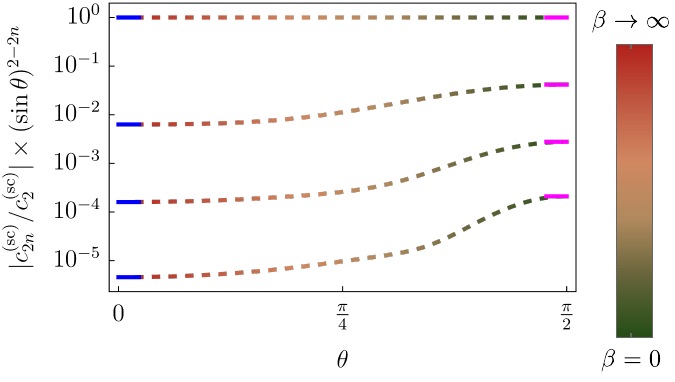}
    \caption{Visualization of the temperature dependence ($\beta$ increasing from zero to $\infty$ from green to red) of the absolute value of the first four coefficients $c_{2n}^{(\mathrm{sc})}(\beta)$ of the classical complexity early-time series from the semiclassical partition function of DSSYK, normalized by $c_{2}^{(\mathrm{sc})}(\beta)$ and rescaled by $(\sin\theta)^{2-2n}$ for visual clarity (dashed). They correspond to $n=\{1,2,3,4\}$ from top to bottom. The first four early-time coefficients of Schwarzian complexity (solid blue) are reproduced at $\beta\to\infty$, and the first four early-time coefficients of $L_{\beta=0}=2\log\cosh t/2$ (solid magenta) at $\beta\to0$.}
    \label{fig:c2_to_c8_sc_all_temp}
\end{figure}

Yet, the results \eqref{eq:complexitycoeff_sc} from \textit{CfZ} applied to the semiclassical saddle in DSSYK reproduce other known behavior in the JT limit: Using again the first two orders as example, the coefficients $c_{2n}^{(\mathrm{sc})}(\beta)$ in this limit go as
 \begin{align}
    &\lim_{\beta\to\infty} c_2^{(\mathrm{sc})}(\beta)\simeq\frac{4\pi^2}{\beta_{BH}^3}\,,\quad\lim_{\beta\to\infty} c_4^{(\mathrm{sc})}(\beta)\simeq-\frac{\pi^2}{\beta_{BH}^5}\,,
 \end{align}
and agree precisely with the classical limit of the Krylov spread complexity early-time expansion for Schwarzian theory with partition function~\cite{Mertens:2022irh}
\begin{align}
\label{eq:schwarzian_Z}
    Z_{\mathrm{sch}}=\frac{1}{4\pi^2}\left(\frac{2\pi c}{\beta}\right)^{3/2}e^{S_0+\frac{2\pi^2 c}{\beta}}\,,
\end{align}
which has the form
\begin{align}
\label{eq:schwarzian_ck_early_time_series}
    \lim_{c\to\infty}\frac{1}{c}C_K(t)_\mathrm{sch}=\frac{2\pi^2}{\beta}\left(2x^2-\frac{x^4}{2}+\frac{x^6}{2}
    +\dots\right)
\end{align}
where 
$c$ here is the Schwarzian coupling inversely proportional to the NAdS$_2$ gravitational constant and we introduced $x=t/\beta$. The complexity associated to this partition function was previously studied in \cite{Balasubramanian:2022tpr} and we will discuss the relation between our analysis and theirs more in part \ref{sec:CK_of_cft2} of this work. For now, we merely want to mention that this shows that at the other end of the temperature spectrum, the complexity prediction from $Z_{\mathrm{sc}}$ via $\textit{CfZ}$ also appears reasonable, as we reproduce Schwarzian complexity, knowing that the low-energy sector of DSSYK is governed by precisely this partition function. Our corroboration is to take this classical early-time series that can be derived from the semiclassical saddle of DSSYK \eqref{eq:Zsc} via \textit{CfZ} seriously for any value of $\beta$, and hence as a novel finite-temperature prediction of Krylov spread complexity in DSSYK at early times in the $q\to1$ limit for a reference state at temperature $\beta$. A summary of aspect \textbf{ii)} is given in Fig. \ref{fig:c2_to_c8_sc_all_temp}. Note that this figure can also be read as visualizing how the complexity prediction for finite temperature reference states deviates from the sine dilaton gravity wormhole length prediction $L_\mathrm{SD}$ given in \eqref{eq:L_SD} with increasing $\beta$. The wormhole prediction corresponds to extending the constant pink lines that are the $\log \cosh$ series coefficients to all $\theta$ as $L_\mathrm{SD}$ is precisely a $\log \cosh$ in $t/\beta_{\mathrm{BH}}$ for all temperatures. These extended lines then clearly mismatch with the actual result (dashed lines) away from $\theta=\pi/2$.\\ 

\newpage
\textbf{iii) Analytic structure of Lanczos coefficients and early-time complexity as series in $\abs{\log q}$.} The analytic structure of the semiclassical Lanczos coefficients at finite temperature $a_n^{(\mathrm{sc})}(\beta)$ and $b_n^{(\mathrm{sc})}(\beta)$ in $\abs{\log q}$ is perhaps unexpected: They themselves have a nontrivial expansion in $\abs{\log q}$ for $q\to1$ for general $\beta$. This raises an important question that the numerical analysis leaves unanswered: What is the role of the subleading orders in $\abs{\log q}$ of these Lanczos coefficients? Can we trust them, given that we only started from the leading saddle $Z_\mathrm{sc}$?  Let us dive deeper into this aspect in the limit $\beta=0$ where we can make contact with the known analytical full result for the Lanczos coefficients at $\beta=0$ given in \eqref{eq:an_bn_fullDSSYK_betaZero} \footnote{A generalization of this behavior to nonzero $\beta$ is likely, but we have not explored this direction further.}. We remind the reader that the coefficients $a_n(\beta=0)$ vanish in this case. What will be of particular use is the expansion of the full DSSYK Lanczos coefficients $b_n(\beta=0)$ around $q\to1$. This reads
\begin{align}
\label{eq:bn_fullbetazero_expansion}
    b_n&(\beta=0)\simeq\frac{1}{2}\sqrt{\frac{n}{\abs{\log q}}}-\frac{\sqrt{n}}{8}(n-1)\sqrt{\abs{\log q} }\nonumber\\&+\frac{\sqrt{n}}{192} (n-1) (5 n-1)  \abs{\log q} ^{\frac{3}{2}}+\mathcal{O}\left(\abs{\log q} ^{\frac{5}{2}}\right)\,.
\end{align}
Let us now compare this expansion in $\abs{\log q}$ around $q\to1$ with the semiclassical prediction for the Lanczos coefficients evaluated at $\beta=0$. The diagonal Lanczos coefficients $a^{(\mathrm{sc})}_n(\beta=0)$ we find in the semiclassical limit vanish to all orders in $\abs{\log q}$, allowing us to solely focus on the off-diagonal coefficients $b_n$. What we find is the following nontrivial observation, checked explicitly at low $n$:
\begin{align}
     b_n(\beta=0)-b_n^{(\mathrm{sc})}(\beta=0)=\mathcal{O}\left(\abs{\log q}^\frac{3}{2}\right)\,.
\end{align}
This shows that both the leading as well as the first subleading order in $\abs{\log q}$ agree between the full and the semiclassical prediction of Lanczos coefficients $ b_n(\beta=0)$ and $b_n^{(\mathrm{sc})}(\beta=0)$. Now comes the crux: If it were true that complexity in the classical limit $q\to1$ only depends on the leading (and potentially first subleading) order behavior in $\abs{\log q}$, the previous argument would justify why the early-time series of $\lim_{q\to1}\abs{\log q}C_K(t)_\beta$ and $\lim_{q\to1}\abs{\log q}C_K^{(\mathrm{sc})}(t)_\beta$ agree. \textit{But this is not a true statement, as we will show now.}\\

Let us assume we plug in a general Lanczos coefficient $b_n$ with expansion in a small parameter $\abs{\log q}$ that reads
\begin{align}
    b_n=\abs{\log q}^{-1/2}\sum_{k=0}^{\infty}\abs{\log q}^{k}b^{[k]}_n
\end{align}
into the Krylov spread complexity early-time expansion \eqref{eq:C_K_early_time_series_explicitly}. We find that, in general, the coefficient $c_{2n}$ in front of $t^{2n}$ is a series in increasing integer powers of $\abs{\log q}$ starting at order $\abs{\log q}^{-n}$
\begin{align}
    c_{2n}=\abs{\log q}^{-n}\sum_{k=0}^\infty \abs{\log q}^{k}c_{2n}^{[k]}\,.
\end{align}
We find that $c_{2n}^{[k]}$ ($k$-th subleading order of the coefficient of $t^{2n}$ of the complexity early-time series) depends on all orders of the Lanczos coefficient up to $b^{[k]}_n$ ($k$-th subleading order of the Lanczos coefficient). There are two highly nontrivial observations that go along with this statement. Let us begin by making the first.
\begin{itemize}
    \item It is nontrivial that for $C^{(\mathrm{sc})}_K(t)_\beta$ from $Z_{\mathrm{sc}}$, all coefficients of the early-time expansion are at most of order $\abs{\log q}^{-1}$, as we stated in \eqref{eq:c_2n_sc_logq_expansion_starting_minusone}. 
\end{itemize}
Connecting this to our previous observation, this means that there is a crucial yet subtle interplay between subleading orders of Lanczos coefficients in the case where we derive them from $Z_{\mathrm{sc}}$, such that all $c_{2n}^{[k]}$ vanish for all $k-n<-1$. This requires cancellation between up to the $(n-2)$-th subleading orders in the Lanczos coefficients for coefficient $c_{2n}$ to have a well-defined classical limit. This shows why the result is of such intricate structure: although we can show that the subleading orders differ between $b_n(\beta=0)$ and $b_n^{(\mathrm{sc})}(\beta=0)$, in both cases the associated complexity series coefficients are finite in the classical limit. In the very same spirit, we also note a second important observation.
\begin{itemize}
    \item Although $b_n(\beta=0)$ and $b_n^{(\mathrm{sc})}(\beta=0)$ have different subleading orders starting from the second subleading order, and the complexity coefficients $c_{2n}$ for $n>2$ depend on these higher orders, we find the same value for $c_{2n}$ in both cases in the classical limit. 
\end{itemize}
Let us demonstrate this second point using the example of the coefficient $c_6$ of the general Krylov spread complexity early-time expansion \eqref{eq:C_K_early_time_series_explicitly}. With $a_n=0$, it reads
\begin{align}
    c_6\big\vert_{a_n=0}=\frac{1}{180} b_1^2 \left(8 b_1^4+b_2^2 b_1^2-7 b_2^4+3 b_2^2 b_3^2\right)\,.
\end{align}
Assuming Lanczos coefficients where the two leading orders are the DSSYK behavior but all further subleading orders are kept general,
\begin{align}
    b_n=&\frac{1}{2}\sqrt{\frac{n}{\abs{\log q}}}-\frac{\sqrt{n}}{8}(n-1)  \sqrt{\abs{\log q} }\nonumber\\&+\abs{\log q}^{-1/2}\sum_{k=2}^{\infty}\abs{\log q}^{k}b^{[k]}_n\,,
\end{align}
we find the associated classical complexity early-time coefficient of $t^6$
\begin{align}
\label{eq:c6_class_limit_dep_on_second_subleading}
    \lim_{q\to1}\abs{\log q}c_6\big\vert_{a_n=0}=\frac{288 b_1^{[2]}-288 \sqrt{2} b_2^{[2]}+96 \sqrt{3} b_3^{[2]}+17}{46080}\,.
\end{align}
Now we can plug in the predictions from full DSSYK following \eqref{eq:bn_fullbetazero_expansion},
\begin{align}
    b_1^{[2]}=0\,,\quad b_2^{[2]}=\frac{3}{32 \sqrt{2}}\,,\quad b_3^{[2]}=\frac{7}{16 \sqrt{3}}\,,
\end{align}
and from the semiclassical partition function,
\begin{align}
    b_1^{(\mathrm{sc})[2]}=0\,,\quad b_2^{(\mathrm{sc})[2]}=-\frac{1}{32 \sqrt{2}}\,,\quad b_3^{(\mathrm{sc})[2]}=\frac{1}{16 \sqrt{3}}\,.
\end{align}
One can immediately check that, while the subleading orders evidently differ in value, the associated Krylov spread complexity coefficient $c_6$ is the same in the classical limit. Indeed, in both cases we get the $t^6$ coefficient familiar from expression \eqref{eq:series_ckfull_betazero},
\begin{align}
     \lim_{q\to1}\abs{\log q}c_6\big\vert_{a_n=0}= \lim_{q\to1}\abs{\log q}c^{(\mathrm{sc})}_6\big\vert_{a_n=0}=\frac{1}{1440}\,.
\end{align}
For completeness, let us also mention that in the case we had kept the leading order of the complexity prediction
\begin{align}
    b_n^{[0]}=\frac{1}{2}\sqrt{\frac{n}{\abs{\log q}}}\,,
\end{align}
and set all subleading orders to zero $b_n^{[k>0]}=0$, then by plugging this into the complexity early-time series in the classical limit we would have recovered the well-known statement that Krylov spread complexity associated to $a_n=0$ and $b_n\sim\sqrt{n}$ is eternally quadratic,
\begin{align}
    \lim_{q\to1}\abs{\log q}C_K(t)=\frac{t^2}{4} \quad\text{(all higher orders vanish)}\,.
\end{align}
This is the third crucial insight: Taking into account only some number of leading orders of Lanczos coefficients alters the complexity prediction, even at early times. The classical limit that seems to retain most information is the one where we keep all orders in the Lanczos coefficients, and only take the classical limit at the level of the complexity itself.\\ 

We have now seen the crucial influence of subleading orders in the Lanczos coefficients for the classical limit of the Krylov complexity early-time expansion. Given this sensitive dependence and the fact that the prediction from $Z_{\mathrm{sc}}$ via \textit{CfZ} satisfies the benchmark of matching $L_{\beta=0}$ at high temperatures and Schwarzian complexity at low temperatures, we conclude that:
\begin{framed}
\noindent 
The semiclassical saddle appears sufficient to predict Krylov spread complexity from the partition function via \textit{CfZ} in the appropriate classical limit, taken at the level of the complexity itself. For $a_n$ and $b_n$, subleading orders in the parameter governing the approach to classicality are essential, and keeping only the leading order alters the prediction significantly. 
\end{framed}

Based on this conclusion, validated in the double-scaled SYK model, in the following section we now derive the prediction for the early-time complexity series that arises from the semiclassical black hole saddle in AdS$_3$/CFT$_2$ and discuss the predictions for complexity we can make from it. 

\section{Krylov spread complexity\\ in 2D holographic CFTs}
\label{sec:CK_of_cft2}

The main takeaway of the previous section is that DSSYK suggests that we can meaningfully predict the early-time series of Krylov complexity in the classical limit from the semiclassical saddle of the partition function, conditional on the fact that we take the classical limit at the level of complexity and not Lanczos coefficients. This motivates us to now apply \textit{CfZ} to the BTZ black hole saddle \cite{Banados:1992wn} of a holographic CFT in two dimensions and derive the associated early-time series.

\subsection{Framework: The BTZ black hole saddle}

Let us concern ourselves with a holographic CFT in two dimensions on $S^1_\beta\cross S^{1}$ where $\beta$ is the circumference of the Euclidean time circle and $V$ the circumference of the spatial circle. As a foundational pillar of $\mathrm{AdS}/\mathrm{CFT}$, it is known that above the Hawking-Page temperature \cite{Hawking:1982dh}, the leading saddle of the CFT partition function $Z(\beta)$ is the AdS-Schwarzschild saddle \cite{Witten:1998zw, Witten:1998qj,Maldacena:1997re}. In \(d=2\), this gives the Cardy scaling
\cite{Cardy:1986ie,Strominger:1997eq} associated with the BTZ black
hole \cite{Banados:1992wn},
\begin{align}
    Z(\beta)
    &=
    \exp\left(\frac{\pi cV}{6\beta}\right)
    =
    \exp\left(\frac{\mathfrak c}{\beta}\right),
    \qquad
    \mathfrak c\equiv\frac{\pi cV}{6}.
    \label{eq:Z_cft2}
\end{align}
Here \(c\) is the CFT central charge, while
\(\mathfrak c\) denotes the full dimensionful Cardy coefficient. Our conventions for $V$ follow \cite{Tong:2009np} and with this convention $\beta_\mathrm{HP}=V$. In the following parts, we will now study the Krylov spread complexity prediction we make via \textit{CfZ} applied to this saddle. 

\subsection{Results}
\label{sec:cft_results}
\subsubsection{Lanczos coefficients and complexity early-time\\ series: naive conclusions from fitting}
\begin{figure}
\label{fig:anbn_cft2}
\subfloat[\label{fig:an_cft2}]{%
  \includegraphics[width=0.95\columnwidth]{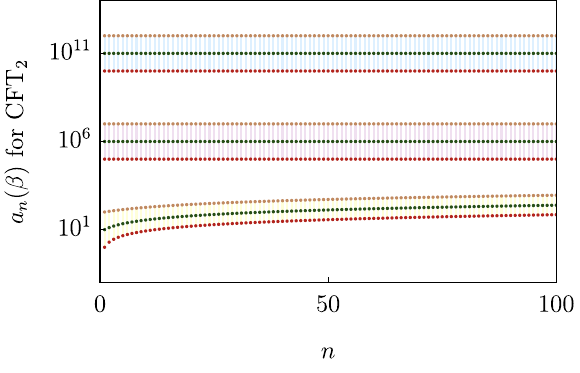}%
}

\medskip

\subfloat[\label{fig:bn_cft2}]{%
  \includegraphics[width=0.95\columnwidth]{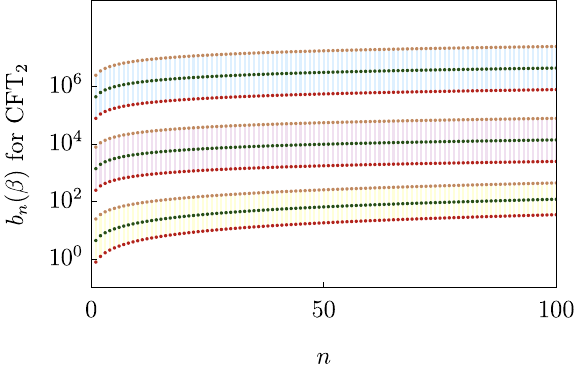}%
}
    
\caption{First 100 Lanczos coefficients $a_n(\beta)$ in \textbf{(a)} and $b_n(\beta)$ in \textbf{(b)} for Krylov spread complexity of finite temperature TFD state in CFT$_2$. The shaded bands indicate $\mathfrak{c}=10^1$ (yellow), $\mathfrak{c}=10^6$ (purple) and $\mathfrak{c}=10^{11}$ (blue). The numerics were carried out at three inverse temperatures $\beta=1$ (green), $\beta=10^{-1/2}$ (orange) and $\beta=10^{1/2}$ (red), respectively as indicated by the color of the data points. It is evident that Lanczos coefficients increase with increasing $c$, $n$ and temperature.} 
\label{fig:cft2_lanczos}
\end{figure}

The application of \textit{CfZ} to the partition function \eqref{eq:Z_cft2} can be straightforwardly carried out on a common laptop. We display numerical results for the first $n=100$ Lanczos coefficients for representative temperatures $\beta=\{10^{-1/2},10^{0},10^{1/2}\}$ \footnote{Indeed, for the standard convention $V=2\pi$, all these satisfy $\beta<\beta_\mathrm{HP}=V=2\pi$.} and $\mathfrak{c}=\{10^1,10^6,10^{11}\}$ in Fig.~\ref{fig:cft2_lanczos}.
These Lanczos coefficients increase with increasing $c$, $n$ and decreasing $\beta$. To quantitatively assess the growth pattern, one could get the idea to fit this data. We present two well-converging fits of this data, but highlight why these fits lead to incorrect conclusions for the complexity, even at the level of the early-time series.\\

We are interested in these Lanczos coefficients in the context of large $c$. Let us therefore naively allow ourselves only to fit a leading-order contribution ($\mathcal{O}(c)$ for $a_n$ and $\mathcal{O}(\sqrt{c})$ for $b_n$ as evident by inspection of the data). We find fits that converge well with increasing $c$
\begin{align}
\label{eq:leading_fit_cft2_a}
    a_n^{[0]}&=\mathfrak{c}\beta^{-2}\,,\\
    b_n^{[0]}&=\sqrt{n}\cdot\sqrt{2\mathfrak{c}\beta^{-3}}\,.\label{eq:leading_fit_cft2_b}
\end{align}
If instead we also allow for a subleading order in $c$, we find the even better converging fits
\begin{align}
\label{eq:subleading_fit_cft2_a}
    a_n^{[0+1]}&=\mathfrak{c} \beta ^{-2}+3n\beta^{-1}\,,\\
    \label{eq:subleading_fit_cft2_b}
    b_n^{[0+1]}&=\sqrt{n\left(\frac{2\mathfrak{c}}{\beta ^{3}}+(n-1)\frac{3}{2\beta^{2}}\right)}\,.
\end{align}
Details on the fitting procedure can be found in Appendix \ref{sec:app_fitting}. There are multiple obvious problems with this approach. One is the effective restriction to the regime $n\ll c$ where the fits model the data well, caused by the implicit assumption of a search for fits that converge well at large $c$, but not scaling the number of coefficients fitted with $c$ in the process. 
In this context we could, in the hope to at least extract something meaningful out of these fits, restrict our attention to low orders of the early-time series, where the answer could at least in theory be unaffected by the assumption of small $n$ compared to $c$, as the order $t^{2n}$ of the early-time series only depends on the first $n$ Lanczos coefficients. 
Yet, even in that regime there is a problem with this approach, namely precisely the observation of the previous section that the classical limit (here $c\to\infty$) should only be taken at the level of the complexity itself, and not the Lanczos coefficients. We illustrate this using the two fits.

It is evident that the leading-order fit vs.~subleading order fit conclude two very different growth patterns: the leading order concludes constant $a_n$ and $b_n$ growing with $\sqrt{n}$. If instead we allow for the subleading order, we conclude linearly growing $a_n$ and a transition from growth as $\sqrt{n}$ to growth as $n$ in $b_n$ with transition point scaling with $c$. Both cases are well-understood, but have crucially different associated complexities. The leading-order Lanczos coefficient prediction represents the Heisenberg-Weyl case with eternally growing quadratic complexity \cite{Balasubramanian:2022tpr}
\begin{align}
    C_K^{[0]}(t)=\frac{\mathfrak{c}}{\beta}\left(2\frac{t^2}{\beta^2}\right) \quad\text{(all higher orders vanish)}\,,
\end{align}
while the subleading order is an instance of the well-studied $SL(2,\mathbb{R})$ Krylov spread complexity \cite{Balasubramanian:2022tpr} with 
\begin{align}
    &a_n=\delta +\gamma (h+n),\\
&b_n=\alpha  \sqrt{n(2 h+n-1)},\,
\end{align}
and associated complexity
\begin{align}
    C_K(t)=2 h \sinh ^2\left(\alpha  t \sqrt{1-\frac{\gamma ^2}{4 \alpha ^2}}\right)/\left(1-\frac{\gamma ^2}{4 \alpha ^2}\right)\,.
\end{align}
The subleading order fit therefore predicts oscillating Krylov spread complexity with early-time expansion
\begin{align}
    C_K^{[0+1]}(t)&=\frac{4\mathfrak{c} }{3\beta}\left[1-\cos \left(\sqrt{3}\cdot t/\beta\right)\right]\nonumber\\
    &=\frac{\mathfrak{c}}{\beta}\left(2\frac{t^2}{\beta ^2}-\frac{t^4}{2 \beta ^4}+\frac{t^6}{20 \beta ^6}+\dots\right)
\end{align}
We see that the leading in $c$ and subleading in $c$ fits already differ in their complexity prediction at the quartic order in the early-time expansion, although they look highly convergent on the first $n\sim\mathcal{O}(100)$ numerical Lanczos coefficients at large $c$. And indeed a full numeric evaluation of the Krylov spread complexity early-time series does not converge towards either of these early-time predictions.
This brief analysis highlights the need to do precisely what we learned from the previous section: Take into account all subleading orders in $c$ at the level of Lanczos coefficients, and only take the classical limit at the level of the complexity itself. 

\subsubsection{Lanczos coefficients and complexity early-time\\ series: analytic and numeric results}

In this paragraph, let us instead begin by obtaining the Lanczos coefficients that arise from the partition function \eqref{eq:Z_cft2} analytically, and as all-order expressions in $c$. As an example, the first two Lanczos coefficients each are
\begin{align}
    a_0&= \mathfrak{c} \beta ^{-2}\,,\\
    a_1&= \mathfrak{c} \beta ^{-2}+3\beta^{-1}\,,
\end{align}
and
\begin{align}
    b_1&= \sqrt{2\mathfrak{c}  \beta ^{-3}}\,,\\
    b_2&= \sqrt{4 \mathfrak{c}  \beta ^{-3}+3\beta^{-2}}\,.
\end{align}
Notice that all four of these follow the subleading order fits \eqref{eq:subleading_fit_cft2_a} and \eqref{eq:subleading_fit_cft2_b} perfectly, but this is merely an artifact of displaying only the first two $a_n$ and $b_n$. Already $a_2$ and $b_3$ are much more lengthy (hence we do not display them here) and disagree with \eqref{eq:subleading_fit_cft2_a} and \eqref{eq:subleading_fit_cft2_b} in their further subleading orders in $c$. More precisely, for generic $n$ we find that the true results differ from the fits as
\begin{align}
    &a_n^{[0]}-a_n=\mathcal{O}(1)\,,&&(b_n^{[0]})^2-b_n^2=\mathcal{O}(1)\,,\\
    &a_n^{[0+1]}-a_n=\mathcal{O}(1/c)\,,&&(b_n^{[0+1]})^2-b_n^2=\mathcal{O}(1/c)\,.
\end{align}
As opposed to working with the fits, we now take the analytically derived coefficients to all orders and plug them in the early-time expansion of Krylov spread complexity \eqref{eq:C_K_early_time_series_explicitly}. After taking the appropriate classical limit that entails the multiplication by $c^{-1}$, we again find cancellations in the differences between Lanczos coefficients exactly such that the limit is finite, but higher orders in $c$ in the Lanczos coefficients contribute nontrivially. Before we move on, let us also state that the natural reference scale for time in this setting is the inverse temperature~$\beta$. Up to the overall prefactor $\beta^{-1}$ (which exactly accounts for the dimension of $\mathfrak{c}$ in our conventions), the series organises in even powers of $x=t/\beta$. Hence, for the following analysis, let us study the Krylov spread complexity as a series in $x$. Explicitly we obtain
\begin{align}
\lim_{c\to\infty}\frac{C_K(x)}{c}=\frac{\pi V}{6\beta}\left(2x^2-\frac{x^4}{2}+\frac{x^6}{2}
    +\dots\right)\,,
    \label{eq:cft2_ck_x_series}
\end{align} 
where we write it such that the overall prefactor is precisely the exponent of the partition function \eqref{eq:Z_cft2}. This quantity naturally turns out to be extensive in $V$. 
At low orders we can analytically derive this series, for example to $t^{10}$ in a couple of seconds. And indeed, this result is the series that the numerics converge to: We obtained the series numerically at $\mathfrak{c}=3.1 \cdot10^{15}$ up to $t^{200}$ in around forty minutes on a common laptop, and it agrees with the low order analytic predictions we derived analytically to very high accuracy. We will make use of this numerical version of \eqref{eq:cft2_ck_x_series} over the course of the rest of this article. The attentive reader will of course point out that this series has not appeared for the first time in this paper. Indeed, this precisely agrees (up to an overall rescaling) with the complexity prediction for Schwarzian theory that we reported in \eqref{eq:schwarzian_ck_early_time_series}. This is logical because the semiclassical saddle of the Schwarzian partition function $Z_{\mathrm{sch}}$ given in \eqref{eq:schwarzian_Z} is precisely of the Cardy form \eqref{eq:Z_cft2}. This can also be viewed as further reassuring evidence that the semiclassical saddle is sufficient to predict the early-time complexity series: 
We see explicitly that in this example the 1-loop determinant prefactor in \eqref{eq:schwarzian_Z} does not influence the complexity prediction. Yet, there is also one crucial difference between our result for Schwarzian complexity \eqref{eq:schwarzian_ck_early_time_series} and CFT$_2$ complexity \eqref{eq:cft2_ck_x_series} in the limit in which they apply: \eqref{eq:schwarzian_ck_early_time_series} is valid in the low temperature limit $\beta\to\infty$ of DSSYK, while \eqref{eq:cft2_ck_x_series} is valid above the Hawking-Page temperature in CFT$_2$.\\

We have also performed two further sanity checks of the prediction \eqref{eq:cft2_ck_x_series}, which we present in App.~\ref{sec:app_cft2_sanitychecks}. This includes the inclusion of the empty AdS saddle as well as the case of a generic 1-loop determinant.\\

\subsubsection{Comparison to the analysis of Schwarzian complexity\\ in \cite{Balasubramanian:2022tpr} by Balasubramanian et al.}

To the best of our knowledge, there has not appeared an in-depth analysis of this early-time series \eqref{eq:cft2_ck_x_series} (and the predictions we can make from it) in the literature. It is a core goal of this work and the following sections in particular to take this result seriously and perform such an analysis.\\

We begin by claiming a refinement of the statement made in \cite{Balasubramanian:2022tpr} that Krylov spread complexity from the Schwarzian partition function \eqref{eq:schwarzian_Z} (and by our previous arguments also structurally equivalently complexity of the Cardy saddle in CFT$_2$ \eqref{eq:Z_cft2}), is eternally quadratic,
\begin{align}
    C_K(t)_\mathrm{Sch}\sim 
    C_K(t)_{\mathrm{CFT}_2} \sim t^2\,.
\end{align}
In their work \cite{Balasubramanian:2022tpr}, this was argued for by investigating the asymptotics at early and late times, and establishing both as quadratic  (we also note that \cite{Balasubramanian:2026azk} showed late-time quadratic growth for a similar case). We would like to clarify that this does not prove eternal quadratic growth, as a non-quadratic intermediate transition regime between early and late-time quadratic growth is not ruled out. In fact, we claim that Schwarzian complexity exhibits precisely this behavior and we can see this from the early-time series.\\

\begin{figure}
    \centering
    \includegraphics[width=0.95\columnwidth]{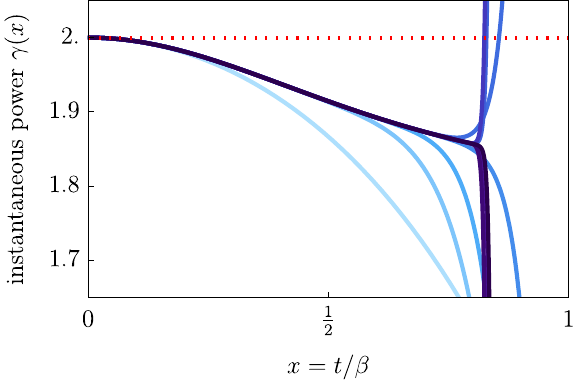}
    \caption{Instantaneous power $\gamma(x)$ of the (numerically obtained) classical Krylov spread complexity early-time series $\lim_{c\to\infty}c^{-1}C_K(t)$ for CFT$_2$ at 
    $\mathfrak{c}=3.1 \cdot10^{15}$. Different order of truncation of the early-time series is indicated by color: $\{4,12,20,28,36\}$ from light to medium blue and $\{168,176,184,192,200\}$ from dark blue to black. We see that the complexity is not eternally quadratic (dotted red), but instead deviates away from early-time quadratic growth within the radius of convergence of the early-time series.}
    \label{fig:cft2_early_time_series}
\end{figure}

It is immediately evident from the first term in \eqref{eq:cft2_ck_x_series} that the complexity is quadratic at early times, as universally true for any Krylov spread complexity. But subsequently, as visualized in Fig.~\ref{fig:cft2_early_time_series}, we see in a plot of the instantaneous power $\gamma(x)$ \eqref{gammafunction}, in the coordinate $x=t/\beta$, that the complexity clearly starts to deviate from pure quadratic growth inside the direct-convergence regime
\[
|x|<R_x\simeq0.8,
\qquad x\equiv t/\beta,
\] of the early-time series.\\

From the perspective of the early-time expansion, no statement can be made whether the complexity grows quadratically (as claimed by \cite{Balasubramanian:2022tpr}) or linearly (as expected for a typical dual of holographic complexity) at late times. Let us investigate this question by analytically continuing beyond early times by again using the Padé method.

\subsection{Beyond early times in CFT$_2$: Padé-approx.}
\label{sec:pade_cft2}

One obvious shortcoming of the early-time analysis we carried out is that, by definition, it concerns only the early-time regime. Yet, motivated by the successes of extracting the exact answer from the early-time series in DSSYK at $\beta=0$ (see Sec.~\ref{sec:warmup_resumming_dssyk_betazero}), we now go beyond early times via the same method.\\

\begin{figure}
    \centering
    \includegraphics[width=0.95\columnwidth]{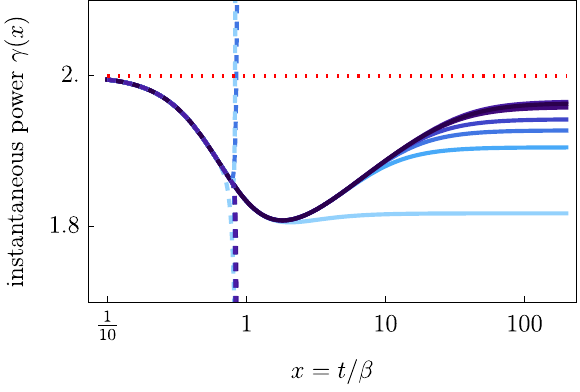}
    \caption{Instantaneous power $\gamma(x)$ of the (numerically obtained) classical Krylov spread complexity early-time series $\lim_{c\to\infty}c^{-1}C_K(t)$ for CFT$_2$ at working precision $5\cdot10^3$ and $\mathfrak{c}=3.1 \cdot10^{15}$ (dashed) and its diagonal Padé approximation (solid). 
     Different order of truncation of the early-time series is indicated by color: from 20 to 200 in steps of 30 from light to dark blue. The associated Padé approximations are of order 10 to 100 in steps of 15 with the same coloring. Both of these are compared to eternal quadratic growth (dotted red). 
     After the initial quadratic regime, the approximants dip below two and then turn upward over the displayed time interval. The order dependence remains visible even among the highest available diagonal approximants, so this behavior should be interpreted as evidence compatible with a return toward quadratic growth, rather than as a controlled determination of the strict late-time limit.}
    \label{fig:cft2_pade}
\end{figure}

Because we do not use a closed-form resummation of the full time
dependence encoded by the coefficients \(c_{2n}\) in
\eqref{eq:cft2_ck_x_series}, we employ Padé approximation to continue
the series beyond its convergence radius. Our findings via this method are visualized in Fig.~\ref{fig:cft2_pade}.
We expect these approximants to be informative for times that do not
scale with \(c\). Within the time interval displayed in
Fig.~\ref{fig:cft2_pade}, the decrease from the early-time quadratic
exponent appears to be intermediate: the higher-order diagonal Padé
approximants dip below two and subsequently turn upward. This behavior
is compatible with the quadratic late-time asymptotics argued for in
Refs.~\cite{Balasubramanian:2022tpr,Balasubramanian:2026azk}. However, the visible order dependence and slow convergence prevent us from using the present Padé sequence alone to establish the strict \(t/\beta\to\infty\) limit. We therefore use it as a finite-time
analytic continuation and as a comparison curve for the bulk
construction, rather than as an independent proof of late-time
quadratic growth.

Importantly, this makes our prediction consistent with the analysis in \cite{Balasubramanian:2022tpr} (see also \cite{Balasubramanian:2026azk}) that predicts quadratic growth at late times for this kind of Krylov spread complexity. The advantage of our construction is that the Padé construction gives an (approximate) prediction for the (instantaneous power of) complexity at all times, and not only in some asymptotic regime. Only this will allow us later to compare this prediction to bulk holographic complexities at generic times, and specifically in the regime of $t/\beta\sim\mathcal{O}(1)$ that goes beyond the radius of convergence of the early-time series. It is nontrivial that the highest-order diagonal Padé
approximants turn upward toward the quadratic value over the accessible range, purely from the early-time series data, only based on the assumption that the late-time behavior should be power-law, but not specifying the power itself. We also would like to emphasize that finding the quadratic growth at late times is by no means built into the method we use, as the same method correctly predicted late-time linear growth in the case of $L_{\beta=0}=2\log \cosh (t/2)$ in DSSYK, see Fig.~\ref{fig:log_cosh_pade}.\\

Let us conclude this section by noting the following tension. The available Padé continuation of the BTZ boundary series exhibits finite-time behavior compatible with the quadratic asymptotics found for related partition functions in \cite{Balasubramanian:2022tpr,Balasubramanian:2026azk}, although the
strict late-time limit is not established by the present Padé sequence
alone. Together with the previous asymptotic results, the continuation is
compatible with quadratic growth in the semiclassical pre-saturation
regime \(1\ll t/\beta\ll c\). This is vastly different from the behavior which is known for \emph{all} standard finite-functional holographic complexity proposals, including
CV and the original CAny class. These are characterized by linear growth that typically onsets at $t/\beta$ of $\mathcal{O}(1)$ (see the most well-studied examples compared in \cite{Carmi:2017jqz}) where $\beta$ here refers to the inverse temperature of the black hole geometries these complexities were computed in. In the next section, we discuss this discrepancy in more detail and provide a mechanism to resolve this tension.

\section{Bulk perspective}
\label{sec:bulk}

\subsection{Comparison to bulk volume}

As shown above, semiclassical Krylov complexity of finite-temperature TFD reference states in two-dimensional holographic CFTs is quadratic at short times and is compatible with quadratic pre-saturation asymptotics. This may come as a bit of a surprise, given that linear late-time growth has often been considered a hallmark of holographic complexity, starting from the CV proposal of \cite{Susskind:2014moa,Susskind:2014rva} and later in the CAny proposal of \cite{Belin:2021bga,Belin:2022xmt}. 
But if \textit{``anything equals complexity"} and complexity was assumed to only grow linearly at late times in \cite{Belin:2021bga,Belin:2022xmt}, what can a bulk dual for Krylov complexity look like in the case investigated in this work? 
Certainly, it cannot be the volume: 
Not only does the volume proposal not have the appropriate late-time behavior, it also deviates significantly from the boundary result already at the level of the early-time expansion. Concretely, we find
\begin{align}
\label{eq:CV_early_time_series}
 \frac{G_N}{V_{\phi}}\mathcal{O}_V(t)=\frac{2 \pi^2}{\beta} \left(2 x^2 -5 x^4 +\frac{37}{2} x^6 
 + \dots \right),
\end{align}
which cannot be matched to \eqref{eq:cft2_ck_x_series} even if we allow for an overall rescaling as well as a rescaling of time. 
Here, we used $x=\tau/\beta$ with the two-sided bulk time $\tau$ to connect to the typical bulk nomenclature in \cite{Belin:2021bga,Belin:2022xmt} 
\footnote{To streamline the comparison between the boundary and bulk observables, we identify the boundary time \(t_{bdry}\) with the total two-sided bulk time \(\tau_{bulk}\). This convention ensures that both quantities are evaluated at the same physical time separation between the two asymptotic boundaries. In terms of the corresponding one-sided bulk time \(t_R=t_L\), this amounts to set \(t_{bdry}=\tau_{bulk}=2 t_R\).}.

How this series is obtained is discussed in detail in the following part of this section.  The undeformed volume therefore does not reproduce the finite-temperature Krylov spread complexity obtained from the BTZ saddle via \textit{CfZ}. This disagreement should not be interpreted as a breakdown of the semiclassical boundary construction or of the BTZ saddle. The semiclassical Krylov calculation continues to determine the boundary expansion systematically, while the maximal slice provides a well-defined geometric observable with the expected qualitative growth properties. 
What fails is the stronger assumption that this particular Krylov complexity must coincide with the undeformed extremal volume. From this perspective, the failure of the undeformed volume is not an obstruction but rather the starting point for identifying the bulk functional naturally associated with the boundary Krylov complexity. This mismatch is consistent with the DSSYK benchmark:
\textit{CfZ} computes the spread of the finite-temperature TFD state
in its temperature-dependent Krylov basis. This Krylov problem need
not coincide with the distinct construction adapted to the wormhole
length.

\subsection{A new complexity functional}

Because of the general construction of the bulk complexity functionals to display linear late-time growth \cite{Belin:2021bga,Belin:2022xmt}, we can in fact deduce that neither the extremal volume nor any simple modification of the volume proposal can provide a bulk dual for Krylov complexity. In this section, we will nevertheless construct a candidate for an appropriate bulk functional in the context of AdS$_3$/CFT$_2$ along the lines of CAny, but deliberately breaking a core assumption of this class in the process. 

Following \cite{Belin:2021bga,Belin:2022xmt}, let us consider an observable 
\begin{equation}
\cO=
 \frac1{G_NL}\int_{\Sigma_{F_2}}\dd^2\sigma\sqrt h\,F_1,
 \label{eq:Cany}
\end{equation}
where the surface $\Sigma_{F_2}$ extremizes a functional $F_2$ and $F_1$ is a different functional depending on the bulk metric $g_{\mu\nu}$ and the embedding of $\Sigma_{F_2}$ (as well as derivatives of both in the most general case). The relation to the central charge of the dual CFT$_2$ lies in the Brown-Henneaux formula $c=3L/2G_N$ \cite{Brown:1986nw} and we set the AdS radius $L$ to one from now on, in accordance with $V=2\pi$ for the spatial circle circumference in the CFT. For convenience, we will take $F_2=1$ so that $\Sigma_{F_2}$ is the extremal volume slice but allow $F_1$ to be more general. How general can it be? As the BTZ black hole \cite{Banados:1992wn} is locally AdS${}_3$, all bulk curvature invariants are constants determined by the cosmological constant.
For $F_1$ to be non-trivial, it will thus have to depend on the extrinsic curvature tensor $K_{ab}$ of $\Sigma_{F_2}$. 

For concreteness, the BTZ black hole metric in ingoing Eddington-Finkelstein coordinates reads
\begin{equation}
    ds^2
    =
    - f(r) dv^2 + 2\,dv\,dr
    + r^2 d\phi^{2},
    \label{eq:metric}
\end{equation}
where, setting the AdS radius to one, the blackening factor is given by:
\begin{equation}
    f(r)=r^2-r_h^2
    \label{eq:blackening}
\end{equation}
and the horizon is located at $r=r_h$. Calculating the extremal volume surface $\Sigma_{F_2}$ and its extrinsic curvature tensor, we find (see Appendix \ref{sec::bulk_d} for details) $K=0$ by definition, but
\begin{equation}
   K_{ab}K^{ab}
 \equiv 2X\neq0.
\end{equation}
Here, $X$ is a function of the spatial coordinate $r$ as well as the conserved momentum $\mathcal{P}_v$ along the extremal volume surface. It will later become important that at late times, the extremal slices $\Sigma_{F_2}$ tend towards a final slice defined by a constant radius $r=r_f$, where $X=1$ exactly.  
All possible higher order contractions of $K_{ab}$ can be written as powers of $X$ (or vanish outright), for instance  
\begin{align}
K^{\alpha}_{\beta}K^{\beta}_{\gamma}K^{\gamma}_{\alpha}&=0,
\\
K^{\alpha}_{\beta}K^{\beta}_{\gamma}K^{\gamma}_{\delta}K^{\delta}_{\alpha} &=2X^2,
\end{align}
and so on. Hence, for our setup, the most generic ansatz without derivatives acting on $K_{ab}$ is 
\begin{equation}
\begin{aligned}
    F_1&= \bar{\lambda}\left(1+ \sum_{n=0}^\infty \lambda_n \left(\frac{K_{\alpha \beta} K^{\alpha \beta}}{2}\right)^{n+1} \right)\,.
\label{eq.functional_F1}
\end{aligned}
\end{equation}

The free parameters of this function are the overall factor
\(\bar{\lambda}\) and the coefficients \(\lambda_n\), where \(n\) is
a non-negative integer. The term linear in
\(K_{\alpha\beta}K^{\alpha\beta}\) is related to the volume term through
the Gauss--Codazzi equation. On the maximal slice, \(K=0\), and in our
conventions
\[
K_{\alpha\beta}K^{\alpha\beta}
=
\mathcal R-2\Lambda
=
\mathcal R+2.
\]
The integral of \(\mathcal R\) is topological only after it is combined
with the appropriate boundary term, as required by the
Gauss--Bonnet theorem, and for fixed topology and anchoring data.
Subject to this qualification, and with the boundary terms and
counterterms treated consistently, the term linear in
\(K_{\alpha\beta}K^{\alpha\beta}\) can be absorbed into the volume
normalization together with topology-dependent contributions.
We therefore set \(\lambda_0=0\).\\

Investigating the functional \eqref{eq.functional_F1}, we find that a
term proportional to
\(X^p=(K_{\alpha\beta}K^{\alpha\beta}/2)^p\) first contributes to
\(\mathcal O(\tau)\) at order \(\tau^{2p}\). Since \(\lambda_n\)
multiplies \(X^{n+1}\) in Eq.~\eqref{eq.functional_F1}, the coefficient
\(\lambda_n\) first contributes at order \(\tau^{2n+2}\) in the early-time expansion of the generalized CAny observable $\cO(\tau)$ in \eqref{eq:Cany} , or equivalently at order \(\tau^{2n+1}\) in
\(d\mathcal O/d\tau\) (see
Appendix~\ref{sec::bulk_d} for the full derivation):
\begin{align} \frac{G_N}{V_{\phi}} \frac{d\mathcal O}{d\tau}
=
\bar{\lambda}\Bigg(
&\frac{8\pi^2}{\beta^3} \tau
+\frac{40\pi^2}{\beta^5}
\left(\lambda_1-1\right)\tau^3
\label{eq:ads3_early_beta_short}\\&-\frac{3\pi^2}{\beta^7}
\left(
85\lambda_1
-63\lambda_2
-74
\right)\tau^5 + \dots \Bigg) \nonumber
\end{align} 

Notice that the previously mentioned volume early-time series \eqref{eq:CV_early_time_series} arises from this by setting $\bar{\lambda}=1$ and $\lambda_n=0$ and then integrating in time once with boundary condition $\mathcal{O}_V(0)=0$. The remaining task is to see whether any choice of the parameters $\bar{\lambda}$ and $\lambda_n$ can match the boundary Krylov complexity, and in particular, whether any choice of $\bar{\lambda}$ and $\lambda_n$ can produce quadratic growth at late times. 

\subsection{Parameter matching and late-time growth}

In the semiclassical regime under consideration, the early-time expansion of the Krylov spread complexity, as extensively discussed in the previous section, is \eqref{eq:cft2_ck_x_series}. For this part, we explicitly make use of its time derivative, which reads 
\begin{equation}
\begin{aligned}
\lim_{c\to\infty}\frac{1}{c}\frac{d C_K(t)}{d t}
=\frac{\pi^2}{3}\left( \frac{4}{\beta^3}t-\frac{2}{\beta^5}t^3 + \frac{3}{\beta^7}t^5
    +\dots\right)\,.
\end{aligned}
\label{eq:ads3_krylov_early_time}
\end{equation}
Now comes the crux: The structure of \eqref{eq:ads3_early_beta_short} enables us to determine all parameters $\bar\lambda$ and $\lambda_n$ uniquely, analytically at low orders and numerically in general, by simply matching the bulk result for complexity \eqref{eq:ads3_early_beta_short} to the desired boundary result \eqref{eq:ads3_krylov_early_time} order by order in the early-time series. The triangular matching fixes one new bulk coefficient at every accessible order and thereby defines the candidate functional
recursively. The first coefficients obtained from this matching are
\begin{equation}
\label{eq.analyiticlambdas}
\bar\lambda=\frac{1}{6},\qquad
\lambda_1=\frac{9}{10},\qquad
\lambda_2=\frac{1}{14},\qquad
\lambda_3=\frac{43}{130}.
\end{equation}

A crucial consistency requirement is that the \(\lambda_n\) characterize the observable itself and therefore remain independent of the black-hole state under consideration. In particular, they should not acquire any explicit dependence on the inverse temperature \(\beta\) of the black hole. Such a dependence would amount to redefining the observable whenever the state is changed, thereby weakening the predictive content of the construction. Instead, all temperature dependence must arise from evaluating the same fixed functional \(F_1\) on different extremal slices. 
It is therefore inherited entirely through the geometry of the slice. The order-by-order matching should consequently yield a single set of state-independent coefficients \(\lambda_n\), valid across the family of thermal states rather than separately tuned for each value of \(\beta\).\\

\begin{figure}[htb]
     \centering
     \includegraphics[width=0.95\columnwidth]{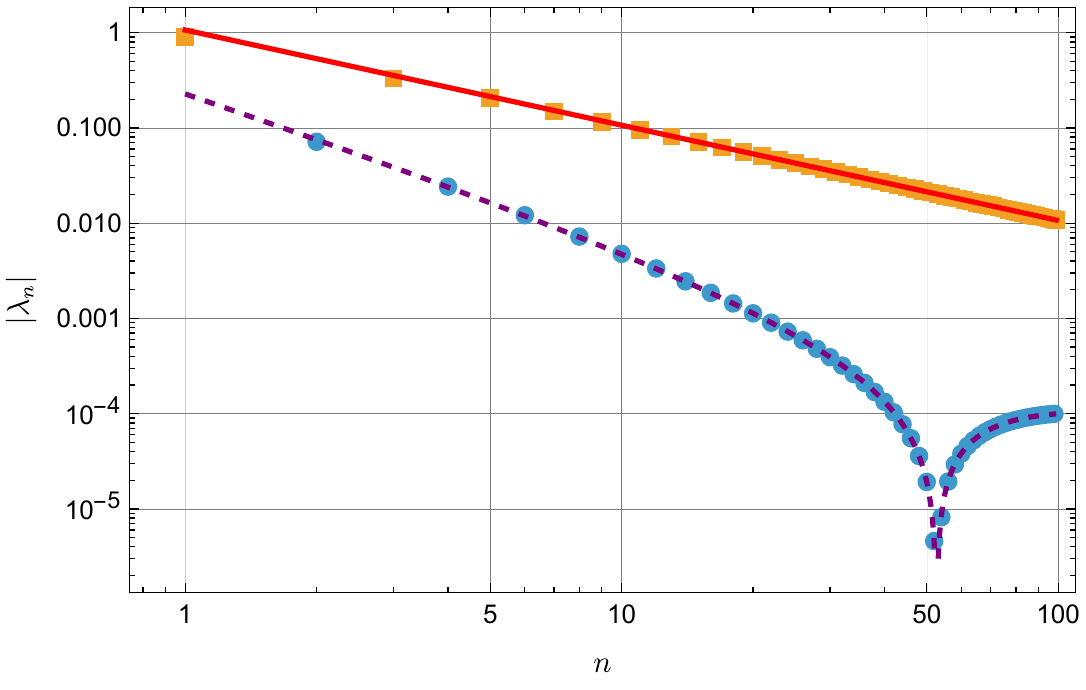}
     \caption{Numerical values of the parameters $\lambda_n$ for $1\leq n\leq 99$, obtained by matching bulk to boundary complexity order by order. There appear to be two different sub-series for even and odd $n$, marked by circles and squares respectively. The solid red and dashed purple lines show curves of the form \eqref{even_model} and \eqref{odd_model_2} respectively, fitted to the large $n$ behavior. The even coefficients switch sign for $n\geq 54 $.
     }
     \label{fig:lambda_n}
 \end{figure}

The boundary expansion through order \(t^{200}\) fixes $100$ bulk
parameters in total,
\(\bar\lambda,\lambda_1,\ldots,\lambda_{99}\), consisting of the
overall normalization and 99 nontrivial coefficients. The nontrivial
coefficients are shown in Fig.~\ref{fig:lambda_n}. We obtained them by utilizing very precise (as we had to take up to 100 derivatives) numerical differentiation applied to the bulk expression evaluated on a pseudospectral grid~\cite{Grandclement:2007sb} with 1400 points in~$t^2$ and matching with early-time Taylor series coefficients of $C_{K}$ on the boundary. We certify the numerical method by reproducing the lowest orders~\eqref{eq.analyiticlambdas} analytically up to the accuracy consistent with the inverse of the central charge and by numerical stability up to this accuracy with respect to changing number of grid points and number of significant digits.\\

Based on structural agreement of \eqref{eq:ads3_early_beta_short} and \eqref{eq:ads3_krylov_early_time}, the lambdas we find are independent of $\beta$. The fact that they source their dimensionality from the AdS curvature scale alone is in our view nontrivial and shows that the bulk object defined in exactly the same covariant way serves as a holographic Krylov spread complexity carrier for all black hole temperatures.\\

Fig.~\ref{fig:lambda_n} clearly illustrates that the $\lambda_n$ separate into two distinct series, one for even and one for odd values of $n$. Over the available large-\(n\) range, the odd-\(n\) coefficients are consistent with the asymptotic form
\begin{align}
\lambda_n\approx \frac{A_o}{n} \text{ for odd $n$ with } A_o\approx 1.066.
\label{even_model}
\end{align}
The even-\(n\) branch changes sign and is less stable under changes of
the fit window. A plausible asymptotic ansatz is
\begin{align}
&\lambda_n\approx \frac{A_e}{n}+\frac{B_e}{n^{3/2}} 
\label{odd_model_2}
\\
&\text{ for even $n$ with }A_e\approx -0.036,\ B_e\approx 0.26.
\nonumber
\end{align}
The exact numerical values are not crucial for the late-time argument. We therefore interpret
these fits as evidence consistent with a parity-resolved harmonic
tail, rather than as a determination of its exact asymptotic form. The relevant feature for the argument is the power-law form described below.\\

What is important to us is the observation that if we were to truncate the series in \eqref{eq.functional_F1} at some finite order $n_{\max}$, the resulting bulk complexity would match the boundary complexity only at early times, but ultimately transition to linear growth at sufficiently late times (as expected from standard CAny \cite{Belin:2021bga,Belin:2022xmt}). 
Let us assume instead that the series in \eqref{eq.functional_F1} is a true infinite series, with the large-$n$ behavior of the parameters $\lambda_n$ described accurately by the models \eqref{even_model} and \eqref{odd_model_2}. The corresponding fitted functional becomes
\begin{align}
F_{\mathrm{fit}}(X)
=
\bar{\lambda}
\Bigg[
1
&+A_o X\,\text{ArcTanh}(X)
-\frac{A_e}{2}X\log\!\left(1-X^2\right)
\nonumber
\\
&+\frac{B_e}{2\sqrt{2}}X\,\operatorname{Li}_{3/2}\!\left(X^2\right)
\Bigg].
\label{eq:afit_resummed}
\end{align}
Here \(\operatorname{Li}_{3/2}\) denotes the polylogarithm of order \(3/2\). We thus see that the functional \eqref{eq:afit_resummed} contains contributions such as 
\begin{align}
-\frac{1}{2}X\log(1-X^2)&=\sum_{n\geq2,\ \text{even}}^{\infty}  \frac{X^{n+1}}{n}, 
\\
X\text{ArcTanh}(X)&=\sum_{n\geq1,\ \text{odd}}^{\infty}  \frac{X^{n+1}}{n}, 
\end{align}
which both diverge as $-\log(1-X)$ for $X\rightarrow1$
\footnote{In general, the polylogarithm function $\text{Li}_{a}(X)=\Sigma_{n=1}^\infty \frac{X^n}{n^a}$ diverges at $X=1$ for $a\leq1$}, corresponding to the late-time limit where the extremal volume surfaces $\Sigma_{F_2}$ asymptote to the final slice at $r=r_f$. 
\\

\begin{figure}[htb]
    \centering 
\includegraphics[width=0.9\columnwidth]{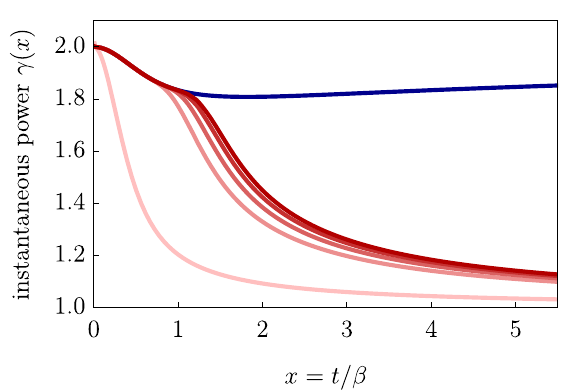}
\includegraphics[width=0.9\columnwidth]{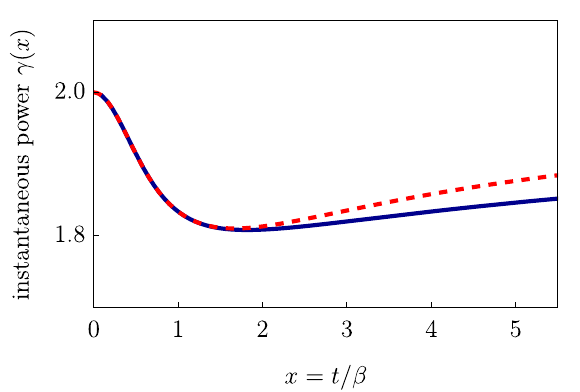}
     \caption{ \emph{Top panel:} The dark blue curve is the diagonal
\([100/100]\) Pad\'e approximant constructed from the boundary
early-time expansion through order \(t^{200}\) (as in Fig.~\ref{fig:cft2_pade}). The red curves show
the bulk result obtained by truncating the series defining the local
functional at
\(
n_{\max}=0,25,50,75,99
\),
with the colour changing continuously from light to dark red as
\(n_{\max}\) increases. \\
\emph{Bottom panel:} Comparison between the same boundary Pad\'e approximant as above (dark blue) and the hybrid infinite-order bulk resummation \eqref{eq:hybrid_functional} (dashed red).
In the hybrid construction, the coefficients
\(\lambda_1,\ldots,\lambda_{99}\) are fixed to their numerically matched values obtained
from the order-by-order bulk-boundary matching, while for \(n\geq100\)
the coefficients are replaced by their asymptotic fits \eqref{even_model} and \eqref{odd_model_2}, whose infinite tail is resummed analytically using the closed-form expression~\eqref{eq:afit_resummed}. 
}    \label{fig:Ads.logder}
\end{figure}

Let us finally understand what precise bulk complexity the $\lambda_n$ we solved for give rise to. The results are shown in Fig.~\ref{fig:Ads.logder}, where we again display the instantaneous power
\(
\gamma(x)=x\,\partial_x\log C_K(x)
\),
where \(x=t/\beta\). Fig.~\ref{fig:Ads.logder} illustrates that increasing the number of matched bulk coefficients extends the interval over which the bulk result follows the boundary continuation. For comparison, we use the highest-order reliable diagonal Pad\'e approximant shown in Fig.~\ref{fig:cft2_pade}. If the full coefficient sequence were known and resummed, the resulting bulk observable would reproduce the boundary early-time series to all orders; the Pad\'e curve is used here only as a finite-time representation of that boundary target. Nevertheless, every finite truncation eventually departs from the Pad\'e curve and approaches
\(\gamma=1\), corresponding to the linear late-time growth expected for
a finite-order \emph{CAny} observable whose evaluated functional never diverges, not even on the accumulation surface. The convergence with increasing truncation order is, however, remarkably slow: increasing \(n_{\max}\) progressively
shifts the onset of the asymptotic linear regime to later times, but no
finite truncation can remove this crossover. This same qualitative behavior is reproduced by the toy model discussed in App.~\ref{subsec:toy_model_bulk_pade}. By contrast,
the hybrid infinite-order resummation shown in the lower panel of Fig.~\ref{fig:Ads.logder}, in which the low-lying coefficients are kept at their numerically matched values while the large-\(n\) tail is completed using the asymptotic fits \eqref{even_model} and \eqref{odd_model_2}, indeed avoids this
finite-order crossover and retains the singular contribution required
to sustain the non-linear late-time behavior suggested by the boundary
Pad\'e continuation. For comparison with the numerical result, we use the hybrid completion
\begin{align}
F_{\mathrm{hyb}}(X)
=
\bar\lambda\left[
1+\sum_{n=1}^{99}\lambda_n X^{n+1}
+\sum_{n=100}^{\infty}\lambda_n^{(\mathrm{fit})}X^{n+1}
\right],
\label{eq:hybrid_functional}
\end{align}
where the first 99 coefficients are the computed values obtained
by order-by-order matching, while only the uncomputed tail is supplied by
the parity-resolved asymptotic fits. Any remaining discrepancy therefore measures the limitations of the Pad\'e approximant on the one hand and of the
fitted continuation of the uncomputed large-\(n\) tail on the other hand, rather than an
approximation of the known low-lying coefficients.\\

The $-\log(1-X)$-like singular behavior of \eqref{eq.functional_F1} is not accidental, but is precisely what makes
quadratic late-time growth possible. For a broad class of observables
defined on codimension-one extremal surfaces \(\Sigma_{F_2}\), the
surfaces approach at late times a stationary final slice at a constant
radial position \(r=r_f\) \cite{Belin:2021bga,Belin:2022xmt}. If the local functional evaluated on this limiting slice remains finite, its
late-time growth rate approaches a constant determined by the
corresponding effective potential, and the observable consequently
grows linearly with time. This is the mechanism responsible for the
limit \(\gamma\to1\) displayed by every finite truncation in
Fig.~\ref{fig:Ads.logder}.\\

The infinite-order functional considered here behaves differently.
The resummation of the large-\(n\) coefficients develops a logarithmic
singularity as the final slice is approached, so that the assumptions underlying the generic linear-growth
argument cease to apply. As discussed in detail in App.~\ref{sec::bulk_d} and in the toy model in App.~\ref{subsec:toy_model_bulk_pade}, the critical asymptotic scaling
\(\lambda_n\sim 1/n\) is singled out by the requirement of late-time quadratic growth. A faster decay
of the coefficients makes
\(F_1\) finite and restores linear growth,
whereas a tail decreasing slower than \(1/n\) generates a singularity stronger than the logarithmic one and leads to faster than quadratic growth at late times. Fig.~\ref{fig:Ads.logder} thus illustrates that the approach to \(\gamma=1\) is a consequence of truncating the bulk functional at finite order, rather than an unavoidable feature of the complete infinite-order observable.\\

These results suggest that the infinite-order CAny functional obtained by matching the boundary expansion provides a concrete candidate for the geometric dual of Krylov spread complexity in AdS$_3$/CFT$_2$ at the semiclassical level. Its state-independent couplings reproduce the early-time boundary data order by order, while their behavior consistent with a parity-resolved harmonic tail
over the computed range supplies the singular resummation required to evade the generic linear-growth regime and support quadratic pre-saturation behavior compatible with the Pad\'e continuation. The proposal should however be regarded not as a unique holographic construction, but as evidence that semiclassical Krylov complexity admits a natural realization in terms of a suitably generalized bulk geometric observable.
\\

\section{Conclusions and Outlook}

The development of holographic complexity was largely built on the
premise that its boundary counterpart is circuit complexity: The
multitude of bulk proposals was taken to mirror the multitude of
admissible gate sets and cost functions, all sharing universal
features such as late-time linear growth in black hole states
\cite{Susskind:2014moa, Stanford:2014jda, Susskind:2014rva,
Belin:2021bga, Belin:2022xmt}. Yet, in the only setting in which a
holographic complexity proposal has been accounted for
microscopically, as it stands for now, the prime boundary candidate turned out to be Krylov
spread complexity \cite{Lin:2022rbf,Rabinovici:2023yex, Heller:2024ldz}, rather than circuit complexity, with the match deeply
rooted in the peculiarities of SYK holography. Our work takes this
finding seriously, with the motivating conjecture that Krylov spread
complexity admits a bulk realization beyond two-dimensional
gravity. Testing this conjecture required two ingredients: the
ability to compute spread complexity in a semiclassical holographic
theory with spacetime dimension larger than two, and a bulk object matching it.\\

The first ingredient is supplied by the \textit{CfZ}
construction~\eqref{eq.CfZ}: For thermofield-double states, all
Hamiltonian moments and through them the complete Lanczos
data follow from derivatives of the thermal partition
function. Black hole thermodynamics, in the guise of the on-shell
action of the dominant saddle, thereby becomes a generator of
boundary Krylov dynamics. Benchmarking against DSSYK,
where the exact partition function and its semiclassical saddle are
simultaneously available, revealed a sharp order-of-limits rule: The
classical limit may be taken only after the complexity has been
assembled, because subleading terms in individual Lanczos
coefficients feed the leading classical complexity through delicate
cancellations. We regard this
lesson as significant well beyond holography, wherever Krylov
methods meet semiclassical, or otherwise asymptotic
input.\\

Applying \textit{CfZ} to the Cardy saddle of a two-dimensional holographic CFT dual to the BTZ black hole produced an explicit short-time prediction for the spread complexity of the thermofield-double state at large central charge. The direct series
establishes universal quadratic short-time growth and a subsequent departure from it inside its convergence domain. Beyond that domain,
the highest available diagonal Padé approximants turn upward toward two over the accessible continuation range. This behavior is compatible with, but does not by itself establish, quadratic
pre-saturation asymptotics. This is consistent with the asymptotic analyses of
\cite{Balasubramanian:2022tpr,Balasubramanian:2026azk}. Quadratic early- and late-time asymptotics do not by themselves imply
eternally quadratic growth, and the crossover we identify is a genuine feature of the semiclassical answer. This also connects to the
quadratic growth of complexity recently found in \cite{Balasubramanian:2026azk} from the Maloney–Witten–Keller partition function \cite{Maloney:2007ud,Keller:2014xba}.\\

An important feature of this crossover is its timescale. Expressed in terms of
\(x=t/\beta\), the departure from the initial quadratic regime, the minimum of
the instantaneous exponent, and the onset of its return toward two all occur at
\(x=O(1)\). The qualitative reorganization of the growth law is therefore
parametrically thermal: it is already present at times of order \(\beta\),
rather than waiting for the scrambling time
\(t_{\mathrm{scr}}\sim\beta\log c\), which is parametrically later in the
semiclassical limit. The final convergence of \(\gamma(x)\) to two remains slow
and is not established by the present Pad\'e sequence alone; nevertheless, the
crossover is initiated and reversed on the thermal scale. This separation
is consistent with the effect being encoded in the equilibrium black-hole saddle and in
the Krylov dynamics of the unperturbed thermofield-double state, rather than
being triggered by the onset of scrambling.\\

In the BTZ setup, CV settles into linear growth at times of order $\beta$, and the standard finite-functional CAny class grows linearly at late times \emph{by design}: As
long as the defining functionals remain finite, in particular on the final slice
inside the horizon, the growth rate asymptotes to a constant set by
the effective potential evaluated on the accumulation surface \cite{Belin:2021bga, Belin:2022xmt,
Carmi:2017jqz}. The quadratic pre-saturation behavior supported by the boundary and bulk analyses therefore indicates that Krylov
spread complexity of the thermofield-double state lies outside the
CAny class as originally envisioned. This prominently includes CV, where we observed this mismatch also by finding non-reconcilable early-time expansions. Identifying the assumption
responsible, namely finiteness on the final slice, also locates the
loophole through which a bulk representation can nevertheless
exist.\\

We exploited this loophole by constructing a generalized
observable: extremal-volume slices weighted by an infinite series
of extrinsic-curvature invariants, with coefficients fixed order by
order by the boundary series, and
with a single temperature-independent set of coefficients
accounting for the entire family of thermal states. Every finite
truncation ultimately reverts to linear growth, while the resummed
infinite series can diverge on the final slice, and this divergence is the mechanism by which the candidate functional can support quadratic pre-saturation growth; we see the divergence explicitly arising, as the odd sequence of coefficients we solved for is consistent with a \(1/n\) tail over the computed range, which resummed, gives rise to a logarithmically diverging functional on the accumulation surface.\\

The distinction between linear and quadratic growth is not merely a difference
in exponent. A finite CAny density on the final slice produces a stationary
complexity flux: its growth rate approaches a constant, and the corresponding
observable grows linearly. By contrast, \(C_K(t)\sim t^2\) means that the
Krylov growth rate itself remains proportional to time, so that the mean
position of the state continues to accelerate along the Krylov chain throughout
the semiclassical late-time window. In the bulk, sustaining this behavior
requires an infinite tower of invariants whose resummation becomes singular on
the accumulation surface. The late-time exponent therefore diagnoses the
nature of the holographic dictionary: linear growth is natural for finite local
geometric densities, whereas quadratic Krylov growth points to a genuinely
resummed observable probing a different aspect of state-space exploration. At
finite~$c$, both regimes must ultimately give way to saturation; the
distinction concerns the parametrically long semiclassical window rather than
literal eternal growth.

Our gravitational construction reproduces the boundary behavior by
construction, and it is certainly not unique. Allowing
analogous infinite series to define both the surface and the
functional evaluated on it would even suggest one real parameter of ambiguity
at each order in both sums. We therefore view our result as an
existence statement, and the pressing question is what constrains
the space of admissible functionals. 
Let us outline one strategy towards tackling this question. Modeled after holographic entanglement
entropy, one key requirement is internal consistency: The same bulk prescription should
compute Krylov spread complexity for different states and
evolutions within one gravity dual. Our functional already passes
the simplest version of this test, since a single
temperature-independent set of coefficients accounts for the whole
family of BTZ black holes. Natural next tests include boosted time
slices, rotating BTZ, and locally perturbed thermofield-double
states of shockwave type, where switchback-like behavior of the
growth would provide a stringent benchmark \cite{Stanford:2014jda,
Ambrosini:2025hvo}. Seeking such internal self-consistency
inevitably retains a flavor of searching for a needle in a
haystack, but at least the haystack is now well defined.\\

Because \textit{CfZ} takes nothing but $Z(\beta)$ as input, our general method is
in principle not confined to the classical limit. As long as we can ensure computational tractability, wherever the
partition function is known beyond the leading saddle, the construction defines
subleading corrections to, and in principle the full finite-$c$
version of, the boundary complexity. We showed that one-loop
factors drop out at leading classical order, but the subleading orders
might be the natural place to search for the onset of complexity
saturation at exponentially late times, which is invisible at
strictly infinite $c$ and to classical
geometric proposals. One interesting possible application of \textit{CfZ} in a similar context for DSSYK might be the partition functions studied in \cite{Beccaria:2026ndg}.\\ 

Finally, essentially none of our analysis is tied to three bulk
dimensions. On the boundary, \textit{CfZ} applies verbatim to
planar AdS$_{d+1}$ saddles, for which we can already derive the
classical early-time series; continuing beyond early times seems to require
different techniques than Padé, on which we plan to report elsewhere
\cite{Bhattacharya_wip}. On the gravity side, the higher-dimensional
problem is richer: The geometry is no longer locally AdS,
so bulk curvature invariants and a larger set of independent
extrinsic invariants become available as building blocks. This
raises the stakes for the existence of a state-independent
functional, thereby sharpening the internal-consistency
requirement. Charged and rotating black holes, with grand-canonical
partition functions feeding a suitably extended \textit{CfZ},
supply further families of states on which any candidate functional
must agree.\\

Taken together, these directions outline a program in
which holographic complexity is no longer postulated, but can potentially be
reconstructed order by order from the partition function.

\vspace{0.2cm}
\begin{acknowledgments}
We would like to thank S. E. Aguilar-Gutierrez, P. Caputa, R. N. Das, K. Ghosh, J. Kastikainen, B. Kent, J. Papalini, A. Sanchez-Garrido, T. Tappeiner and T. G. Mertens for insightful discussions and/or comments on the draft and V. Patil for collaboration during earlier stages of this project. T.S. is supported by the Research Foundation - Flanders (FWO) doctoral fellowship 11I5425N. T.S. acknowledges the hospitality of the University of California at Berkeley where a core part of this work was carried out, and financial support through the FWO grant for a long stay abroad V436025N and a UGent SIP grant during this stay. This work was partially supported by the FWO-NWO Weave project G0ADL26N and by the Priority Research Area Digiworld under the program Excellence Initiative - Research University at the Jagiellonian University in Krakow. The work of both MF and ER as well as, up until 31.08.2025, AB was supported by the Polish National Science Centre (NCN) grant $2021/42/E/ST2/00234$. AB acknowledges support from United Kingdom Research and Innovation (UKRI) under the UK government’s Horizon Europe guarantee ($EP/Y00468X/1$) (since 01.09.2025). We benefited from the use of ChatGPT 5.4 and 5.6 and Claude 4.8 and 5.0 running on the highest available settings to tighten our arguments, in particular regarding late time behavior on both the boundary and the bulk, and to improve the manuscript presentation. All analytic
derivations, numerical checks, physical interpretations, and
conclusions were independently verified by the authors.

\quad The data and codes that
support the findings of this article are openly available at \cite{UJ/CESIIM_2026}. For the purpose of open
access, the authors have applied a Creative Commons Attribution (CC BY 4.0) licence to any
Author Accepted Manuscript version arising from this submission.
\end{acknowledgments}

\bibliographystyle{bibstyl}
\bibliography{refs}

\providecommand{\href}[2]{#2} \providecommand{\beforedoihref}{} \providecommand{\afterdoihref}{}\begingroup\raggedright\begin{thebibliography}{10}

\bibitem{Maldacena:1997re}
J.~M. Maldacena, {\it {The Large $N$ limit of superconformal field theories and supergravity}},  \beforedoihref\href{http://dx.doi.org/10.4310/ATMP.1998.v2.n2.a1}{Adv. Theor. Math. Phys.}\afterdoihref\  {\bf 2} (1998) 231--252 [\href{http://arXiv.org/abs/hep-th/9711200}{{arXiv:hep-th/9711200}}].

\bibitem{Gubser:1998bc}
S.~S. Gubser, I.~R. Klebanov and A.~M. Polyakov, {\it {Gauge theory correlators from noncritical string theory}},  \beforedoihref\href{http://dx.doi.org/10.1016/S0370-2693(98)00377-3}{Phys. Lett. B}\afterdoihref\  {\bf 428} (1998) 105--114 [\href{http://arXiv.org/abs/hep-th/9802109}{{arXiv:hep-th/9802109}}].

\bibitem{Witten:1998qj}
E.~Witten, {\it {Anti de Sitter space and holography}},  \beforedoihref\href{http://dx.doi.org/10.4310/ATMP.1998.v2.n2.a2}{Adv. Theor. Math. Phys.}\afterdoihref\  {\bf 2} (1998) 253--291 [\href{http://arXiv.org/abs/hep-th/9802150}{{arXiv:hep-th/9802150}}].

\bibitem{Ryu:2006bv}
S.~Ryu and T.~Takayanagi, {\it {Holographic derivation of entanglement entropy from AdS/CFT}},  \beforedoihref\href{http://dx.doi.org/10.1103/PhysRevLett.96.181602}{Phys. Rev. Lett.}\afterdoihref\  {\bf 96} (2006) 181602 [\href{http://arXiv.org/abs/hep-th/0603001}{{arXiv:hep-th/0603001}}].

\bibitem{Casini:2011kv}
H.~Casini, M.~Huerta and R.~C. Myers, {\it {Towards a derivation of holographic entanglement entropy}},  \beforedoihref\href{http://dx.doi.org/10.1007/JHEP05(2011)036}{JHEP}\afterdoihref\  {\bf 05} (2011) 036 [\href{http://arXiv.org/abs/1102.0440}{{arXiv:1102.0440}}].

\bibitem{Lewkowycz:2013nqa}
A.~Lewkowycz and J.~Maldacena, {\it {Generalized gravitational entropy}},  \beforedoihref\href{http://dx.doi.org/10.1007/JHEP08(2013)090}{JHEP}\afterdoihref\  {\bf 08} (2013) 090 [\href{http://arXiv.org/abs/1304.4926}{{arXiv:1304.4926}}].

\bibitem{Dong:2016hjy}
X.~Dong, A.~Lewkowycz and M.~Rangamani, {\it {Deriving covariant holographic entanglement}},  \beforedoihref\href{http://dx.doi.org/10.1007/JHEP11(2016)028}{JHEP}\afterdoihref\  {\bf 11} (2016) 028 [\href{http://arXiv.org/abs/1607.07506}{{arXiv:1607.07506}}].

\bibitem{Susskind:2014moa}
L.~Susskind, {\it {Entanglement is not enough}},  \beforedoihref\href{http://dx.doi.org/10.1002/prop.201500095}{Fortsch. Phys.}\afterdoihref\  {\bf 64} (2016) 49--71 [\href{http://arXiv.org/abs/1411.0690}{{arXiv:1411.0690}}].

\bibitem{Stanford:2014jda}
D.~Stanford and L.~Susskind, {\it {Complexity and Shock Wave Geometries}},  \beforedoihref\href{http://dx.doi.org/10.1103/PhysRevD.90.126007}{Phys. Rev. D}\afterdoihref\  {\bf 90} (2014), no.~12 126007 [\href{http://arXiv.org/abs/1406.2678}{{arXiv:1406.2678}}].

\bibitem{Susskind:2014rva}
L.~Susskind, {\it {Computational Complexity and Black Hole Horizons}},  \beforedoihref\href{http://dx.doi.org/10.1002/prop.201500092}{Fortsch. Phys.}\afterdoihref\  {\bf 64} (2016) 24--43 [\href{http://arXiv.org/abs/1403.5695}{{arXiv:1403.5695}}]. [Addendum: Fortsch.Phys. 64, 44--48 (2016)].

\bibitem{Belin:2021bga}
A.~Belin, R.~C. Myers, S.-M. Ruan, G.~S{\'a}rosi and A.~J. Speranza, {\it {Does Complexity Equal Anything?}},  \beforedoihref\href{http://dx.doi.org/10.1103/PhysRevLett.128.081602}{Phys. Rev. Lett.}\afterdoihref\  {\bf 128} (2022), no.~8 081602 [\href{http://arXiv.org/abs/2111.02429}{{arXiv:2111.02429}}].

\bibitem{Belin:2022xmt}
A.~Belin, R.~C. Myers, S.-M. Ruan, G.~S{\'a}rosi and A.~J. Speranza, {\it {Complexity equals anything II}},  \beforedoihref\href{http://dx.doi.org/10.1007/JHEP01(2023)154}{JHEP}\afterdoihref\  {\bf 01} (2023) 154 [\href{http://arXiv.org/abs/2210.09647}{{arXiv:2210.09647}}].

\bibitem{Brown:2015bva}
A.~R. Brown, D.~A. Roberts, L.~Susskind, B.~Swingle and Y.~Zhao, {\it {Holographic Complexity Equals Bulk Action?}},  \beforedoihref\href{http://dx.doi.org/10.1103/PhysRevLett.116.191301}{Phys. Rev. Lett.}\afterdoihref\  {\bf 116} (2016), no.~19 191301 [\href{http://arXiv.org/abs/1509.07876}{{arXiv:1509.07876}}].

\bibitem{Brown:2015lvg}
A.~R. Brown, D.~A. Roberts, L.~Susskind, B.~Swingle and Y.~Zhao, {\it {Complexity, action, and black holes}},  \beforedoihref\href{http://dx.doi.org/10.1103/PhysRevD.93.086006}{Phys. Rev. D}\afterdoihref\  {\bf 93} (2016), no.~8 086006 [\href{http://arXiv.org/abs/1512.04993}{{arXiv:1512.04993}}].

\bibitem{Couch:2016exn}
J.~Couch, W.~Fischler and P.~H. Nguyen, {\it {Noether charge, black hole volume, and complexity}},  \beforedoihref\href{http://dx.doi.org/10.1007/JHEP03(2017)119}{JHEP}\afterdoihref\  {\bf 03} (2017) 119 [\href{http://arXiv.org/abs/1610.02038}{{arXiv:1610.02038}}].

\bibitem{Sachdev:1992fk}
S.~Sachdev and J.~Ye, {\it {Gapless spin fluid ground state in a random, quantum Heisenberg magnet}},  \beforedoihref\href{http://dx.doi.org/10.1103/PhysRevLett.70.3339}{Phys. Rev. Lett.}\afterdoihref\  {\bf 70} (1993) 3339 [\href{http://arXiv.org/abs/cond-mat/9212030}{{arXiv:cond-mat/9212030}}].

\bibitem{kitaevvideo}
A.~Kitaev, ``A simple model of quantum holography.'' \url{http://online.kitp.ucsb.edu/online/entangled15/kitaev/} and \url{http://online.kitp.ucsb.edu/online/entangled15/kitaev2/}, 2015.

\bibitem{Maldacena:2016hyu}
J.~Maldacena and D.~Stanford, {\it {Remarks on the Sachdev-Ye-Kitaev model}},  \beforedoihref\href{http://dx.doi.org/10.1103/PhysRevD.94.106002}{Phys. Rev. D}\afterdoihref\  {\bf 94} (2016), no.~10 106002 [\href{http://arXiv.org/abs/1604.07818}{{arXiv:1604.07818}}].

\bibitem{Harlow:2018tqv}
D.~Harlow and D.~Jafferis, {\it {The Factorization Problem in Jackiw-Teitelboim Gravity}},  \beforedoihref\href{http://dx.doi.org/10.1007/JHEP02(2020)177}{JHEP}\afterdoihref\  {\bf 02} (2020) 177 [\href{http://arXiv.org/abs/1804.01081}{{arXiv:1804.01081}}].

\bibitem{Lin:2022rbf}
H.~W. Lin, {\it {The bulk Hilbert space of double scaled SYK}},  \beforedoihref\href{http://dx.doi.org/10.1007/JHEP11(2022)060}{JHEP}\afterdoihref\  {\bf 11} (2022) 060 [\href{http://arXiv.org/abs/2208.07032}{{arXiv:2208.07032}}].

\bibitem{Rabinovici:2023yex}
E.~Rabinovici, A.~S{\'a}nchez-Garrido, R.~Shir and J.~Sonner, {\it {A bulk manifestation of Krylov complexity}},  \beforedoihref\href{http://dx.doi.org/10.1007/JHEP08(2023)213}{JHEP}\afterdoihref\  {\bf 08} (2023) 213 [\href{http://arXiv.org/abs/2305.04355}{{arXiv:2305.04355}}].

\bibitem{Heller:2024ldz}
M.~P. Heller, J.~Papalini and T.~Schuhmann, {\it {Krylov Spread Complexity as Holographic Complexity beyond Jackiw-Teitelboim Gravity}},  \beforedoihref\href{http://dx.doi.org/10.1103/spcr-jgm6}{Phys. Rev. Lett.}\afterdoihref\  {\bf 135} (2025), no.~15 151602 [\href{http://arXiv.org/abs/2412.17785}{{arXiv:2412.17785}}].

\bibitem{Balasubramanian:2022tpr}
V.~Balasubramanian, P.~Caputa, J.~M. Magan and Q.~Wu, {\it {Quantum chaos and the complexity of spread of states}},  \beforedoihref\href{http://dx.doi.org/10.1103/PhysRevD.106.046007}{Phys. Rev. D}\afterdoihref\  {\bf 106} (2022), no.~4 046007 [\href{http://arXiv.org/abs/2202.06957}{{arXiv:2202.06957}}].

\bibitem{Parker:2018yvk}
D.~E. Parker, X.~Cao, A.~Avdoshkin, T.~Scaffidi and E.~Altman, {\it {A Universal Operator Growth Hypothesis}},  \beforedoihref\href{http://dx.doi.org/10.1103/PhysRevX.9.041017}{Phys. Rev. X}\afterdoihref\  {\bf 9} (2019), no.~4 041017 [\href{http://arXiv.org/abs/1812.08657}{{arXiv:1812.08657}}].

\bibitem{Caputa:2021ori}
P.~Caputa and S.~Datta, {\it {Operator growth in 2d CFT}},  \beforedoihref\href{http://dx.doi.org/10.1007/JHEP12(2021)188}{JHEP}\afterdoihref\  {\bf 12} (2021) 188. [Erratum: JHEP 09, 113 (2022)].

\bibitem{Dymarsky:2021bjq}
A.~Dymarsky and M.~Smolkin, {\it {Krylov complexity in conformal field theory}},  \beforedoihref\href{http://dx.doi.org/10.1103/PhysRevD.104.L081702}{Phys. Rev. D}\afterdoihref\  {\bf 104} (2021), no.~8 L081702 [\href{http://arXiv.org/abs/2104.09514}{{arXiv:2104.09514}}].

\bibitem{Kundu:2023hbk}
A.~Kundu, V.~Malvimat and R.~Sinha, {\it {State dependence of Krylov complexity in 2d CFTs}},  \beforedoihref\href{http://dx.doi.org/10.1007/JHEP09(2023)011}{JHEP}\afterdoihref\  {\bf 09} (2023) 011 [\href{http://arXiv.org/abs/2303.03426}{{arXiv:2303.03426}}].

\bibitem{Erdmenger:2023wjg}
J.~Erdmenger, S.-K. Jian and Z.-Y. Xian, {\it {Universal chaotic dynamics from Krylov space}},  \beforedoihref\href{http://dx.doi.org/10.1007/JHEP08(2023)176}{JHEP}\afterdoihref\  {\bf 08} (2023) 176 [\href{http://arXiv.org/abs/2303.12151}{{arXiv:2303.12151}}].

\bibitem{Bhattacharya:2023zqt}
A.~Bhattacharya, P.~Nandy, P.~P. Nath and H.~Sahu, {\it {On Krylov complexity in open systems: an approach via bi-Lanczos algorithm}},  \beforedoihref\href{http://dx.doi.org/10.1007/JHEP12(2023)066}{JHEP}\afterdoihref\  {\bf 12} (2023) 066 [\href{http://arXiv.org/abs/2303.04175}{{arXiv:2303.04175}}].

\bibitem{Baggioli:2024wbz}
M.~Baggioli, K.-B. Huh, H.-S. Jeong, K.-Y. Kim and J.~F. Pedraza, {\it {Krylov complexity as an order parameter for quantum chaotic-integrable transitions}},  \beforedoihref\href{http://dx.doi.org/10.1103/PhysRevResearch.7.023028}{Phys. Rev. Res.}\afterdoihref\  {\bf 7} (2025), no.~2 023028 [\href{http://arXiv.org/abs/2407.17054}{{arXiv:2407.17054}}].

\bibitem{Caputa:2024sux}
P.~Caputa, B.~Chen, R.~W. McDonald, J.~Sim{\'o}n and B.~Strittmatter, {\it {Spread complexity rate as proper momentum}},  \beforedoihref\href{http://dx.doi.org/10.1103/7zs8-9zpg}{Phys. Rev. D}\afterdoihref\  {\bf 113} (2026), no.~4 L041901 [\href{http://arXiv.org/abs/2410.23334}{{arXiv:2410.23334}}].

\bibitem{Jeong:2024jjn}
H.-S. Jeong, A.~Kundu and J.~F. Pedraza, {\it {Brickwall one-loop determinant: spectral statistics {\&} Krylov complexity}},  \beforedoihref\href{http://dx.doi.org/10.1007/JHEP05(2025)154}{JHEP}\afterdoihref\  {\bf 05} (2025) 154 [\href{http://arXiv.org/abs/2412.12301}{{arXiv:2412.12301}}].

\bibitem{Bhattacharya:2023yec}
A.~Bhattacharya, R.~N. Das, B.~Dey and J.~Erdmenger, {\it {Spread complexity for measurement-induced non-unitary dynamics and Zeno effect}},  \beforedoihref\href{http://dx.doi.org/10.1007/JHEP03(2024)179}{JHEP}\afterdoihref\  {\bf 03} (2024) 179 [\href{http://arXiv.org/abs/2312.11635}{{arXiv:2312.11635}}].

\bibitem{Das:2024tnw}
R.~N. Das, S.~Demulder, J.~Erdmenger and C.~Northe, {\it {Spread complexity for the planar limit of holography}},  \beforedoihref\href{http://dx.doi.org/10.1007/JHEP06(2025)166}{JHEP}\afterdoihref\  {\bf 06} (2025) 166 [\href{http://arXiv.org/abs/2412.09673}{{arXiv:2412.09673}}].

\bibitem{Basu:2024tgg}
R.~Basu, A.~Ganguly, S.~Nath and O.~Parrikar, {\it {Complexity growth and the Krylov-Wigner function}},  \beforedoihref\href{http://dx.doi.org/10.1007/JHEP05(2024)264}{JHEP}\afterdoihref\  {\bf 05} (2024) 264 [\href{http://arXiv.org/abs/2402.13694}{{arXiv:2402.13694}}]. [Erratum: JHEP 03, 202 (2026)].

\bibitem{Aguilar-Gutierrez:2025kmw}
S.~E. Aguilar-Gutierrez, H.~A. Camargo, V.~Jahnke, K.-Y. Kim and M.~Nishida, {\it {Krylov operator complexity in holographic CFTs: Smeared boundary reconstruction and the dual proper radial momentum}},  \beforedoihref\href{http://dx.doi.org/10.1103/6bgg-vglp}{Phys. Rev. D}\afterdoihref\  {\bf 112} (2025), no.~12 126014 [\href{http://arXiv.org/abs/2506.03273}{{arXiv:2506.03273}}].

\bibitem{Ambrosini:2024sre}
M.~Ambrosini, E.~Rabinovici, A.~S{\'a}nchez-Garrido, R.~Shir and J.~Sonner, {\it {Operator K-complexity in DSSYK: Krylov complexity equals bulk length}},  \beforedoihref\href{http://dx.doi.org/10.1007/JHEP08(2025)059}{JHEP}\afterdoihref\  {\bf 08} (2025) 059 [\href{http://arXiv.org/abs/2412.15318}{{arXiv:2412.15318}}].

\bibitem{Fu:2025kkh}
Y.~Fu, H.-S. Jeong, K.-Y. Kim and J.~F. Pedraza, {\it {Toward Krylov-based holography in double-scaled SYK}},  \beforedoihref\href{http://dx.doi.org/10.1007/JHEP05(2026)056}{JHEP}\afterdoihref\  {\bf 05} (2026) 056 [\href{http://arXiv.org/abs/2510.22658}{{arXiv:2510.22658}}].

\bibitem{Angelinos:2025drf}
N.~Angelinos, D.~Chakraborty and A.~Dymarsky, {\it {Temperature dependence in Krylov space}},  \beforedoihref\href{http://dx.doi.org/10.1088/1751-8121/ae7167}{J. Phys. A}\afterdoihref\  {\bf 59} (2026), no.~23 235204 [\href{http://arXiv.org/abs/2508.19233}{{arXiv:2508.19233}}].

\bibitem{Ambrosini:2025hvo}
M.~Ambrosini, E.~Rabinovici and J.~Sonner, {\it {Holography of K-complexity: switchbacks and shockwaves}},  \beforedoihref\href{http://dx.doi.org/10.1007/JHEP06(2026)060}{JHEP}\afterdoihref\  {\bf 06} (2026) 060 [\href{http://arXiv.org/abs/2510.17975}{{arXiv:2510.17975}}].

\bibitem{Fatemiabhari:2025cyy}
A.~Fatemiabhari, H.~Nastase and D.~Roychowdhury, {\it {Holographic Krylov complexity in N=4 SYM theory}},  \beforedoihref\href{http://dx.doi.org/10.1103/6999-p31b}{Phys. Rev. D}\afterdoihref\  {\bf 113} (2026), no.~10 106033 [\href{http://arXiv.org/abs/2511.19286}{{arXiv:2511.19286}}].

\bibitem{Heller:2025ddj}
M.~P. Heller, F.~Ori, J.~Papalini, T.~Schuhmann and M.-T. Wang, {\it {De Sitter holographic complexity from Krylov complexity in DSSYK}},  \href{http://arXiv.org/abs/2510.13986}{{arXiv:2510.13986}}.

\bibitem{Nandy:2024evd}
P.~Nandy, A.~S. Matsoukas-Roubeas, P.~Mart{\'\i}nez-Azcona, A.~Dymarsky and A.~del Campo, {\it {Quantum dynamics in Krylov space: Methods and applications}},  \beforedoihref\href{http://dx.doi.org/10.1016/j.physrep.2025.05.001}{Phys. Rept.}\afterdoihref\  {\bf 1125-1128} (2025) 1--82 [\href{http://arXiv.org/abs/2405.09628}{{arXiv:2405.09628}}].

\bibitem{Baiguera:2025dkc}
S.~Baiguera, V.~Balasubramanian, P.~Caputa, S.~Chapman, J.~Haferkamp, M.~P. Heller and N.~Y. Halpern, {\it {Quantum complexity in gravity, quantum field theory, and quantum information science}},  \beforedoihref\href{http://dx.doi.org/10.1016/j.physrep.2025.11.001}{Phys. Rept.}\afterdoihref\  {\bf 1159} (2026) 1--77 [\href{http://arXiv.org/abs/2503.10753}{{arXiv:2503.10753}}].

\bibitem{Rabinovici:2025otw}
E.~Rabinovici, A.~S{\'a}nchez-Garrido, R.~Shir and J.~Sonner, {\it {Krylov Complexity}},  \href{http://arXiv.org/abs/2507.06286}{{arXiv:2507.06286}}.

\bibitem{Berkooz:2018qkz}
M.~Berkooz, P.~Narayan and J.~Simon, {\it {Chord diagrams, exact correlators in spin glasses and black hole bulk reconstruction}},  \beforedoihref\href{http://dx.doi.org/10.1007/JHEP08(2018)192}{JHEP}\afterdoihref\  {\bf 08} (2018) 192 [\href{http://arXiv.org/abs/1806.04380}{{arXiv:1806.04380}}].

\bibitem{Berkooz:2018jqr}
M.~Berkooz, M.~Isachenkov, V.~Narovlansky and G.~Torrents, {\it {Towards a full solution of the large N double-scaled SYK model}},  \beforedoihref\href{http://dx.doi.org/10.1007/JHEP03(2019)079}{JHEP}\afterdoihref\  {\bf 03} (2019) 079 [\href{http://arXiv.org/abs/1811.02584}{{arXiv:1811.02584}}].

\bibitem{Berkooz:2024lgq}
M.~Berkooz and O.~Mamroud, {\it A cordial introduction to double scaled syk},  \beforedoihref\href{http://dx.doi.org/10.1088/1361-6633/ada889}{Reports on Progress in Physics}\afterdoihref\  {\bf 88} (feb, 2025) 036001.

\bibitem{Erdos:2014zgc}
L.~Erd{\H{o}}s and D.~Schr{\"o}der, {\it {Phase Transition in the Density of States of Quantum Spin Glasses}},  \beforedoihref\href{http://dx.doi.org/10.1007/s11040-014-9164-3}{Math. Phys. Anal. Geom.}\afterdoihref\  {\bf 17} (2014), no.~3-4 441--464 [\href{http://arXiv.org/abs/1407.1552}{{arXiv:1407.1552}}].

\bibitem{Goel:2023svz}
A.~Goel, V.~Narovlansky and H.~Verlinde, {\it {Semiclassical geometry in double-scaled SYK}},  \beforedoihref\href{http://dx.doi.org/10.1007/JHEP11(2023)093}{JHEP}\afterdoihref\  {\bf 11} (2023) 093 [\href{http://arXiv.org/abs/2301.05732}{{arXiv:2301.05732}}].

\bibitem{Banados:1992wn}
M.~Banados, C.~Teitelboim and J.~Zanelli, {\it {The Black hole in three-dimensional space-time}},  \beforedoihref\href{http://dx.doi.org/10.1103/PhysRevLett.69.1849}{Phys. Rev. Lett.}\afterdoihref\  {\bf 69} (1992) 1849--1851 [\href{http://arXiv.org/abs/hep-th/9204099}{{arXiv:hep-th/9204099}}].

\bibitem{Balasubramanian:2026azk}
V.~Balasubramanian, R.~N. Das, J.~Erdmenger, J.~Karl and H.~Verlinde, {\it {Complexity and the Hilbert space dimension of 3D gravity}},  \href{http://arXiv.org/abs/2602.02645}{{arXiv:2602.02645}}.

\bibitem{Maldacena:2001kr}
J.~M. Maldacena, {\it {Eternal black holes in anti-de Sitter}},  \beforedoihref\href{http://dx.doi.org/10.1088/1126-6708/2003/04/021}{JHEP}\afterdoihref\  {\bf 04} (2003) 021 [\href{http://arXiv.org/abs/hep-th/0106112}{{arXiv:hep-th/0106112}}].

\bibitem{Note1}
Restoring the coupling scale, the leading half-integer behavior takes the dimensional form \(b_n\sim J/\protect \sqrt {|\log q|}\). This estimate makes the overall scale transparent, but does not explain the full Laurent hierarchy or the cancellations that occur after the complexity is assembled.

\bibitem{Blommaert:2024ydx}
A.~Blommaert, T.~G. Mertens and J.~Papalini, {\it {The dilaton gravity hologram of double-scaled SYK}},  \href{http://arXiv.org/abs/2404.03535}{{arXiv:2404.03535}}.

\bibitem{Mertens:2022irh}
T.~G. Mertens and G.~J. Turiaci, {\it {Solvable models of quantum black holes: a review on Jackiw{\textendash}Teitelboim gravity}},  \beforedoihref\href{http://dx.doi.org/10.1007/s41114-023-00046-1}{Living Rev. Rel.}\afterdoihref\  {\bf 26} (2023), no.~1 4 [\href{http://arXiv.org/abs/2210.10846}{{arXiv:2210.10846}}].

\bibitem{Blommaert:2023opb}
A.~Blommaert, T.~G. Mertens and S.~Yao, {\it {Dynamical actions and q-representation theory for double-scaled SYK}},  \beforedoihref\href{http://dx.doi.org/10.1007/JHEP02(2024)067}{JHEP}\afterdoihref\  {\bf 02} (2024) 067 [\href{http://arXiv.org/abs/2306.00941}{{arXiv:2306.00941}}].

\bibitem{Blommaert:2024whf}
A.~Blommaert, A.~Levine, T.~G. Mertens, J.~Papalini and K.~Parmentier, {\it {An entropic puzzle in periodic dilaton gravity and DSSYK}},  \beforedoihref\href{http://dx.doi.org/10.1007/JHEP07(2025)093}{JHEP}\afterdoihref\  {\bf 07} (2025) 093 [\href{http://arXiv.org/abs/2411.16922}{{arXiv:2411.16922}}].

\bibitem{Blommaert:2025avl}
A.~Blommaert, A.~Levine, T.~G. Mertens, J.~Papalini and K.~Parmentier, {\it {Wormholes, branes and finite matrices in sine dilaton gravity}},  \beforedoihref\href{http://dx.doi.org/10.1007/JHEP09(2025)123}{JHEP}\afterdoihref\  {\bf 09} (2025) 123 [\href{http://arXiv.org/abs/2501.17091}{{arXiv:2501.17091}}].

\bibitem{Balasubramanian:2024lqk}
V.~Balasubramanian, J.~M. Magan, P.~Nandi and Q.~Wu, {\it {Spread complexity and the saturation of wormhole size}},  \beforedoihref\href{http://dx.doi.org/10.1103/vpyr-b3fb}{Phys. Rev. D}\afterdoihref\  {\bf 113} (2026), no.~4 046004 [\href{http://arXiv.org/abs/2412.02038}{{arXiv:2412.02038}}].

\bibitem{Note2}
Lanczos coefficients of related models, for example Random Matrix Theory, are widely studied and well-known. An initial analysis appeared in \cite {Balasubramanian:2022tpr} and for further works we refer to the excellent reviews \cite {Nandy:2024evd,Baiguera:2025dkc,Rabinovici:2025otw} and the references on Krylov spread complexity of TFD states therein.

\bibitem{Note3}
A generalization of this behavior to nonzero $\beta $ is likely, but we have not explored this direction further.

\bibitem{Hawking:1982dh}
S.~W. Hawking and D.~N. Page, {\it {Thermodynamics of Black Holes in anti-De Sitter Space}},  \beforedoihref\href{http://dx.doi.org/10.1007/BF01208266}{Commun. Math. Phys.}\afterdoihref\  {\bf 87} (1983) 577.

\bibitem{Witten:1998zw}
E.~Witten, {\it {Anti-de Sitter space, thermal phase transition, and confinement in gauge theories}},  \beforedoihref\href{http://dx.doi.org/10.4310/ATMP.1998.v2.n3.a3}{Adv. Theor. Math. Phys.}\afterdoihref\  {\bf 2} (1998) 505--532 [\href{http://arXiv.org/abs/hep-th/9803131}{{arXiv:hep-th/9803131}}].

\bibitem{Cardy:1986ie}
J.~L. Cardy, {\it {Operator Content of Two-Dimensional Conformally Invariant Theories}},  \beforedoihref\href{http://dx.doi.org/10.1016/0550-3213(86)90552-3}{Nucl. Phys. B}\afterdoihref\  {\bf 270} (1986) 186--204.

\bibitem{Strominger:1997eq}
A.~Strominger, {\it {Black hole entropy from near horizon microstates}},  \beforedoihref\href{http://dx.doi.org/10.1088/1126-6708/1998/02/009}{JHEP}\afterdoihref\  {\bf 02} (1998) 009 [\href{http://arXiv.org/abs/hep-th/9712251}{{arXiv:hep-th/9712251}}].

\bibitem{Tong:2009np}
D.~Tong, {\it {String Theory}},  \href{http://arXiv.org/abs/0908.0333}{{arXiv:0908.0333}}.

\bibitem{Note4}
Indeed, for the standard convention $V=2\pi $, all these satisfy $\beta <\beta _\protect \mathrm {HP}=V=2\pi $.

\bibitem{Carmi:2017jqz}
D.~Carmi, S.~Chapman, H.~Marrochio, R.~C. Myers and S.~Sugishita, {\it {On the Time Dependence of Holographic Complexity}},  \beforedoihref\href{http://dx.doi.org/10.1007/JHEP11(2017)188}{JHEP}\afterdoihref\  {\bf 11} (2017) 188 [\href{http://arXiv.org/abs/1709.10184}{{arXiv:1709.10184}}].

\bibitem{Note5}
To streamline the comparison between the boundary and bulk observables, we identify the boundary time \(t_{bdry}\) with the total two-sided bulk time \(\tau _{bulk}\). This convention ensures that both quantities are evaluated at the same physical time separation between the two asymptotic boundaries. In terms of the corresponding one-sided bulk time \(t_R=t_L\), this amounts to set \(t_{bdry}=\tau _{bulk}=2 t_R\).

\bibitem{Brown:1986nw}
J.~D. Brown and M.~Henneaux, {\it {Central Charges in the Canonical Realization of Asymptotic Symmetries: An Example from Three-Dimensional Gravity}},  \beforedoihref\href{http://dx.doi.org/10.1007/BF01211590}{Commun. Math. Phys.}\afterdoihref\  {\bf 104} (1986) 207--226.

\bibitem{Grandclement:2007sb}
P.~Grandclement and J.~Novak, {\it {Spectral methods for numerical relativity}},  \beforedoihref\href{http://dx.doi.org/10.12942/lrr-2009-1}{Living Rev. Rel.}\afterdoihref\  {\bf 12} (2009) 1 [\href{http://arXiv.org/abs/0706.2286}{{arXiv:0706.2286}}].

\bibitem{Note6}
In general, the polylogarithm function $\protect \text {Li}_{a}(X)=\Sigma _{n=1}^\infty \protect \frac {X^n}{n^a}$ diverges at $X=1$ for $a\leq 1$.

\bibitem{Maloney:2007ud}
A.~Maloney and E.~Witten, {\it {Quantum Gravity Partition Functions in Three Dimensions}},  \beforedoihref\href{http://dx.doi.org/10.1007/JHEP02(2010)029}{JHEP}\afterdoihref\  {\bf 02} (2010) 029 [\href{http://arXiv.org/abs/0712.0155}{{arXiv:0712.0155}}].

\bibitem{Keller:2014xba}
C.~A. Keller and A.~Maloney, {\it {Poincare Series, 3D Gravity and CFT Spectroscopy}},  \beforedoihref\href{http://dx.doi.org/10.1007/JHEP02(2015)080}{JHEP}\afterdoihref\  {\bf 02} (2015) 080 [\href{http://arXiv.org/abs/1407.6008}{{arXiv:1407.6008}}].

\bibitem{Beccaria:2026ndg}
M.~Beccaria and E.~Alfinito, {\it {Modular structures in the DSSYK partition function}},  \href{http://arXiv.org/abs/2607.11828}{{arXiv:2607.11828}}.

\bibitem{Bhattacharya_wip}
A.~Bhattacharya, M.~Flory, M.~P. Heller, E.~Rizza and T.~Schuhmann, {\it {To appear}}, .

\bibitem{UJ/CESIIM_2026}
A.~Bhattacharya, M.~Flory, M.~P. Heller, E.~Rizza and T.~Schuhmann, ``{The quadratic growth of Krylov spread complexity in the BTZ black hole - replication data}.'' \url{https://doi.org/10.57903/UJ/CESIIM}, 2026.

\end{thebibliography}\endgroup

\clearpage

\onecolumngrid

\setcounter{figure}{0}
\renewcommand{\thefigure}{A\arabic{figure}}

\appendix
\begin{center}
    \large{APPENDIX}
\end{center}

\section{More on the moment method}
\label{app:moment method details}
In this section, we summarize the moment-to-Lanczos recursion used in our calculations and specify its implementation. The moment method is standard, and we do not repeat its derivation here; see, for example ~\cite{Parker:2018yvk, Nandy:2024evd}. Our purpose is to fix conventions and explain how the finite set of moments generated by the partition function is converted into Lanczos coefficients and, subsequently, into the early-time expansion of Krylov spread complexity.

\subsubsection{Finite moment data and Lanczos coefficients}
\label{app:moment-lanczos}
Consider a normalized reference state $\ket{R}$ and a time-independent Hermitian Hamiltonian $H$. We define the Hamiltonian moments
\begin{equation}
\label{eq:app-moments}
    m_k \equiv \langle R|H^k|R \rangle,\qquad m_0=1.    
\end{equation}
A finite set of moments through order $2N+1$ determines the diagonal Lanczos coefficients $\{a_0,\ldots,a_N\}$ and the off-diagonal coefficients $\{b_1,\ldots,b_N\}$:
\begin{equation}
\label{eq:app-finite-moment-map}
    \left\{m_0,m_1,\ldots,m_{2N+1}\right\} \quad\longrightarrow\quad \left\{a_0,\ldots,a_N;\, b_1,\ldots,b_N\right\}.
\end{equation}
In the thermofield-double applications considered in the main text, these moments are generated by the thermal partition function,
\begin{equation}
\label{eq:app-moments-from-Z}
    m_k(\beta) = \frac{(-\partial_\beta)^k Z(\beta)}{Z(\beta)}.
\end{equation}
Here, $m_k(\beta)$ are normalized raw moments. They should not be confused with thermal cumulants, which are instead generated by derivatives of $\log Z(\beta)$. A derivation of Eq.~\eqref{eq:app-moments-from-Z} is given in Sec.~\ref{app:TFD-moments}.

For completeness, we now state the recursive moment algorithm used in our calculation. Introduce two triangular arrays $M_k^{(n)}$ and $L_k^{(n)}$, initialized by
\begin{align}
\label{eq:app-M-initial}
    M_k^{(0)} &= (-1)^k m_k,\\
    \label{eq:app-L-initial}
    L_k^{(0)}&=(-1)^{k+1}m_{k+1}.
\end{align}
For $n\geq 1$ and $k\geq n$, the higher rows are generated recursively according to
\begin{align}
\label{eq:app-M-recursion}
M_k^{(n)}&=L_k^{(n-1)}-L_{n-1}^{(n-1)}\frac{M_k^{(n-1)}}{M_{n-1}^{(n-1)}},\\
\label{eq:app-L-recursion}
L_k^{(n)} &= \frac{M_{k+1}^{(n)}}{M_n^{(n)}}-\frac{M_k^{(n-1)}}{M_{n-1}^{(n-1)}}.
\end{align}
The Lanczos coefficients are read off from the diagonal elements of these arrays:
\begin{equation}
\label{eq:app-lanczos-from-arrays}
    b_n = \sqrt{M_n^{(n)}}, \qquad a_n = -L_n^{(n)}.
\end{equation}
For a positive moment functional, we choose the conventional branch $b_n>0$. The first few coefficients provide a useful check of the conventions:
\begin{align}
    \label{eq:app-a0}
    a_0 &= m_1,\\
    \label{eq:app-b1}
    b_1^2 &= m_2-m_1^2,\\
    \label{eq:app-a1}
    a_1 &= \frac{m_3-2m_1m_2+m_1^3}{m_2-m_1^2}.
\end{align}
Equivalently, the first moments reconstructed from the Lanczos coefficients are
\begin{align}
\label{eq:app-first-reconstructed-moments}
m_0 &= 1, \nonumber\\ m_1 &= a_0, \nonumber\\ m_2 &= a_0^2+b_1^2, \nonumber\\ m_3 &= a_0^3+2a_0b_1^2+a_1b_1^2.
\end{align}
Unlike in moment problems with an even spectral measure, the odd moments in the present finite-temperature construction do not generically vanish. Consequently, both sets of Lanczos coefficients, $a_n$ and $b_n$, must be retained.

\subsubsection{Finite-order construction of spread complexity}
\label{app:finite-complexity-construction}

The Lanczos coefficients define the Jacobi matrix representation of the Hamiltonian in the Krylov basis,
\begin{equation}
\label{eq:app-jacobi-matrix}
    T_N = \begin{pmatrix}
    a_0 & b_1 & 0 & \cdots & 0 \\
    b_1 & a_1 & b_2 & \ddots & \vdots \\
    0 & b_2 & a_2 & \ddots & 0 \\
    \vdots & \ddots & \ddots & \ddots & b_N \\ 0 & \cdots & 0 & b_N & a_N
    \end{pmatrix}.
\end{equation}
The first Krylov vector is represented by
\begin{equation}
\ket{K_0} = \begin{pmatrix} 1&0&\cdots&0
\end{pmatrix}^{\mathsf T}.
\end{equation}
The Krylov amplitudes associated with the truncated matrix $T_N$ are therefore
\begin{equation}
\label{eq:app-truncated-amplitudes}
\phi_j^{(N)}(t) = \langle K_j|e^{-itT_N}|K_0\rangle =\left(e^{-itT_N}\right)_{j0}, \qquad 0\leq j\leq N.  
\end{equation}
The corresponding finite-chain spread complexity is
\begin{equation}
\label{eq:app-truncated-complexity}
    C_K^{(N)}(t) =\sum_{j=0}^{N}j\, \left|\left(e^{-itT_N}\right)_{j0}\right|^2.
\end{equation}
Expanding around $t=0$ gives
\begin{equation}
\label{eq:app-CK-series}
    C_K(t) = \sum_{n=1}^{\infty} c_{2n}t^{2n}.
\end{equation}
Only even powers of time occur. Moreover, the coefficient $c_{2n}$ depends only on
\begin{equation}
\label{eq:app-finite-data-c}
    \left\{a_0,\ldots,a_{n-1}; \, b_1,\ldots,b_n \right\}.
\end{equation}
It follows that the $(N+1)\times(N+1)$ matrix $T_N$ reproduces the early-time expansion of the full Krylov problem exactly through order $t^{2N}$. Although $T_N$ truncates the full Krylov chain, paths contributing through this order cannot reach sites beyond the $N$-th Krylov site. Operationally, our reconstruction therefore takes the form
\begin{equation}
\label{eq:app-computational-chain}
\left\{m_0,\ldots,m_{2N+1}\right\} \longrightarrow \left\{a_0,\ldots,a_N;\, b_1,\ldots,b_N\right\} \longrightarrow T_N \longrightarrow \left\{c_2,\ldots,c_{2N}\right\}.
\end{equation}
This finite-data property is what makes it possible to reconstruct a large number of early-time coefficients without constructing the Krylov basis vectors explicitly.

\subsubsection{Order of asymptotic expansion and Krylov reconstruction}

\label{app:order-of-limits}

A point of particular importance for the applications in the main text is that the operations entering the reconstruction are nonlinear. In particular, the moment recursion contains ratios of combinations of moments, while the coefficients $c_{2n}$ contain nonlinear combinations of the resulting $a_j$ and $b_j$.

Suppose that the partition function, moments, and Lanczos
coefficients depend on a parameter $\lambda$ controlling a semiclassical limit, with $\lambda\rightarrow 0$. Schematically, the construction is
\begin{equation}
\label{eq:app-parametric-construction}
    Z(\beta;\lambda) \longrightarrow m_k(\beta;\lambda) \longrightarrow \left\{a_n(\beta;\lambda), b_n(\beta;\lambda) \right\}
    \longrightarrow c_{2n}(\beta;\lambda).
\end{equation}
In our implementation, the moment recursion and the construction of
$c_{2n}$ are carried out at finite $\lambda$. The asymptotic limit is
taken only after the complete coefficient $c_{2n}$ has been assembled:
\begin{equation}
\label{eq:app-limit-after-complexity}
c_{2n}^{\mathrm{cl}}(\beta)= \lim_{\lambda\rightarrow 0} c_{2n}(\beta;\lambda).
\end{equation}
In general, this procedure is not equivalent to first truncating the individual Lanczos coefficients,
\begin{equation}
 \label{eq:app-truncated-lanczos}
    a_n(\beta;\lambda) \longrightarrow a_n^{\mathrm{trunc}}(\beta;\lambda),
    \qquad b_n(\beta;\lambda) \longrightarrow b_n^{\mathrm{trunc}}(\beta;\lambda),
\end{equation}
and then using the truncated data to construct the complexity. Terms that appear subleading in individual $a_n$ or $b_n$ can combine with parametrically large terms elsewhere and contribute to the leading semiclassical part of $c_{2n}$. Consequently,
\begin{equation}
\label{eq:app-noncommuting-limitt}
\lim_{\lambda\rightarrow 0} \mathcal{F}_{2n} \left[\{a_j(\lambda),b_j(\lambda)\}\right]\neq \mathcal{F}_{2n} \left[\left\{\lim_{\lambda\rightarrow 0}a_j(\lambda), \lim_{\lambda\rightarrow 0}b_j(\lambda)\right\}\right]
\end{equation} in general, where $\mathcal{F}_{2n}$ denotes the nonlinear function of Lanczos coefficients that produces $c_{2n}$. It is also useful to distinguish two different truncations. A \emph{moment-order truncation},
\begin{equation}
\{m_0,\ldots,m_{2N+1}\},
\end{equation}
is controlled and determines the complexity exactly through $t^{2N}$. By contrast, a truncation of a semiclassical, high-temperature, or low-temperature expansion of $Z(\beta)$ changes the input moments themselves and does not generically commute with the moment-to-Lanczos reconstruction. The validity of such an
asymptotic truncation must therefore be examined separately in each regime.

\subsubsection{Symbolic implementation and consistency checks}
\label{app:implementation-checks}

We implemented Eqs.~\eqref{eq:app-M-initial}-\eqref{eq:app-lanczos-from-arrays} as a triangular recursion. Starting from the required set of moments, the rows of $M_k^{(n)}$ and $L_k^{(n)}$ were generated successively in $n$.

At each step, $a_n$ and $b_n$ were extracted from the corresponding diagonal elements, after which the Jacobi matrix \eqref{eq:app-jacobi-matrix} was constructed and Eq.~\eqref{eq:app-truncated-complexity} was expanded in time.
The calculation was performed in \textsc{Mathematica}. Whenever possible, the moment recursion was carried out symbolically. For
numerical evaluations, arbitrary-precision arithmetic was used
instead of machine precision, since the recursion involves repeated divisions and cancellations between parametrically large
expressions. Stability was tested by increasing the working
precision and, when asymptotic series were used, by increasing the
number of retained terms.

We performed the following consistency checks. First, the input moments were reconstructed from the Jacobi matrix according to
\begin{equation}
\label{eq:app-moment-check}
m_k=\langle{K_0}|{T_N^k}|{K_0}\rangle=\left(T_N^k\right)_{00}.
\end{equation}
For all moments that can be determined from the available finite Lanczos data, this reproduces the original input expressions. Second, for exact positive thermal measures, we verified the positivity condition
\begin{equation}
\label{eq:app-b-positivity}
b_n^2\geq 0.
\end{equation}
Third, the first few coefficients $c_{2n}$ were checked against a direct expansion of the time-dependent Krylov amplitudes. Finally, the calculation was repeated after increasing the symbolic series order or numerical working precision, and the reported coefficients were found to be stable.

\section{Moments in the TFD state}\label{app:TFD-moments}

In this section, we derive the relation between Hamiltonian moments
in the thermofield-double state and derivatives of the thermal
partition function. Let $h$ denote the Hamiltonian of a single copy,
with spectral decomposition
\begin{equation}
    h=\sum_n E_n \ket{n}\bra{n},
    \qquad
    h\ket{n}=E_n\ket{n},
    \label{eq:app-one-copy-hamiltonian}
\end{equation}
where possible degeneracies are included in the label $n$. On the
doubled Hilbert space $\mathcal{H}_L\otimes\mathcal{H}_R$, we define
\begin{equation}
    h_L=h\otimes\mathbb{1},
    \qquad
    h_R=\mathbb{1}\otimes h.
    \label{eq:app-doubled-hamiltonians}
\end{equation}
The normalized thermofield-double state at inverse temperature
$\beta$ is
\begin{equation}
    \ket{TFD(\beta)}
    =
    \frac{1}{\sqrt{Z(\beta)}}
    \sum_n
    e^{-\beta E_n/2}
    \ket{n}_L\otimes\ket{n}_R,
    \label{eq:app-normalized-tfd}
\end{equation}
where
\begin{equation}
    Z(\beta)
    =
    \Tr\!\left(e^{-\beta h}\right)
    =
    \sum_n e^{-\beta E_n}.
    \label{eq:app-partition-function}
\end{equation}
For the Krylov construction we may take, without loss of generality,
the Hamiltonian to be
\begin{equation}
    H_K=h_L,
    \label{eq:app-krylov-hamiltonian}
\end{equation}
with the choice $H_K=h_R$ giving identical moments in the TFD state.
Indeed,
\begin{equation}
    h_L^k
    \bigl(\ket{n}_L\otimes\ket{n}_R\bigr)
    =
    E_n^k
    \ket{n}_L\otimes\ket{n}_R,
\end{equation}
and therefore the $k$-th Hamiltonian moment is
\begin{align}
\label{eq:app-tfd-moment-spectral}
    \langle TFD(\beta)|H_K^k|TFD(\beta)\rangle
    &=
    \frac{1}{Z(\beta)}
    \sum_{m,n}
    e^{-\beta(E_m+E_n)/2}
    E_n^k\,
    {}_L\!\braket{m}{n}_L\,
    {}_R\!\braket{m}{n}_R
    \nonumber\\
    &=
    \frac{1}{Z(\beta)}
    \sum_n E_n^k e^{-\beta E_n}.
\end{align}
On the other hand, differentiating the partition function $k$ times
gives
\begin{equation}
    \partial_\beta^k Z(\beta)
    =
    \sum_n (-E_n)^k e^{-\beta E_n},
\end{equation}
and hence
\begin{equation}
    (-\partial_\beta)^k Z(\beta)
    =
    \sum_n E_n^k e^{-\beta E_n}.
    \label{eq:app-Z-derivative}
\end{equation}
Combining Eqs.~\eqref{eq:app-tfd-moment-spectral} and
\eqref{eq:app-Z-derivative}, we obtain
\begin{equation}
    \boxed{
    \langle TFD(\beta)|H_K^k|TFD(\beta)\rangle
    =
    \frac{(-\partial_\beta)^k Z(\beta)}{Z(\beta)}
    }.
    \label{eq:app-tfd-moment-result}
\end{equation}
The same result follows if one chooses $H_K=h_R$. One may alternatively
consider the symmetric doubled Hamiltonian
\begin{equation}
    H_{\mathrm{sym}}
    =
    \frac{1}{2}\left(h_L+h_R\right).
    \label{eq:app-symmetric-hamiltonian}
\end{equation}
On the full doubled Hilbert space, however, this is a distinct
operator: its eigenstates are $\ket{m}_L\otimes\ket{n}_R$ with
\begin{equation}
    H_{\mathrm{sym}}
    \bigl(\ket{m}_L\otimes\ket{n}_R\bigr)
    =
    \frac{E_m+E_n}{2}
    \ket{m}_L\otimes\ket{n}_R.
    \label{eq:app-symmetric-spectrum}
\end{equation}
It agrees with $h_L$ and $h_R$ only on the energy-paired subspace
spanned by $\ket{n}_L\otimes\ket{n}_R$, which is precisely the
subspace supporting the TFD state:
\begin{equation}
    H_{\mathrm{sym}}\bigl(\ket{n}_L\otimes\ket{n}_R\bigr)
    =
    h_L
    \bigl(\ket{n}_L\otimes\ket{n}_R\bigr)
    =
    h_R
    \bigl(\ket{n}_L\otimes\ket{n}_R\bigr)
    =
    E_n
    \ket{n}_L\otimes\ket{n}_R.
\end{equation}
Thus all three choices give the same moments when acting on the TFD
state, although they should not be identified as operators on the
full doubled Hilbert space. Accordingly, in the main text
Eq.~\eqref{eq:momentsfromZ_generalformula} we suppress the left/right
label and write
\begin{equation}
    m_k(\beta)
    =
    \langle TFD(\beta)|H_K^k|TFD(\beta)\rangle
    =
    \frac{(-\partial_\beta)^kZ(\beta)}{Z(\beta)}.
\end{equation}

\section{Inverse temperature conventions for DSSYK}

We follow the DSSYK conventions of \cite{Blommaert:2024ydx} up to the replacement $q^2\to q$. Explicitly, we work with 
\begin{align}
    Z_\mathrm{DSSYK}(\beta)=\int_0^{\pi}d\theta\, \abs{(e^{2i\theta},q)_\infty}^2 e^{-\beta E(\theta)}\,,
\end{align}
where
\begin{align}
    E(\theta)=\frac{-\cos \theta}{\abs{\log q}}\,.
\end{align}
For clarity, let us contrast this with the conventions of \cite{Goel:2023svz}. Other than a less relevant different normalization of the partition function, in the semiclassical limit $q\to1$ we differ in the conventions for the energy by a factor of $2$ as well as $\theta=\pi-\theta_{\text{Goel et al.}}$. This implies that the inverse temperature we use is not identical to the one of \cite{Goel:2023svz} but instead related by these replacements. To convert the results of \cite{Goel:2023svz} that are mostly phrased in terms of $v_{\text{Goel et al.}}=\frac{2\theta_{\text{Goel et al.}}}{\pi}-1\in[-1,1]$ to our conventions, we explicitly need to invoke the replacement rule
\begin{align}
    \beta&=2\beta_{\text{Goel et al.}}\nonumber\\&=2\frac{\pi v_{\text{Goel et al.}}}{\cos\frac{\pi v_{\text{Goel et al.}}}{2}}=2\frac{-\pi+2\theta_{\text{Goel et al.}}}{\sin \theta_{\text{Goel et al.}}}\\&=\frac{2\pi-4\theta}{\sin \theta}\,.\nonumber
\end{align}
In our conventions, the low energy regime lies at $\beta\to\infty$ or $\theta\to0$ while for \cite{Goel:2023svz} it is consequently $\beta_{\text{Goel et al.}}\to\infty$ or $\theta_{\text{Goel et al.}}\to\pi$ or $ v_{\text{Goel et al.}}\to1$. Notice also that we work in conventions where the DSSYK coupling constant $J$ (denoted as $\mathbb{J}$ in \cite{Goel:2023svz}) is set to one and hence we do not display it in the above equations. It can always be restored through dimensional analysis.

\section{Lanczos coefficients from the semiclassical saddle: the role of temperature limits and truncations}
\label{sec:app_temp_limits}

To investigate the interplay of temperature limits and expansion at the level of the partition function with \textit{CfZ}, we emphasize two special limits of the semiclassical partition function $Z_\mathrm{sc}$ of DSSYK given in \eqref{eq:Zsc}, namely the high temperature limit ($\beta\to0\Leftrightarrow \theta\to\pi/2$) and the low temperature limit ($\beta\to\infty\Leftrightarrow \theta\to0$). If we want to obtain an approximate semiclassical partition function in $\beta$ in both temperature limits respectively we can sequentially i) find the series expansion of $\beta(\theta)$ in \eqref{eq:dssyk_beta} up to desired order in that limit, ii) invert this series to get an approximate relation $\theta(\beta)$, iii) plug this relation back into $Z_\mathrm{sc}$, and iv) expand the resulting exponent around the desired limit in $\beta$ up to order consistent with the initial truncation. Concretely, this yields low- and high-temperature approximations of the semiclassical partition function of the structure
\begin{align}
\label{eq:Zsc_hightemp}
    Z_\mathrm{sc}^{(\beta\to0)}&\simeq \exp\left[\frac{1}{\abs{\log q}}\sum_{k=0}^{k_\mathrm{cut}}h_k\beta^{2+2k}\right]\,,\\
 \label{eq:Zsc_lowtemp}   Z_\mathrm{sc}^{(\beta\to\infty)}&\simeq\exp\left[\frac{1}{\abs{\log q}}\sum_{k=0}^{k_\mathrm{cut}}l_k\beta^{1-k}\right]\,.
\end{align}
The numerical values of the coefficients $h_k$ and $l_k$ are known precisely through the previously described algorithm. The high-temperature approximation carries an exponent which is a series in powers of $\beta^2$ starting at quadratic order, while the low-temperature limit organizes itself in increasing powers of $\beta^{-1}$ starting at order $\beta$. We will discuss the impact of the order of truncation on the resulting complexity prediction in both limits below. In general, it can be drastic.\\

Concretely, we will compare the prediction from \textit{CfZ} applied to $Z_\mathrm{sc}$ in \eqref{eq:Zsc} of the main text with the results of \textit{CfZ} applied to the high-temperature and low-temperature approximations $Z_\mathrm{sc}^{(\beta\to0)}$ in \eqref{eq:Zsc_hightemp} and $Z_\mathrm{sc}^{(\beta\to\infty)}$ in \eqref{eq:Zsc_lowtemp}. Interestingly, these two limits behave differently, and together make the general behavior clear.\\

\textbf{High temperatures.} Let us start by applying \textit{CfZ} to $Z_{sc}$ expanded around $\beta\to0$, provided in \eqref{eq:Zsc_hightemp}, and vary the order $k_\mathrm{cut}$ at which we choose to cut off the expansion. The associated early-time series in the $q\to1$ limit after multiplication with $\abs{\log q}$ and afterwards evaluating at $\beta=0$ yields
\begin{align}
    &k_\mathrm{cut}=0:\quad \frac{t^2}{4}\quad\textcolor{blue}{\text{(all higher terms vanish)}}\,,\nonumber\\
    &k_\mathrm{cut}=1:\quad \frac{t^2}{4}-\frac{t^4}{96}\textcolor{blue}{-\frac{t^6}{1152}-\frac{t^8}{4608}+\dots}\,,\nonumber\\
    &k_\mathrm{cut}=2:\quad \frac{t^2}{4}-\frac{t^4}{96}+\frac{t^6}{1440}\textcolor{blue}{-\frac{7t^8}{23040}+\dots}\,,\nonumber\\
    &k_\mathrm{cut}=3:\quad \frac{t^2}{4}-\frac{t^4}{96}+\frac{t^6}{1440}-\frac{17t^8}{322560}\textcolor{blue}{+\dots}\,.
\end{align}
Comparing this to the answer from full DSSYK \eqref{eq:series_ckfull_betazero} that we restate up to order $t^8$ here for clarity,
\begin{align}
    \lim_{q\to1}\abs{\log q}C_K(t)_{\beta=0}\simeq\frac{t^2}{4}-\frac{t^4}{96}+\frac{t^6}{1440}-\frac{17\, t^8}{322560}+\dots\,,
\end{align}
it is clear that we get wrong predictions starting at order $t^{2(k_\mathrm{cut}+2)}$ for a fixed truncation $k_\mathrm{cut}$ (disagreeing coefficients in comparison with \eqref{eq:series_ckfull_betazero} highlighted in blue). In particular, if we had kept the leading order in temperature in the $\beta\to0$ limit, leading to a purely quadratic exponential in the partition function 
\begin{align}
    Z_{\mathrm{sc}}^{(\beta\to0)}\big\vert_{k_\mathrm{cut}=0}=\exp\left[\frac{\beta^2}{8\abs{\log q}}\right]\,,
\end{align}
this indeed is exactly the Heisenberg-Weyl case \cite{Balasubramanian:2022tpr} with eternally growing quadratic complexity. In this crude approximation, we get the first term of the early-time series correct, but miss all higher order terms of the early-time expansion. From this analysis it becomes clear, that in general, \textit{temperature expansions up to finite order do not commute with \textit{CfZ}}. Especially, in general it is insufficient to consider only the leading temperature behavior. This is not surprising, as to derive the early-time series of complexity via \textit{CfZ} up to order $t^{2n}$, we take as input the first $2n$ moments $m_k$ of the Hamiltonian in the reference state, which requires us to take $2n$ derivatives in $\beta$ of the partition function. The finite truncation in the beta expansion does not commute with taking these derivatives. To our knowledge, this critical aspect of \textit{CfZ} has received limited attention, and hence we care to point it out here.\\

\textbf{Low temperatures.} Let us continue with the opposite limit of low energies $\beta\to\infty$, where the partition function organizes itself in descending integer powers of $\beta$ starting from linear order. A priori, we expect the same non-commutation to appear as in the previous case. To our surprise, we find that the Schwarzian prediction is reproduced for all $k_\mathrm{cut}>1$, where for any coefficient of the complexity series we always retain only the leading term in $\beta\to\infty$. The result reads
\begin{align}
    &k_\mathrm{cut}=0:\quad\textcolor{blue}{0}\,,\nonumber\\
    &k_\mathrm{cut}=1:\quad \textcolor{blue}{0}\,,\nonumber\\
    &k_\mathrm{cut}=2:\quad \frac{4 \pi^2   t^2}{\beta ^3}-\frac{\pi^2  t^4}{\beta ^5}+\frac{\pi^2  t^6}{\beta ^7}-\frac{9 \pi^2  t^8}{8 \beta ^9}+\dots\,,\nonumber\\
    &k_\mathrm{cut}=3:\quad \frac{4 \pi^2   t^2}{\beta ^3}-\frac{\pi^2  t^4}{\beta ^5}+\frac{\pi^2  t^6}{\beta ^7}-\frac{9 \pi^2  t^8}{8 \beta ^9}+\dots\,,
\end{align}
and as compared to the Schwarzian prediction \eqref{eq:schwarzian_ck_early_time_series} which we also restate here for convenience
\begin{align}
    \lim_{c\to\infty}\frac{1}{c}C_K(t)_\mathrm{sch}\simeq\frac{4 \pi^2   t^2}{\beta ^3}-\frac{\pi^2  t^4}{\beta ^5}+\frac{\pi^2  t^6}{\beta ^7}-\frac{9 \pi^2  t^8}{8 \beta ^9}+\dots\,,
\end{align}
matches for $k_\mathrm{cut}>1$ (where again mismatches are highlighted in blue). This is a particular instance where keeping only a finite amount of leading order terms in the expansion of the partition function is sufficient to reproduce the full classical complexity behavior in a certain temperature limit. This can happen case-by-case, but is not guaranteed to work in general as we have seen in the previous limit.\\

\textbf{General temperature dependence.} Let us conclude by providing the constructive understanding why these two limits work differently in the case where we apply \textit{CfZ} to a finite truncation of the temperature-expanded semiclassical partition function. For this, let us study a saddle with generic temperature-dependence specified by the function $f(\beta)$
\begin{align}
    Z(\beta)=\exp\left[\frac{f(\beta)}{\abs{\log q}}\right]\,.
\end{align}
Applying \textit{CfZ} to this partition function and taking the classical limit $\abs{\log q}$ including again the multiplication by $\abs{\log q}$, we find for the early-time series
\begin{align}
    \lim_{q\to1}&\abs{\log q}C_K(t)_{f(\beta)}= f''(\beta ) t^2 + \left(\frac{1}{6} f''''(\beta )-\frac{f'''(\beta )^2}{4 f''(\beta )}\right) t^4\nonumber+\dots\,.\nonumber
\end{align}
This is an alternative manifestation of the fact that the early-time series coefficient $c_{2n}$ depends on the Lanczos coefficients up to $a_{n-1}$ and $b_n$, hence on the first $2n$ moments, which in turn precisely requires $2n$ derivatives of the partition function.\\
Now let us compare the two temperature limits we discussed before: In the high temperature limit \eqref{eq:Zsc_hightemp}, $f(\beta)$ is a series in increasing integer powers of $\beta^2$ starting at quadratic order. Neglecting higher orders in this case impacts the result significantly, because a higher order term, with the appropriate amount of derivatives applied to it, contributes a constant to the result that remains when we take the $\beta\to0$ at the end. This is different in the low-temperature limit \eqref{eq:Zsc_lowtemp}, because here the dominant order in the derivatives (which is the second-to-leading order proportional to $\beta^{-1}$) is always the dominant contribution in the $\beta\to\infty$ limit. The derivatives of the higher order terms with larger inverse powers of $\beta$ go to zero faster in the limit, and hence ultimately never contribute if we only keep the leading behavior in $\beta\to\infty$ (as we do).

\section{Fitting the Lanczos coefficients of CFT$_2$ at large $c$}
\label{sec:app_fitting}

In the main text, we present the leading order in $c$ fits 
\begin{align}
\label{eq:app_leading_fit_cft2_a}
    a_n^{[0]}&=\mathfrak{c}\beta^{-2}\,,\\
    b_n^{[0]}&=\sqrt{n}\cdot\sqrt{2\mathfrak{c}\beta^{-3}}\,,\label{eq:app_leading_fit_cft2_b}
\end{align}
and subleading order in $c$ fits 
\begin{align}
\label{eq:app_subleading_fit_cft2_a}
    a_n^{[0+1]}&=\mathfrak{c} \beta ^{-2}+3n\beta^{-1}\,,\\
    \label{eq:app_subleading_fit_cft2_b}
    b_n^{[0+1]}&=\sqrt{n\left(\frac{2\mathfrak{c}}{\beta ^{3}}+(n-1)\frac{3}{2\beta^{2}}\right)}\,,
\end{align}
at large $c$ for low orders of Lanczos coefficients derived via the moment method from the partition function \eqref{eq:Z_cft2}. In this Appendix, we show how these fits arise. As a disclaimer, it is important to state that these are fits, and hence approximations to the given data. We merely want to show that the data is reasonably approximated by these fits which give rise to very different associated complexity. Of course, there exist better fits to this data that are unknown to us that would give a more accurate complexity prediction, but that is not the point of this analysis.\\

The data we fit are the first 80 Lanczos coefficients, numerically derived from the partition function via the moment method at a given $\mathfrak{c}=\{10^{1},10^{3},10^{5},10^{7},10^{9},10^{11}\}$. We do so at three representative temperatures $\beta=\{2^{-1},2^0,2^1\}$ to get an understanding about the temperature dependence. The goal is to find convergence in the fit parameters with increasing $c$. The choice of fitting only the first 80 coefficients is arbitrary. We consider it large enough to derive analytic behavior of low lying Lanczos coefficients at large $c$ in $n$, but small enough we could still generate the data in a short time on a common laptop.\\

\textbf{Leading order.} By inspection of the data, we see that at large $c$, the leading behavior of the Lanczos coefficients is $a_n\propto c$ and $b_n\propto \sqrt{c \,n}$. Hence we fit to the previously mentioned data sets the laws $a_n=\alpha_1$ and $b_n=\sqrt{\delta_1 n}$. We find that the fit parameters $\alpha_1$, presented in Tab. \ref{tab:app_an_lead}, converge towards $\alpha_1=\mathfrak{c}\beta^{-2}$ and $\delta_1$, presented in Tab. \ref{tab:app_bn_lead}, converge towards $\delta_1=2\mathfrak{c}\beta^{-3}$ with increasing $c$.\\

\textbf{Subleading order.} We can repeat the previous procedure, but now while allowing for every fit to have two free parameters, one for the leading and one for the first subleading order in $c$. The laws we deduce by inspection are $a_n\propto\alpha_1+\alpha_2 n$ and $b_n\propto \sqrt{n(\delta_1+(n-1)\delta_2)}$. Let us highlight that by choosing this functional form of the fit we implicitly assume these Lanczos coefficients to be well approximated by the $SL(2,\mathbb{R})$ form. As we find convergence, this assumption is reasonable in the range we consider. Yet, of course, this is by no means unique and a different choice of functional form of the subleading fit would lead to a different complexity prediction. We find that the fit parameters $\{\alpha_1,\alpha_2\}$, presented in Tab. \ref{tab:app_an_sub}, converge towards $\{\mathfrak{c}\beta^{-2},3\beta^{-1}\}$ and $\{\delta_1,\delta_2\}$, presented in Tab. \ref{tab:app_bn_sub}, converge towards $\{2\mathfrak{c}\beta^{-3},3\beta^{-2}/2\}$ with increasing $c$.\\

\begin{table}[H]
    \centering
    \begin{tabular}{c||c |c |c}
    $\alpha_1$&$\beta=2^{-1}$&$\beta=2^{0}$&$\beta=2^{1}$\\
    \hline
    \hline
        $\mathfrak{c}=10^1$&247.35 & 108.77 & 49.706 \\
        $\mathfrak{c}=10^3$&4239.2 & 1119.2 & 309.25\\
        $\mathfrak{c}=10^5$&400240. & 100120. & 25060. \\
        $\mathfrak{c}=10^7$&$4.0000\times 10^7 $& $1.0000\times 10^7$ & $2.5001\times 10^6$ \\
        $\mathfrak{c}=10^9$&$4.0000\times 10^9$ & $1.0000\times 10^9 $& $2.5000\times 10^8$ \\
        $\mathfrak{c}=10^{11}$&$ 4.0000\times 10^{11}$ &$ 1.0000\times 10^{11}$ & $2.5000\times 10^{10} $\\
    \end{tabular}
    \caption{Fit parameter results for leading order fit of $a_n=\alpha_1$ for CFT$_2$.}
    \label{tab:app_an_lead}
\end{table}

\begin{table}[H]
    \centering
    \begin{tabular}{c||c |c |c}
    $\delta_1$&$\beta=2^{-1}$&$\beta=2^{0}$&$\beta=2^{1}$\\
    \hline
    \hline
        $\mathfrak{c}=10^1$&451.36 & 89.347 & 18.917 \\
        $\mathfrak{c}=10^3$&16316. & 2078.9 & 269.69 \\
        $\mathfrak{c}=10^5$&$1.6003\times10^{6}$ & $2.0008\times10^{5}$ & $2.5020\times10^{4}$ \\
        $\mathfrak{c}=10^7$&$1.6000\times10^{8}$ & $2.0000\times10^{7}$ & $2.5000\times10^{6}$ \\
        $\mathfrak{c}=10^9$&$1.6000\times10^{10}$ & $2.0000\times10^{9}$ & $2.5000\times10^{8}$ \\
        $\mathfrak{c}=10^{11}$&$1.6000\times10^{12}$ & $2.0000\times10^{11}$ & $2.5000\times10^{10}$ \\
    \end{tabular}
    \caption{Fit parameter results for leading order fit of $b_n=\sqrt{n\delta_1}$ for CFT$_2$.}
    \label{tab:app_bn_lead}
\end{table}

\begin{table}[H]
    \centering
    \begin{tabular}{c||c |c |c}
    $\{\alpha_1,\alpha_2\}$&$\beta=2^{-1}$&$\beta=2^{0}$&$\beta=2^{1}$\\
    \hline
    \hline
        $\mathfrak{c}=10^1$&\{44.585,4.9454\} & \{12.444,2.3494\} & \{3.5375,1.1261\} \\
        $\mathfrak{c}=10^3$& \{3994.4,5.9709\} & \{997.41,2.9714\} & \{248.89,1.4724\} \\
        $\mathfrak{c}=10^5$&\{399990.,5.9997\} & \{99997.,2.9997\} & \{24999.,1.4997\} \\
        $\mathfrak{c}=10^7$&$\{4.0000\times 10^7,6.0000\} $& $\{1.0000\times 10^7,3.0000\}$ & $\{2.5000\times 10^6,1.5000\}$ \\
        $\mathfrak{c}=10^9$&$\{4.0000\times 10^9,6.0000\} $&$ \{1.0000\times 10^9,3.0000\}$ &$ \{2.5000\times 10^8,1.5000\}$ \\
        $\mathfrak{c}=10^{11}$&$ \{4.0000\times 10^{11},6.0000\} $&$ \{1.000\times 10^{11},3.0000\}$ &$ \{2.5000\times 10^{10},1.5000\}$ \\
    \end{tabular}
    \caption{Fit parameter results for subleading order fit of $a_n=\alpha_1+\alpha_2\,n$ for CFT$_2$.}
    \label{tab:app_an_sub}
\end{table}

\begin{table}[H]
    \centering
    \begin{tabular}{c||c |c |c}
    $\{\delta_1,\delta_2\}$&$\beta=2^{-1}$&$\beta=2^{0}$&$\beta=2^{1}$\\
    \hline
    \hline
        $\mathfrak{c}=10^1$& \{169.19,5.4822\} & \{22.949,1.2968\} & \{3.2796,0.30664\} \\
        $\mathfrak{c}=10^3$& \{16000.,5.9997\} & \{2000.0,1.4997\} & \{250.01,0.37469\} \\
        $\mathfrak{c}=10^5$& $\{1.6000\times 10^6,6.0000\} $& \{200000.,1.5000\} & \{25000.,0.37500\} \\
        $\mathfrak{c}=10^7$&$\{1.6000\times 10^8,6.0000\} $& $\{2.0000\times 10^7,1.5000\}$ &$ \{2.5000\times 10^6,0.37500\} $\\
        $\mathfrak{c}=10^9$&$\{1.6000\times 10^{10},6.0000\}$ & $\{2.0000\times 10^9,1.5000\} $&$ \{2.5000\times 10^8,0.37500\} $\\
        $\mathfrak{c}=10^{11}$&$\{1.6000\times 10^{12},6.0000\} $& $\{2.0000\times 10^{11},1.5000\}$ & $\{2.5000\times 10^{10},0.37500\}$ \\
    \end{tabular}
    \caption{Fit parameter results for subleading order fit of $b_n=\sqrt{n(\delta_1+(n-1)\delta_2)}$ for CFT$_2$.}
    \label{tab:app_bn_sub}
\end{table}

\section{Further sanity checks for CFT$_2$ classical Krylov spread complexity early-time series}
\label{sec:app_cft2_sanitychecks}

In Sec. \ref{sec:cft_results} in the main text, we derived the Krylov spread complexity prediction from the Cardy partition function 
\begin{align}
\label{eq:appendix_Z_cft2}
    Z(\beta)=\exp\left(\frac{\pi c\,V}{6\beta}\right)=\exp\left(\frac{\mathfrak{c}}{\beta}\right)\,.
\end{align}
via \textit{CfZ}. The early-time series in the appropriate classical limit which involves the factor $c^{-1}$ included in the limit reads
\begin{align}
\label{eq:appendix_cft2_CK_series}
    \lim_{c\to\infty}\frac{C_K(t)}{c}=\frac{\pi V}{6\beta}\left(\frac{2 t^2}{\beta ^2}-\frac{t^4}{2 \beta ^4}+\frac{t^6}{2 \beta ^6}-\frac{9 t^8}{16 \beta ^8}
    +\frac{21 t^{10}}{32 \beta ^{10}}
    +\dots\right)
\end{align}
and will be again subject of interest in this Appendix. Concretely, we test the stability of this prediction versus the approximations that go into using purely the partition function \eqref{eq:appendix_Z_cft2}. To carry this out explicitly, we include first subleading terms in the approximation we make into the partition function and ask whether the prediction \eqref{eq:appendix_cft2_CK_series} stays invariant. We verified invariance of the series analytically at low orders in $n$.\\

\textbf{Multiple saddles: The Hawking-Page transition.} Let us for simplicity set $V=2\pi$. We have stated before that our derivation only applies above the Hawking-Page temperature \cite{Hawking:1982dh}. Let us show explicitly that the empty AdS$_3$ saddle indeed does not spoil our results as long as we are above the Hawking-Page temperature $\beta=2\pi$ and take the classical limit $c\to \infty$. For this reason we apply \textit{CfZ} to
\begin{align}
Z(\beta)=\exp\left(\frac{\mathfrak{c}}{\beta}\right)+\exp\left(\frac{\mathfrak{c}\beta}{4\pi^2}\right)
\end{align}
and study the associated Krylov spread complexity of early-time series. Before taking the classical limit and at general $\beta$, we find (we only display the quadratic coefficient as an example)
\begin{align}
    C_K(t)=t^2\cdot\frac{\mathfrak{c} e^{\left(\frac{\beta }{4 \pi ^2}+\frac{1}{\beta }\right) \mathfrak{c}} \left(32 \pi ^4 \beta +\left(\beta ^2+4 \pi ^2\right)^2 \mathfrak{c}\right)+32 \pi ^4 \beta  \mathfrak{c} e^{\frac{2 \mathfrak{c}}{\beta }}}{16 \pi ^4 \beta ^4 \left(e^{\mathfrak{c}/\beta }+e^{\frac{\beta \mathfrak{c}}{4 \pi ^2}}\right)^2}+\dots
\end{align}
where it is evident that taking the limit $c\to\infty$ requires specification of whether $\beta$ is larger or smaller than $2\pi$ because of the question of dominance of the appearing exponentials. Under the assumption that we consider the regime above the Hawking-Page temperature $\beta<2\pi$, we then take the classical limit and indeed recover the correct term
\begin{align}
    \lim_{c\to\infty}\frac{C_K(t)}{c}= t^2\cdot\frac{\pi V}{6}\frac{ 32 \pi ^4 \beta   e^{\frac{2 c}{\beta }}}{16 \pi ^4 \beta ^4 \left(e^{c/\beta }\right)^2}+\dots=\frac{\pi V}{6\beta}\left(\frac{2t^2}{\beta^2}+\dots\right)
\end{align}
and similarly for the next terms. This shows, analytically checked explicitly for low orders, that the empty AdS$_3$ saddle does not influence the result for the Krylov spread complexity early-time series in the classical limit $c\to \infty$, as long as we are above the Hawking-Page temperature.
\\

\textbf{Corrections from a generic 1-loop determinant.} In our analysis, we focus on a semiclassical saddle. For consistency, we therefore want to show that the classical limit of the early-time series does not depend on any 1-loop determinant around this saddle. This analysis includes for example 1-loop corrections that arise when employing the modular transform of the Virasoro vacuum character. In order to carry out this analysis, we apply \textit{CfZ} to the partition function
\begin{align}
Z(\beta)=f(\beta)\exp\left(\frac{\mathfrak{c}}{\beta}\right)\,,
\end{align}
with general 1-loop contribution $f(\beta)$. The associated complexity of early-time series before the classical limit is (where again we just show the first coefficient as an example)
\begin{align}
    C_K(t)=t^2 \left(\frac{2  \mathfrak{c}}{\beta ^3}+\frac{f(\beta ) f''(\beta )-f'(\beta )^2}{f(\beta )^2}\right)+\dots\,.
\end{align}
We see the strong influence the 1-loop determinant has, but notice that it is subleading in $c$ as expected. This is precisely the reason that in the classical limit $c\to\infty$ this contribution is subdominant and we recover our known result
\begin{align}
    \lim_{c\to\infty}\frac{C_K(t)}{c}=\frac{\pi V}{6\beta}\left(\frac{2t^2}{\beta^2}+\dots\right)
\end{align}
We have verified the validity of this argument also for multiple next orders. It therefore shows for the low orders and suggests for all orders that the classical result for the Krylov spread complexity early-time series is indeed unaffected by a 1-loop determinant contribution.

\section{Bulk calculations}
\label{sec::bulk_d}

In this Appendix, we give additional details for the calculations shown in section \ref{sec:bulk}, considering again a BTZ black hole in ingoing Eddington-Finkelstein coordinates \eqref{eq:metric}, \eqref{eq:blackening} and largely following \cite{Belin:2021bga,Belin:2022xmt}. For the volume functional $F_2= 1$, the extremal surface problem can be mapped to the problem of a particle moving in an effective potential
\begin{align}
\dot{r}^2+ U_2(r)=\mP_v^2   , \qquad     U_2(r)
    =
    -f(r)r^{2}
    =
    r^2\left(r_h^2-r^2\right).
    \label{eq:U2}
\end{align}
where the dot denotes differentiation with respect to the world-volume coordinate \(\sigma\) while $\mathcal{P}_v$ is the momentum conserved on the extremal surfaces. The radial turning point \(r_{\min}\) is determined by 
\begin{equation}
    \mathcal{P}_v^2=U_2(r_{\min})
    \label{eq:turning_condition}.
\end{equation}
and corresponds to the minimal radial coordinate that the extremal surface will attain for a given $\mathcal{P}_v$. Using \eqref{eq:U2}, this gives
\begin{equation}
    r_{\min}(\mathcal{P}_v)
    =
    \sqrt{
    \frac{r_h^2+\sqrt{r_h^{4}-4\mathcal{P}_v^2}}{2}
    },
    \label{eq:rmin}
\end{equation}
where we have picked the positive root as the physical branch. The allowed range is therefore $  0\leq \mathcal{P}_v < \frac{r_h^2}{2}.$ The maximum of the effective potential is located at $r_f=\frac{r_h}{\sqrt{2}},$ and determines the final slice to which extremal slices will asymptote at late boundary times. \\

A translation invariant codimension-one spacelike hypersurface \(\Sigma\) is parametrized by
\begin{equation}    X^\mu(\sigma,\phi)=\big(v(\sigma),r(\sigma),\phi\big),
\end{equation}
where $\sigma$ can be introduced as a new world-volume coordinate on $\Sigma$. The parametrization of the hypersurface is invariant under
redefinitions \(\sigma\rightarrow \sigma'(\sigma)\). For the reference
functional \(F_2=1\), this freedom has been fixed by imposing \(    \sqrt{-f(r)\dot v^{\,2}+2\dot v\dot r} = r\) \cite{Belin:2021bga,Belin:2022xmt}. Using this gauge and \eqref{eq:U2}, the induced metric $h_{ab}$ can be written as
\begin{equation}
    ds^2_\Sigma
    =
    \frac{r^{2}}{\mathcal P_v^2-U_2(r)}\,dr^2
    +r^2 d\phi^{2}.
\end{equation}
The hypersurface may equivalently be described by an implicit equation
\(\mathcal{F}(v,r)=v-v(r)=0\),  whose normal covector is proportional to
\(\partial_\mu \mathcal{F}\). After imposing the timelike normalization condition $n_\mu n^\mu=-1,$ one can compute the extrinsic curvature
\begin{equation}
    K_{ab}
    =
    -P_a{}^\mu P_b{}^\nu \nabla_\mu n_\nu,
    \qquad
    P_{\mu\nu}=g_{\mu\nu}+n_\mu n_\nu .
\end{equation}
Since the hypersurface is extremal, its mean curvature vanishes by definition,
\begin{equation}
    K\equiv h^{ab}K_{ab}=0 .
\end{equation}
The nontrivial scalar entering the local functional is therefore
\(K_{ab}K^{ab}\). For the above minimal slice one finds 
\begin{equation}
    K_{ab}K^{ab}
    =
    2\frac{\mathcal P_v^2}{r^{4}},
\end{equation}
and using \eqref{eq:rmin} it is easy to see that $K_{ab}K^{ab}\rightarrow 2$ at late times where $\mP_v\rightarrow \frac{r_h^2}{2}$. 
\\

The boundary time \(\tau\) is determined by the conserved momentum \(\mathcal{P}_v\). The relation is
\begin{equation}
    \tau(\mathcal{P}_v)
    =
    -2\mathcal{P}_v
    \int_{r_{\min}(\mathcal{P}_v)}^{\infty}
    \frac{dr}
    {f(r)\sqrt{\mathcal{P}_v^2-U_2(r)}}.
    \label{eq:tauP}
\end{equation}
This equation implicitly defines \(\mathcal{P}_v=\mathcal{P}_v(\tau)\). Near \(\tau=0\), one expands Eq.~\eqref{eq:tauP} around \(\mathcal{P}_v=0\), then inverts the resulting series. For example,
\begin{equation}
\begin{aligned}
    &\tau(\mathcal{P}_v)
    =
    \alpha_1 \mathcal{P}_v +\alpha_3 \mathcal{P}_v ^3+\alpha_5\mathcal{P}_v^5+\cdots,
    \\
    &\mathcal{P}_v(\tau)
    = \beta_1\tau+\beta_3\tau^3+\beta_5\tau^5+\cdots .    
\end{aligned}
    \label{eq:P_tau_expansion}
\end{equation}
In the specific case of the BTZ black hole, the relation between the boundary time and the conserved momentum \eqref{eq:tauP} admits a systematic analytic expansion around \(\mathcal{P}_v=0\), which can subsequently be inverted order by order to express the conserved momentum as a power series in the boundary time \eqref{eq:P_tau_expansion}.
\\

For general codimension-one observables, the volume integrand is replaced by some local scalar functional of the embedding data \cite{Belin:2021bga,Belin:2022xmt}. In the present construction \eqref{eq.functional_F1}, we specialize to the case in which the reference surface $\Sigma_{F_2}$ remains the extremal volume surface, while the deformed observable is weighted by a function \(a_1(\mathcal{P}_v,r)\), which is the local scalar \(F_1\) evaluated on the reference hypersurface selected by \(F_2=1\). The effective potential is thus written as
\begin{equation}
    \widetilde U_1(\mathcal{P}_v,r)
    =
    a_1(\mathcal{P}_v,r)^2 U_2(r),
    \label{eq:U1}
\end{equation}
where as explained in section \ref{sec:bulk}, we assume the function \(a_1(\mathcal{P}_v,r)\) to be constructed perturbatively from scalars of the extrinsic curvature of the maximal slice, i.e.~powers of the dimensionless ratio $\frac{\mathcal{P}_v^2}{r^{4}}$. Thus we write
\begin{equation}
    a_1(\mathcal{P}_v,r)
    = \bar\lambda
\left[
    1+\sum_{n=1}^{\infty}
    \lambda_n
    \left(
        \frac{K_{ab}K^{ab}}{2}
    \right)^{n+1} \right]
    =  \bar\lambda
\left[
    1+\sum_{n=1}^{\infty}
    \lambda_n
    \left(
        \frac{\mathcal{P}_v^2}{r^{4}}
    \right)^{n+1} \right].
    \label{eq:a1_ansatz}
\end{equation}
The presence of the additional coefficient \(\bar\lambda\) absorbs the overall normalization constant of the boundary observable, and makes the subsequent coefficients
\(\lambda_1,\lambda_2,\ldots\) to be determined solely by the relative
early-time coefficients of the boundary complexity. 
The relation \eqref{eq:P_tau_expansion} can be inserted into the generalized codimension-one observable introduced in \eqref{eq:Cany}. In the construction considered here, the ordinary maximal-volume hypersurface is used as the carrier, and the generalized observable takes the form
\begin{equation}
    \mathcal{O}(\mathcal{P}_v)
    =
    -\frac{2V_\phi}{G_N}
    \int_{r_{\min}(\mathcal{P}_v)}^{r_{\max}}
    dr\,
    \frac{a_1(\mathcal{P}_v,r)U_2(r)}
    {f(r)\sqrt{\mathcal{P}_v^2-U_2(r)}}.
    \label{eq:eq40_a1}
\end{equation} where the integrand is evaluated between $r_{\min}$, fixed by the turning point condition \eqref{eq:rmin}, and $r_{\max}$ that acts as a UV cutoff. $V_\phi$ is the volume of the angular direction contribution.
When \(a_1\) depends only on \(r\), the derivative reproduces the standard result of the codimension-one construction appearing in \cite{Belin:2021bga,Belin:2022xmt}. In the present case, however, \(a_1(\mathcal{P}_v,r)\) contains an additional explicit-\(\mathcal{P}_v\) contribution. The result is 
\begin{align}
    \frac{G_N}{V_\phi}\frac{d\mathcal{O}}{d\tau}
    =
   \mathcal{P}_v\,a_1(\mathcal{P}_v,r_{\min})
   -
    2\mathcal{P}_v\frac{d\mathcal{P}_v}{d\tau}
    \int_{r_{\min}}^{\infty}
    dr\,
    \frac{
    r^2
    \left[
        a_1(\mathcal{P}_v,r)-a_1(\mathcal{P}_v,r_{\min})
    \right]
    }
    {\left(\mathcal{P}_v^2-U_2(r)\right)^{3/2}}
    +
    2\frac{d\mathcal{P}_v}{d\tau}
    \int_{r_{\min}}^{\infty}
    dr\,
    \frac{
    r^2\,
    \partial_\mathcal{P} a_1(\mathcal{P}_v,r)
    }
    {\sqrt{\mathcal{P}_v^{2}-U_2(r)}}.
\label{eq:dOdtau_three_terms}
\end{align}
The evaluation of these terms requires the control of the turning-point singularity which can be handled by the change of variables \(r^2=r_{\min}^2+r_h^2 u^2\); the removable horizon singularity is solved by using the regular form of the integrand, and the UV divergence is regularized by subtracting the divergent part and adding the finite counterterm. \\

In order to study how the local functional \(a_1(\mathcal P_v,r)\) affects the time behavior of complexity, it is useful to introduce the dimensionless quantity evaluated at the turning point,
\begin{equation}
    Q(\tau)
    \equiv
    \frac{\mathcal P_v(\tau)^2}{r_{\min}(\tau)^{4}}
    =
    \frac{r_h^2}{r_{\min}(\tau)^2}-1 .
    \label{eq:Q_tau}
\end{equation}
At early times \(r_{\min}\to r_h\), hence \(Q(\tau)\to0\), whereas at late
times \(r_{\min}\to r_f=r_h/\sqrt{2}\), where \(r_f\) is the
maximum of the carrier potential \(U_2(r)\). The position of this final
slice is fixed by the surface-defining functional \(F_2=1\), rather
than by the weighted potential \(\widetilde U_1\). At the turning point,
\(Q(\tau)\to1\).
Near \(\tau=0\), the conserved momentum is analytic and odd in time,
as in Eq.~\eqref{eq:P_tau_expansion}. Hence, for any positive integer
\(p\),
\begin{equation}
    \left(\frac{\mathcal P_v^2}{r^4}\right)^p
    =O(\tau^{2p}).
\end{equation}
Since \(\lambda_n\) multiplies
\((\mathcal P_v^2/r^4)^{n+1}\), it first contributes at order
\(\tau^{2n+1}\) in \(d\mathcal O/d\tau\), and therefore at order
\(\tau^{2n+2}\) in \(\mathcal O(\tau)\). This explains the triangular structure of the early-time expansion.
The early-time expansion obtained from Eq.~\eqref{eq:dOdtau_three_terms} substituting the inverse momentum-energy series \(\mathcal{P}_v\) \eqref{eq:P_tau_expansion} in \eqref{eq:dOdtau_three_terms}, re-expanding at small \(\tau\), and using \(r_h=2\pi/\beta\),
takes the form
\begin{equation} \begin{aligned}  \frac{G_N}{V_{\phi}} \frac{d\mathcal O}{d\tau}
=
\bar\lambda \Bigg[
&\frac{8\pi^2}{\beta^3} \tau
+\frac{40\pi^2}{\beta^5}
\left(\lambda_1-1\right)\tau^3
\\
&-\frac{3\pi^2}{\beta^7}
\left(
85\lambda_1
-63\lambda_2
-74
\right)\tau^5
\\
&+\frac{6\pi^2}{\beta^9}
\left(
183\lambda_1
-259\lambda_2
+143\lambda_3
-195
\right)\tau^7
\Bigg]
+\mathcal O(\tau^9). \end{aligned} \label{eq:ads3_early_beta} \end{equation}
This expression makes the triangular hierarchy manifest.
\(\lambda_i\) first enters at order \(\tau^{2i+1}\) in
\(d\mathcal O/d\tau\), and therefore at order \(\tau^{2i+2}\) in
\(\mathcal O(\tau)\). In particular, \(\lambda_1,\lambda_2,\lambda_3\)
enter the displayed derivative at orders \(\tau^3,\tau^5,\tau^7\). Hence the low-\(i\) terms determine the first departure from the pure volume result, while higher powers of \(K_{ab}K^{ab}\) are invisible until correspondingly higher orders in the early-time expansion.\\

At intermediate times the expansion is no longer controlled only by the
small parameter \(\tau/\beta\). The relevant quantity is instead \(Q(\tau)\) in \eqref{eq:Q_tau}. As the turning point moves from the horizon toward \(r_f\), \(Q(\tau)\) increases from \(0\) to \(1\). Therefore higher-\(i\) corrections remain suppressed in the early part of the evolution, but become increasingly important as the surface approaches the critical radius. This is the mechanism by which
the higher powers of \(K_{ab}K^{ab}\) influence the transition toward the late-time regime. At any finite truncation, and more generally whenever \(a_1(\mathcal P_\infty,r_f)\) remains finite, the endpoint contribution approaches a finite constant:
\begin{equation}
\frac{d\mathcal{O}}{d\tau} = \frac{V_\phi}{G_N}
\sqrt{\widetilde U_1(\mathcal P_v,r_{\min})}
+\mathcal O(1),
\label{eq:CEA_growth}
\end{equation}
An infinite resummation for which
\(a_1(\mathcal P_\infty,r_f)\) diverges on the final slice evades this conclusion and may therefore support faster-than-linear growth. This term is the endpoint contribution, which will determine the late-time behavior of the derivative of complexity when $\mathcal{P}_v$ reaches the value $\mathcal{P}^{2}_{\infty}=U_2(r_f)$.
In fact, when $\tau \rightarrow \infty $, then $r_{min}$ reaches the critical value $r_f$. Therefore, the linear growth of complexity is related to the existence of an extremal surface at late times. Since \(\frac{\mathcal P_\infty^2}{r_f^{4}}=1\) at that limit, all powers in the \(a_1\)-series become equally unsuppressed at late time:
\begin{equation}
    a_1(\mathcal P_\infty,r_f)
    =
    \bar\lambda\left(1+\sum_{n=1}^{\infty}\lambda_n\right) .
\end{equation}
At a finite truncation order \(N\), the late-time slope is consequently
controlled by
\begin{equation}
    \left.
    \frac{G_N}{V_\phi}\frac{d\mathcal O_N}{d\tau}
    \right|_{\tau\to\infty}
    =
    \mathcal P_\infty a_1(\mathcal P_\infty,r_f),
    \qquad
    \mathcal P_\infty=\frac{r_h^2}{2}.
    \label{eq:late_slope_truncated}
\end{equation}
Thus a bulk functional whose sum over $\lambda_n$ diverges when evaluated on the final slice can lead to a faster than linear late-time growth.  
We can relate more quantitatively the large-order behavior of the coefficients \(\lambda_n\) to the late-time growth of the bulk observable through the general late-time
analysis in \cite{Belin:2021bga}. 
For a nondegenerate maximum of \(U_2(r)\), the conserved momentum obeys \cite{Belin:2021bga}
\begin{equation}
    \mathcal P_\infty-\mathcal P_v(\tau)
    =
    C_{\mathcal P}\,e^{-\kappa\tau}
    +\cdots,
    \qquad
    \kappa
    =
    \frac{-f(r_f)
    \sqrt{-2 U_2''(r_f)}}{2\mathcal P_\infty}
    >0.
    \label{eq:P_exponential_approach}
\end{equation}
The factor of \(1/2\) follows from our convention
\(\tau=2t_R\), where \(\tau\) is the total two-sided boundary time.
The expression without this factor gives the decay rate with respect
to the one-sided time \(t_R\). Expanding the potential around its maximum gives
\begin{equation}
    \mathcal P_\infty^2-\mathcal P_v^2
    =
    -\frac{1}{2} U_2''(r_f)
    \left(r_{\min}-r_f\right)^2
    +\mathcal O\!\left(
        \left(r_{\min}-r_f\right)^3
    \right).
    \label{eq:P_rmin_relation}
\end{equation}
Combining Eqs.~\eqref{eq:P_exponential_approach} and
\eqref{eq:P_rmin_relation} therefore yields
\begin{equation}
    r_{\min}(\tau)-r_f
    =
    C_r\,e^{-\kappa\tau/2}
    +\cdots.
    \label{eq:rmin_exponential_approach}
\end{equation}
Moreover, expanding Eq.~\eqref{eq:Q_tau} around \(r_{\min}=r_f\) gives an exponential approach to one,
\begin{equation}
    1-Q(\tau)
    =
    \frac{4\sqrt{2}}{r_h}
    \left(r_{\min}(\tau)-r_f\right)
    +\mathcal O\!\left(
        \left(r_{\min}-r_f\right)^2
    \right)
    =
    C_Q\,e^{-\kappa\tau/2}
    +\cdots.
    \label{eq:q_exponential_approach}
\end{equation}
For the BTZ black hole, using
\(r_f=r_h/\sqrt{2}\),
\(\mathcal P_\infty=r_h^2/2\), and
\(U_2''(r_f)=-4r_h^2\), we obtain
\[
\kappa=\sqrt{2}\,r_h.
\]
To obtain quadratic late-time growth, the resummed function
\(a_1(Q)\), rather than \(Q(\tau)\) itself, must develop a logarithmic
singularity as \(Q\to1^{-}\). Such a singularity is generated by a
harmonic large-order tail. As a particular example, if the leading
contribution lies in the odd subsequence,
\begin{equation}
    \lambda_{2n+1}
    =
    \frac{A_o}{2n+1}
    +
    \mathcal O\!\left(
        (2n+1)^{-1-\epsilon}
    \right),
    \qquad
    \lambda_{2n}
    =
    \mathcal O\!\left(n^{-1-\epsilon}\right),
    \qquad
    \epsilon>0,
    \label{eq:critical_lambda_tail}
\end{equation}
then
\begin{equation}
    \sum_{m=0}^{\infty}
    \frac{A_o}{2m+1}Q^{2m+2}
    =
    A_o Q\,\operatorname{arctanh}(Q)
    =
    -\frac{A_o}{2}\log(1-Q)
    +\mathcal O(1).
    \label{eq:lambda_log_resummation}
\end{equation}
Using Eq.~\eqref{eq:q_exponential_approach}, the functional evaluated at
the endpoint consequently behaves as
\begin{equation}
    a_1\!\left(Q(\tau)\right)
    =
    \bar\lambda
    \left[
        \frac{A_o\kappa}{4}\tau
        +\mathcal O(1)
    \right].
    \label{eq:a1_linear_time}
\end{equation}
At the turning point, the endpoint contribution to the growth rate is given by \eqref{eq:late_slope_truncated}. Thus, one obtains a linear time growth for the derivative of complexity, and a corresponding local logarithmic slope that satisfies
\begin{equation}
    \lim_{\tau\to\infty}
    \tau\frac{d}{d\tau}\log\mathcal O(\tau)
    =
    2.
    \label{eq:bulk_log_slope_two}
\end{equation}
Thus the critical scaling \(\lambda_n\sim 1/n\) is singled out by the requirement that the resummed bulk functional reproduce the asymptotic quadratic behavior suggested by the Pad\'e continuation. Faster decay
renders \(a_1(\mathcal P_\infty,r_f)\) finite and restores linear growth, whereas a tail
slower than \(1/n\) produces a singularity stronger than the logarithmic one. For the parity-resolved fits \eqref{even_model}--\eqref{odd_model_2},
including the even harmonic contribution replaces \(A_o\) by
\(A_o+A_e\) in the leading logarithmic coefficient, without changing
the conclusion \(\gamma\to2\).

\section{A toy model for the interplay between finite bulk truncations and Pad\'e continuation}
\label{subsec:toy_model_bulk_pade}
\setcounter{figure}{0}
\renewcommand{\thefigure}{H\arabic{figure}}
The comparison between the boundary and bulk constructions reveals a characteristic tension beyond the early-time regime. On the boundary side, the Pad\'e continuation of the Krylov complexity is built from a finite number of early-time coefficients and its local logarithmic slope \eqref{gammafunction} decreases below the quadratic value $\gamma=2$ at intermediate times and subsequently turns upward, apparently approaching $2$ again. 
On the bulk side, the generalized \emph{CAny} observable is evaluated at successive truncation orders of the local functional $a_1$. The corresponding logarithmic slopes agree remarkably well with the boundary Pad\'e approximant up to times of order $t\sim\beta$, but eventually depart from it and approach $\gamma=1$, as expected from the linear late-time growth of any finite truncation. This raises the question whether this late-time disagreement arises because the bulk functional is truncated at finite order or, conversely, since the asymptotic behavior of a  Pad\'e approximant can be strongly influenced by the relative degrees of its numerator and denominator, whether the return of the Pad\'e logarithmic slope toward $2$ is itself trustworthy. 
To isolate these two effects, we introduce a solvable toy model designed to reproduce the qualitative structure of this comparison. We want to incorporate the two properties suggested by the BTZ construction and mentioned in the Appendix \ref{sec::bulk_d}. From the numerical fit of the \(\lambda_n\) in the previous section, we know that the bulk data suggest an odd-sector tail proportional to \(1/n\) which is in agreement with the late quadratic time growth of complexity and the fact that the maximal BTZ slice approaches its final value exponentially in time. Let
\begin{equation}
    h(t)
    =
    \sqrt{\mu_1^2+t^2}-\mu_1,
    \qquad
    g(t)
    =
    \sqrt{\mu_2^2+t^2}+\mu_1,
    \label{eq:toy_h_g_definitions}
\end{equation}
with \(\mu_1,\mu_2>0\), and define
\begin{equation}
    u(t)
    =
    1-e^{-\kappa h(t)},
    \qquad
    \kappa>0.
    \label{eq:toy_compact_variable}
\end{equation}
such that the approach to the endpoint is exponential, in direct analogy with the approach of the BTZ extremal surface to the final slice. Since \(h(t)\geq0\), the compact variable satisfies
\begin{equation}
    0\leq u(t)<1,
    \qquad
    u(0)=0,
    \qquad
    \lim_{t\to\infty}u(t)=1.
    \label{eq:toy_q_range}
\end{equation}
We now introduce the basis functions
\begin{equation}
    B_n(t)
    =
    g(t)u(t)^n.
    \qquad
    n=1,2,\ldots .
    \label{eq:toy_exponential_basis}
\end{equation}
This basis is chosen to reproduce precisely the two structural properties relevant for the bulk problem. At early times, the $n$th basis element first contributes at order $t^{2n}$, reproducing the triangular structure of the higher-curvature expansion. At late times, however, every individual basis element grows only linearly. We can now define the \(N\)th truncation target function as
\begin{equation}
    f_N(t)
    = \sum_{n=1}^{N} \alpha_n B_n(t).
    \label{eq:toy_finite_truncation}
\end{equation}
where the coefficients $\alpha_n$ are fixed by requiring the early-time expansion of $f_N(t)$ to agree with that of $f_{\mathrm{test}}(t)$ up to the corresponding order. The latter can be obtained by sending \(N \rightarrow \infty\) and resumming the series. Choosing the coefficients \(\alpha_n = \frac{1}{n},\) the fully resummed target function is then
\begin{equation}
\begin{aligned}
     f_{test}(t)
    &=
    \sum_{n=1}^{\infty}
    \alpha_n B_n(t)
    \\
    &=
    g(t)
    \sum_{n=1}^{\infty}
    \frac{u(t)^n}{n}
    \\
    &=
    \kappa g(t)h(t).
\end{aligned}
\label{eq:toy_exact_resummation}
\end{equation}
We notice that the target function grows quadratically both at early and late times, such that its logarithmic slope develops a broad minimum below $2$ before slowly returning to its asymptotic value. Indeed, calling $\gamma_{\mathrm{test}}(t)$ the exact logarithmic slope and \( \gamma_N(t) = t\frac{f_N'(t)}{f_N(t)}\) the truncated one, then for every finite $N$, \(u(t)\to1\) while the harmonic
partial sum remains finite. Therefore,
\begin{equation}
    f_N(t)
    \underset{t\to\infty}{\sim}
    H_N t,
    \qquad
    H_N
    =
    \sum_{n=1}^{N}\frac{1}{n},
    \label{eq:toy_truncated_late}
\end{equation}
which implies \( \lim_{t\to\infty}
    \gamma_N(t)
    =
    1
    \) for every finite \(N\). Nevertheless, the exact target function behaves as 
\begin{equation}
    f_{test}(t)
    \underset{t\to\infty}{\sim}
    \kappa t^2
    \label{eq:toy_exact_late}
\end{equation}
at late times. Hence \(\lim_{t\to\infty}  \gamma_{test}(t)= 2.\) The apparent contradiction is resolved by first observing that the two limits $N\to\infty$ and $t\to\infty$ do not commute:
\begin{equation}
    \lim_{N\to\infty}
    \lim_{t\to\infty}
    t\frac{f_N'(t)}{f_N(t)}
    =
    1,
    \qquad
    \lim_{t\to\infty}
    \lim_{N\to\infty}
    t\frac{f_N'(t)}{f_N(t)}
    =
    2.
    \label{eq:toy_noncommuting_limits}
\end{equation}
Therefore, the infinite coefficient sequence \(\alpha_n\) must generate a sufficiently strong singularity at the late-time limit.
The mechanism responsible for the additional late-time power is the logarithmic singularity generated by the harmonic coefficient tail,
\begin{equation}
    \sum_{n=1}^{\infty}
    \frac{u^n}{n}
    =
    -\log(1-u).
    \label{eq:toy_logarithmic_singularity}
\end{equation}
Since \(1-u(t)\sim e^{-\kappa t}\), this singularity becomes linear in
time,
\begin{equation}
    -\log\!\left[1-u(t)\right]
    =
    \kappa h(t)
    \sim
    \kappa t.
    \label{eq:toy_log_linear_time}
\end{equation}
Multiplication by \(g(t)\sim t\) then produces the quadratic asymptotic growth of the exact resummation. If the sequence \(\alpha_n\) were
absolutely summable, the generating function would remain finite at \(u=1\), and the linear late-time growth would persist even after the
infinite-order limit. 
This toy model therefore provides a concrete example in which a sequence of finite-order observables, each possessing linear late-time growth, can approximate over an increasingly large time interval a function whose fully resummed asymptotics is instead quadratic. It follows that the tendency of the finite-order bulk curves toward $\gamma=1$ is not, by itself, sufficient to establish that the complete bulk functional must have the same asymptotic behavior. A non-summable \(1/n\) tail can generate a logarithmic final-slice singularity whose resummation supplies the additional power of time required for \(\gamma\to2\). In this way, we are showing that the return of Pad\'e continuation toward the quadratic value is compatible with an infinite-order bulk completion even though every finite truncation remains asymptotically linear. Fig.~\ref{fig:toymodel} illustrates this mechanism.  At low and intermediate times, the truncated curves approach the exact result increasingly well as the order of truncation is raised, and the point at which they depart from the target is pushed toward later times. Nevertheless, every finite truncation eventually turns toward $\gamma_N=1$, whereas the exact function returns to $\gamma_{\mathrm{test}}=2$. The quadratic asymptotics is recovered only after the complete infinite series is resummed.
\begin{figure}[h]
    \centering    \includegraphics[width=0.7\columnwidth]{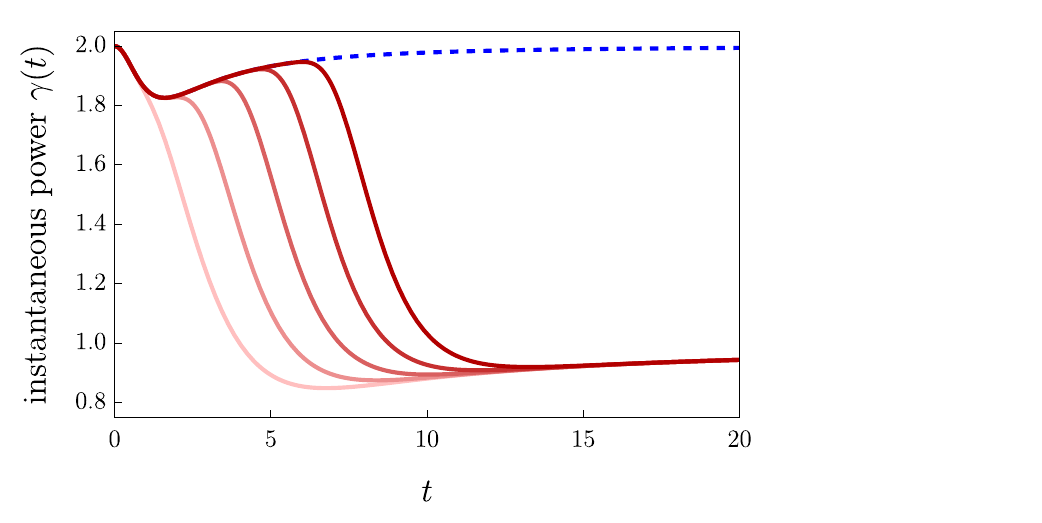}
    \caption{Comparison between the exact resummed toy-model target and its finite truncations. We use the parameter choice $\mu_1=1$, $\mu_2=19/10$, and $\kappa=1$, for which the exact curve develops a broad dip and then slowly returns toward the quadratic value $\gamma=2$. The exact resummation is shown in dashed blue, while the truncations $N=4,16,64,256,1024$ are shown by the red curves with the colour changing continuously from light to dark red as
 \(N\) increases. Each finite truncation initially follows the exact curve but eventually turns toward the late-time value $\gamma=1$, illustrating the non-commutativity of the large-$N$ and late-time limits.}
    \label{fig:toymodel}
\end{figure}
\end{document}